**Title: Reconstruction of Global Groundwater Storage Anomaly from GRACE Satellite Gravimetry Using a Physics-Constrained Decomposition LSTM**

**Authors:** Arpon Sarkar[1*], Showmitra Kumar Sarkar[2], Mafrid Haydar[3]

[1]Graduate Student, Department of Urban and Regional Planning, Khulna University of Engineering & Technology (KUET), Khulna-9203, Bangladesh. [arghosarkar79@gmail.com] (https://orcid.org/0009-0006-1722-4782)

[*]Corresponding author

[2]Assistant Professor, Department of Urban and Regional Planning, Khulna University of Engineering & Technology (KUET), Khulna-9203, Bangladesh. [mail4dhrubo@gmail.com, https://orcid.org/0000-0001-5928-7577]

[3]Assistant Professor, Department of Urban and Regional Planning, Khulna University of Engineering & Technology (KUET), Khulna-9203, Bangladesh.
[mafrid@urp.kuet.ac.bd] (https://orcid.org/0009-0003-9229-0171)

**Corresponding author details**

Email: arghosarkar79@gmail.com

ORCID: https://orcid.org/0009-0006-1722-4782

**Declarations:**

***Funding:*** None

***Conflicts of interest:*** The authors declare that they have no known competing financial interests or personal relationships that could have appeared to influence the work reported in this paper.

***Ethical Approval:*** The authors declare that no ethical approval required. Ethical approval for this type of study is not required by our institute.

# Reconstruction of Global Groundwater Storage Anomaly from GRACE Satellite Gravimetry Using a Physics-Constrained Decomposition LSTM

**Abstract:**

Groundwater is the largest accessible store of freshwater on Earth, yet its depletion has been difficult to quantify globally, because satellite gravimetry senses total water storage rather than groundwater alone, the record is coarse and interrupted, and the models used to isolate groundwater underestimate the largest trends. Here, a continuous and physically bounded record of global groundwater storage change was reconstructed for April 2002 to December 2022 at half-degree resolution. Total water storage from the GRACE and GRACE-FO missions was separated from its land-surface components, and a physics-constrained decomposition network learned the groundwater signal as the sum of a trend, a seasonal cycle, and a residual. The trend of every cell was held within the net moisture that the atmosphere can supply, and the mass balance of each river basin was enforced during training, so the learned field could not violate water availability. The reconstruction reproduced withheld observations with a root mean squared error of 6.97 centimeters and a Nash-Sutcliffe efficiency of 0.69, and against an independent network of monitoring wells it achieved a lower trend error than an independent hydrological model and a data-assimilation product at every aggregation scale tested. Gross groundwater depletion over significantly declining cells was estimated at 358 ± 12 cubic kilometers per year, while net storage change across all land cells was -36 ± 18 cubic kilometers per year, equivalent to 0.10 ± 0.05 millimeters per year of sea-level rise. The associated mass redistribution implies a drift of the Earth's rotational pole of 6.9 centimeters per year toward 68.6 degrees East, within 4.5 degrees of the direction inferred in an earlier study.

## 1. Introduction

Groundwater is the largest accessible store of unfrozen freshwater on Earth, and it underpins the water security of a large share of humanity. It supplies drinking water to billions of people, sustains rivers and groundwater-dependent ecosystems through dry seasons, and provides the buffer that keeps irrigated agriculture productive when rainfall fails (Famiglietti, 2014; Taylor et al., 2013). The resource responds slowly and is hidden from view, and for that reason it has often been treated as if it were inexhaustible, so that in many of the world's most productive agricultural regions it has been withdrawn faster than it is replenished (Bierkens & Wada, 2019; Wada et al., 2010). Water-balance accounting has estimated that non-renewable abstraction rose steeply through the late twentieth century, concentrated in a small number of heavily irrigated aquifers, and the groundwater footprint of those systems, the area required to sustain their abstraction, was shown to exceed their actual extent many times over, which is a direct measure of how far use has moved beyond supply (Gleeson et al., 2012).

The consequences of sustained over-abstraction are now visible across continents. Water tables have fallen by more than a meter per year in parts of India, the North China Plain, the Middle East, and the western United States, and the decline has accelerated across a substantial fraction of the world's aquifers during the twenty-first century (Jasechko et al., 2024; Richey et al., 2015). Falling water tables dry wells, drive land subsidence, allow seawater to intrude into coastal aquifers, and reduce the baseflow that carries rivers through the dry season (Konikow, 2011; Rodell et al., 2018). The lost water does not disappear. Much of it is delivered in the end to the ocean, so groundwater depletion has become a measurable term in the global sea-level budget (Konikow, 2011; Wada et al., 2017). An accurate and continuous account of where groundwater is being lost, and how fast, is therefore a prerequisite for water management and for Earth-system understanding alike.

That account has been hard to produce, because in-situ monitoring networks are sparse, unevenly distributed, and rarely coordinated at the global scale (Jasechko et al., 2024). The route around this limitation came from orbit. It was shown before launch that time variations in the Earth's gravity field could reveal the redistribution of water mass across the land and the ocean (J. Wahr et al., 1998), and the Gravity Recovery and Climate Experiment (GRACE) and its Follow-On mission turned that principle into measurement from 2002 onward, providing the first direct record of month-to-month mass change in the Earth system and, with it, an estimate of total terrestrial water storage over large regions (Tapley et al., 2004, 2019). The early spherical-harmonic solutions were later reprocessed into mass-concentration, or mascon, products that suppress leakage and restore the amplitude lost in processing (Save et al., 2016; Watkins et al., 2015; Wiese et al., 2016), and gain factors were derived to correct for the smoothing that the processing imposes (Landerer & Swenson, 2012). The instrument does not see groundwater directly, however. It senses the combined mass of every store in a vertical column, so the groundwater component must be separated from soil moisture, snow, surface water, and canopy storage before it can be read as an aquifer signal. That separation has usually been made by subtracting the other stores as estimated by land-surface models, an approach established when groundwater loss in northern India was recovered by removing modeled soil moisture from GRACE total water storage (Rodell et al., 2009) and since applied and refined across many settings (Chen et al., 2016; Frappart & Ramillien, 2018). Comparisons in intensively monitored regions such as the Central Valley of California proved the value of the method and exposed its weakness at the same time, because the recovered groundwater signal is only ever as trustworthy as the modeled stores taken away from it (Döll et al., 2012; Scanlon et al., 2012).

Three limitations follow from this design and bear directly on the present work. The first is resolution. The native footprint of GRACE is of the order of a few hundred kilometers, which blurs the boundaries of individual aquifers and mixes their signals with those of their surroundings. The second is continuity. A gap of roughly one year separates the GRACE and GRACE-FO records, and it falls precisely where a long-term trend is most vulnerable to interruption. The third, and the most consequential for interpretation, is bias in the subtraction itself. A systematic comparison found that global hydrological models substantially underestimate the largest declining and rising storage trends relative to the satellite observations, so any groundwater estimate built on a model subtraction inherits that error (Scanlon et al., 2018). Reconstructions have been built to close the gap and extend the record, yet they have reconstructed total water storage rather than groundwater. Statistical approaches have tied total storage to climate forcing and produced gap-free fields at half-degree resolution that match or exceed global models (Humphrey et al., 2016; Humphrey & Gudmundsson, 2019), and deep-learning approaches have reconstructed total storage either by learning the mismatch between a land-surface model and GRACE (A. Y. Sun et al., 2019) or by predicting storage directly over the global land surface and its major basins (Z. Sun et al., 2020). These studies established that data-driven models can reproduce total storage with real skill, but none isolated the groundwater component, none wrote the physical limits on storage change into the model as structural constraints, and none carried the uncertainty of the storage conversion through to the trends themselves. The long-running debate over whether early gravimetric estimates overstated depletion in northern India made the last of these points plainly, because it showed that the step from a mass change to an aquifer storage change is itself a leading source of uncertainty (Long et al., 2016).

Machine learning offered the flexibility such a reconstruction needs, because the links between storage and its climatic and physical drivers are strongly non-linear (Reichstein et al., 2019). The long short-term memory network, introduced to capture long-range dependence in sequential data (Hochreiter & Schmidhuber, 1997), has been especially effective in hydrology, where it has matched or surpassed calibrated process models for rainfall-runoff prediction and has learned behavior that transfers across large samples of catchments (Kratzert et al., 2018, 2019). Its ability to carry the memory of earlier conditions forward in time suits the delayed response of storage to past moisture supply, which is central to the problem addressed here. A purely data-driven model can fit the observed record and still drift into physically implausible states once it is pushed beyond that record, and physics-informed and physics-guided learning arose to prevent exactly this, by embedding conservation laws or process knowledge so that predictions stay consistent with known physics (Jiang et al., 2020; Read et al., 2019). Within the GRACE setting, deep learning has been used to fuse the satellite data with land-surface models and to bridge the mission gap (A. Y. Sun et al., 2019), and a self-supervised data-assimilation model with a physically motivated loss has produced high-resolution total water storage while closing the water balance at the basin scale (Gou & Soja, 2024). Data assimilation provides a parallel path, folding GRACE into land-surface models to sharpen groundwater and drought estimates (Girotto et al., 2016; Li et al., 2019; Zaitchik et al., 2008). The reconstruction reported here draws on both traditions, but it enforces the physical limits on storage change as explicit structural constraints rather than as soft penalties alone.

Whatever the model, the final test of a reconstruction rests on independent observation, and the raw material for that test has improved sharply. A global analysis of about 170,000 monitoring wells showed that rapid groundwater decline is widespread and often accelerating, and it supplied aquifer-system boundaries useful for spatial aggregation (Jasechko et al., 2024), while the release of a large, harmonized, analysis-ready compilation of groundwater time series with matched Earth-system variables made a genuinely global, quality-controlled validation network available for the first time (Bäthge et al., 2026). Turning a measured water-level change into a storage change requires the specific yield of the aquifer, and that conversion remains one of the largest and least constrained terms in groundwater accounting, because specific yield varies with lithology, depth, and the state of the water table and is seldom measured at the scale of a satellite footprint (Long et al., 2016; Scanlon et al., 2012). Gridded global datasets have only recently begun to make the conversion tractable at continental scale (T. Sun et al., 2026). Beyond the wells, the redistribution of water mass leaves fingerprints in the planet itself that offer a further basis for comparison. Land-water storage change modulates the rate of global mean sea-level rise (Reager et al., 2016; Wada et al., 2017), and it perturbs the Earth's rotation, since the excitation of polar motion by a surface load is well understood and climate-driven water redistribution has been shown to drive a measurable drift of the rotational pole (Adhikari & Ivins, 2016). A later study argued that the observed pole drift confirms groundwater depletion as a significant contributor to sea-level rise, closing the loop between aquifer loss and a planetary geodetic signal (Seo et al., 2023).

A clear gap remained across all of this. No global reconstruction had isolated groundwater storage as a continuous and physically constrained field, one in which the long-term trend of each location was held inside the water that the atmosphere can supply and the mass balance of each river basin was respected, while a transparent budget separated the errors of the satellite measurement, the learning model, and the storage properties of the aquifer. Nor had any reconstruction been carried through to its planetary consequences within a single internally consistent framework. Here, groundwater storage anomalies were isolated and reconstructed across the global land surface at half-degree resolution for April 2002 to December 2022, by combining GRACE-derived total water storage with land-surface stores in a physics-constrained decomposition network. The network learned the signal as the sum of a trend, a seasonal cycle, and a residual, so that each part could be constrained and read on its own. The trend of every cell was bounded by the net atmospheric water supply and the mean of each basin was held to observed storage change, so that the learned field could not violate water availability. The product was validated against an independent global network of monitoring wells and against independent hydrological benchmarks, none of which entered the learning, and a full uncertainty budget was propagated that separated the observational, model, and specific-yield contributions and tested the assumption that dominates it. The recovered depletion was then translated into its consequences for the sea-level budget and for the drift of the Earth's rotational pole, so that satellite gravimetry, land-surface physics, in-situ hydrology, and geodesy were made to converge on a single hidden quantity.

# 2. Methods and materials

## 2.1 Study Domain

The study was set over the global land surface, because the question it addressed, how much groundwater has been lost and where, is inherently planetary in scope and cannot be answered from any single region (Famiglietti, 2014). Within that global frame, the analysis was concentrated on the small number of aquifer systems that dominate the world's groundwater loss and that therefore carry most of the signal the study sought to recover. These included the alluvial aquifers of the Indus and Ganges plains of northern India and Pakistan, where abstraction for irrigation is among the most intense on Earth (Rodell et al., 2009); the North China Plain, where decades of pumping have drawn down water tables across a major grain-producing region; the aquifers of the Middle East and the Caspian margin; and the High Plains and Central Valley of the United States (Scanlon et al., 2012). Global assessments have repeatedly identified these systems as being under high or extreme stress, with abstraction outpacing recharge (Jasechko et al., 2024; Richey et al., 2015; Wada et al., 2010). The physical character of these settings shaped the analysis. They are predominantly semi-arid to arid, so recharge is limited and episodic while irrigation demand is high, and they are underlain by thick and permeable aquifers whose storage responds to abstraction over years to decades rather than over single seasons (Bierkens & Wada, 2019; Taylor et al., 2013). This slow, cumulative response is what a long satellite record is able to resolve, and it is what makes a physically constrained reconstruction meaningful in these regions.

The observational setting was defined by the availability of two independent data streams over a common period. Satellite gravimetry from the GRACE and GRACE-FO missions provided continuous, near-global coverage of storage change from 2002 onward (Tapley et al., 2004), while ground-based monitoring, drawn together in the GROW compilation, provided the in-situ records against which the reconstruction was later tested (Bäthge et al., 2026). The analysis window of April 2002 to December 2022 was chosen to coincide with the interval of densest well reporting and with the common availability of all input datasets, so that the satellite and ground-based evidence overlapped as fully as possible. The global reconstruction domain, the monitoring networks, and the basis of the analysis window are shown in Fig. 1.

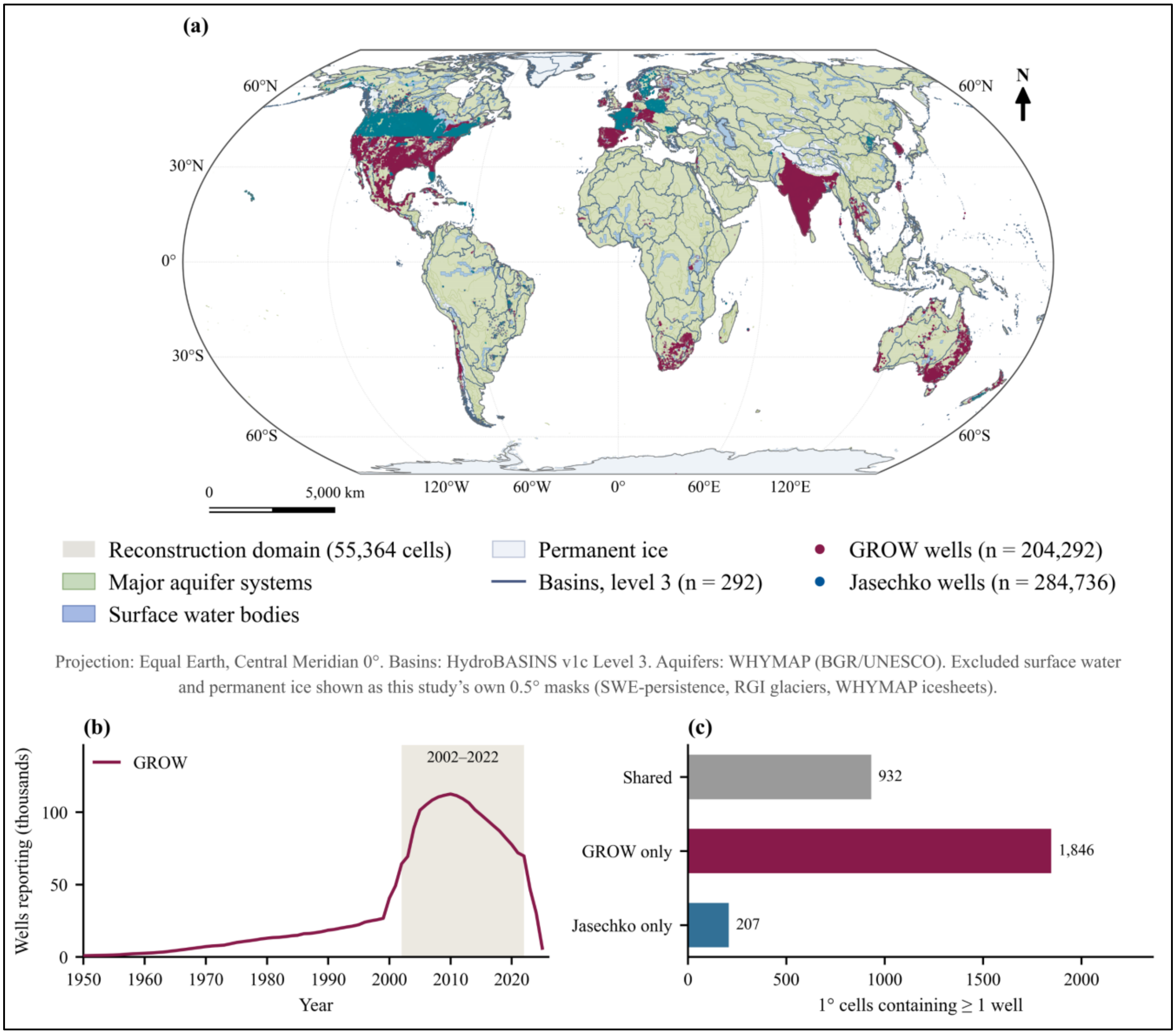

**Figure 1: Study domain, monitoring network and the basis of the analysis window.** The global reconstruction domain of 55,364 half-degree cells is shown in Equal Earth projection over the major aquifer systems from WHYMAP and the HydroBASINS level-3 boundaries, with the permanent-ice and surface-water exclusions drawn from this study's own half-degree masks, shown in panel a. The two candidate monitoring compilations are plotted separately, and neither entered the model; the GROW network was adopted for validation because it covered 1,846 one-degree cells that the alternative did not, against only 207 unique to the alternative, in panel c. The shaded band marks the 2002 to 2022 window, which encloses the maximum of the well-reporting curve and is the empirical justification for the period in panel b.

## 2.2 Data acquisition

The observational core of the reconstruction was the Level-3 mascon solution from the Gravity Recovery and Climate Experiment and its Follow-On mission (GRACE and GRACE-FO). The JPL RL06.3Mv04 CRI-filtered product was used, expressed as liquid water equivalent thickness on a half-degree grid together with its accompanying gain factors and formal measurement uncertainty (Landerer & Swenson, 2012; Watkins et al., 2015; Wiese et al., 2016). This field carried the total change in stored water, from which the groundwater component was later separated.

The non-groundwater stores were taken from the GLDAS Noah land-surface model, version 2.1, which supplied four soil-moisture layers, snow water equivalent, and canopy interception (Rodell et al., 2004). The atmospheric and vegetation forcings that carry the climatic signals relevant to storage change were drawn from several products. Precipitation, shallow soil water, and total runoff came from the ERA5-Land reanalysis, and two-meter air temperature from ERA5 (Hersbach et al., 2020; Muñoz-Sabater et al., 2021). Actual evaporation was taken from GLEAM version 4.3a (Miralles et al., 2011, 2025), and the normalized difference vegetation index from the MODIS MOD13C2 Collection 6.1 product with its pixel-reliability screening applied (Didan, 2021).

The static descriptors located each cell in its physical setting. Subsurface permeability and porosity were taken from GLHYMPS 2.0 (Gleeson et al., 2014; Huscroft et al., 2018), the equipped-irrigation area and the groundwater-irrigated fraction from the FAO Global Map of Irrigation Areas version 5 (Siebert et al., 2005, 2010), surface elevation and slope from ETOPO 2022 (MacFerrin et al., 2025; NOAA NCEI, 2022), and an urban fraction with base year 2000 to represent the built environment (Gao & O'Neill, 2020). Basin and aquifer boundaries for aggregation and for the mass-conservation target were taken from HydroBASINS level 3 (Lehner & Grill, 2013), from the (Jasechko et al., 2024) aquifer delineation, and from the WHYMAP Groundwater Resources of the World map (Richts et al., 2011). The complete inventory of inputs, with sampling grids, periods, and references, is given in Table 1, which should be read as the single authoritative record of provenance for the study.

**Table 1: Datasets used in the reconstruction, validation, and uncertainty analysis.** Sampling grids and periods are as acquired; all fields were harmonized to a 0.5 degree monthly grid spanning April 2002 to December 2022.

| # | Dataset and version | Variable(s) or role | Sampling grid | Period | Reference and data identifier |
|---|---|---|---|---|---|
| **Time-varying model inputs** | | | | | |
| 1 | JPL GRACE / GRACE-FO mascon, CRI-filtered, RL06.3Mv04 | Liquid water equivalent thickness | 0.5° grid (3° equal-area mascons), monthly | Apr 2002 – Dec 2022 | (Landerer & Swenson, 2012; Watkins et al., 2015; Wiese et al., 2016). PO.DAAC 10.5067/TEMSC-3JC634 |
| 2 | GLDAS Noah L4, v2.1 | Four soil-moisture layers, snow water equivalent, canopy water | 0.25°, monthly | 2000 – 2022 | (Rodell et al., 2004);<br>GES DISC 10.5067/SXAVCZFAQLNO |
| 3 | ERA5-Land monthly averaged reanalysis | Total precipitation, volumetric soil water layer 1 | 0.1°, monthly | 2000 – 2022 | (Muñoz-Sabater et al., 2021); CDS 10.24381/cds.68d2bb30 |
| 4 | ERA5-Land monthly averaged reanalysis | Total runoff | 0.1°, monthly | 2002 – 2022 | (Muñoz-Sabater et al., 2021); CDS 10.24381/cds.68d2bb30 |
| 5 | ERA5 monthly averaged reanalysis | 2 m air temperature | 0.25°, monthly | 2002 – 2022 | (Hersbach et al., 2020); CDS 10.24381/cds.f17050d7 |
| 6 | GLEAM v4.3a | Actual evaporation | 0.1°, monthly | 2002 – 2022 | (Miralles et al., 2011, 2025) |
| 7 | MODIS MOD13C2, Collection 6.1 | NDVI, pixel-reliability screened | 0.05°, monthly | 2002 – 2022 | (Didan, 2021).<br>10.5067/MODIS/MOD13C2.061 |
| **Static model inputs** | | | | | |
| 8 | GLHYMPS 2.0 | Log-permeability, porosity | polygon to 0.5° | static | (Gleeson et al., 2014; Huscroft et al., 2018). Borealis 10.5683/SP2/TTJNIU |
| 9 | FAO Global Map of Irrigation Areas v5 | Equipped-irrigation area, groundwater-irrigated fraction | 5 arc-min to 0.5° | static | (Siebert et al., 2005, 2010) |
| 10 | ETOPO 2022 | Elevation, slope | 1 arc-minute to 0.5° | static | (NOAA NCEI, 2022), 10.25921/fd45-gt74; (MacFerrin et al., 2025) |
| 11 | Urban fraction, base year 2000 | Urban fraction | 0.125° to 0.5° | static | (Gao & O'Neill, 2020), 10.1038/s41467-020-15788-7 |
| **Aggregation and masking layers** | | | | | |
| 12 | HydroBASINS v1c, level 3 | 292 basins globally; 233 carried a mass-conservation target | polygon | — | (Lehner & Grill, 2013) |
| 13 | WHYMAP Groundwater Resources of the World (2008 ed., 1:25,000,000) | Aquifer polygons (3,424 features) | polygon | — | (Richts et al., 2011);<br>[BGR & UNESCO (2008)] |
| 14 | Randolph Glacier Inventory, version 6.0 | Glacier outlines for the permanent-ice mask (4,620 cells) | polygon | — | (RGI Consortium, 2017),10.7265/N5-RGI-60 |
| 15 | Jasechko et al. (2024) aquifer boundaries | 1,693 aquifer-system polygons for spatial aggregation | polygon | — | (Jasechko et al., 2024), 10.1038/s41586-023-06879-8; Zenodo 10.5281/zenodo.10003697 |
| **Validation and evaluation** | | | | | |
| 16 | GROW | Primary monitoring-well source (204,292 sites) | point | — | (Bäthge et al., 2026), 10.1038/s41597-026-06966-1; Zenodo 10.5281/zenodo.15149480 |
| 17 | Jasechko et al. (2024) annual depth to groundwater | Coverage comparison in Figure 1 only; not used for validation | point | — | (Jasechko et al., 2024); Zenodo 10.5281/zenodo.10003697 |
| 18 | Global specific-yield field | Converts well-level change to | gridded | — | (T. Sun et al., 2026),<br>10.1038/s41467-026-73521-2; |
| | | storage for temporal validation | | | figshare 10.6084/m9.figshare.31669225 |
| 19 | WaterGAP 2.2d (histsoc) | Independent benchmark (groundwater storage) | 0.5°, monthly | 1901 – 2016 | (Müller Schmied et al., 2021) |
| 20 | GRACE-SeDA (self-supervised data assimilation) | Processing-consistency check, not independent | 0.5°, monthly | 2002 – 2019 | (Gou & Soja, 2024), 10.1038/s44221-024-00194-w; data: 10.3929/ethz-b-000648738 |
| **Cartography** | | | | | |
| 21 | Natural Earth | Coastlines and surface water in Figure 1 | vector | — | Natural Earth (public domain), https://www.naturalearthdata.com |
| 22 | Esri World Countries (Generalized) | Country outlines in figures | vector | — | ESRI |

*Note. All native resolutions were regridded to the 0.5 degree monthly analysis grid before use.*

## 2.3 Reliability masking and the analysis domain

Not every land cell carries a groundwater signal that a gravity mission can resolve, so a reliability mask was constructed and applied before any learning took place. Permanent ice was flagged as the union of three independent criteria: 451 cells whose minimum monthly snow water equivalent across the study period remained persistently high, 4,620 cells intersecting glacier outlines from the Randolph Glacier Inventory, and 27,682 cells lying within the WHYMAP ice-sheet polygons (RGI Consortium, 2017). Masking, rather than subtraction, was used, so that flagged cells were removed from analysis rather than adjusted, which avoided introducing a model-derived correction into the observational record. Although the union of the ice criteria covered 31.4 per cent of the land surface, only 3,366 of those cells fell inside the resolvable gravimetric domain, so the effective cost of the ice mask on the analysis was 4.85 per cent.

A land template derived from the ETOPO surface confined every mask to land. Of roughly 56,000 land cells, 55,364 held at least sixty months of usable GRACE observation and formed the reconstruction domain. A further set of cells was flagged as permanently inundated surface water, where the gravimetric signal reflects lakes, reservoirs, or wetlands rather than aquifer storage; of these, 2,937 lay inside the modelling domain and were excluded, leaving 52,427 cells for training and evaluation. Spatial aggregation later used two polygon layers, the HydroBASINS level-3 basins, of which 233 of the 292 global polygons intersected the domain and carried a mass-conservation target, and the WHYMAP aquifer outlines. The spatial extent of the domain, together with the monitoring networks and the analysis window, is shown in Figure 1.

## 2.4 Isolating the groundwater storage signal

The groundwater storage anomaly was isolated from total water storage by removing the stores that GRACE senses but that are not groundwater. For each month the total water storage anomaly was formed from the GRACE liquid water equivalent thickness after application of the mascon gain factor (Eq. 1), where s is the per-cell gain factor that restores the amplitude lost to spherical-harmonic processing (Landerer & Swenson, 2012).

$$\mathrm{TWSA}(t) = \eta(t) \cdot s \tag{1}$$

where TWSA(t) is the total water storage anomaly of the cell in month t (cm), η(t) is the GRACE liquid water equivalent thickness (cm), and s is the dimensionless per-cell gain factor that restores the amplitude lost to spherical-harmonic processing (Landerer & Swenson, 2012).

Every input was expressed as an anomaly about a common 2004 to 2009 climatology, so that a field was compared only against its own average behavior for the same calendar month (Eq. 2).

$$X'(t) = X(t) - X_{\mathrm{clim}}\big(m(t)\big) \tag{2}$$

where X(t) is any storage field in month t, X′(t) is its anomaly, X_clim(m) is the mean of X over the 2004 to 2009 baseline for calendar month m, and m(t) ∈ {1, …, 12} is the calendar month of time step t.

The groundwater component was then defined as the total anomaly minus the anomalies of the four soil-moisture layers, snow water equivalent, and canopy interception (Eq. 3). This closed-water-column subtraction is the standard basis for recovering groundwater from gravimetry (Rodell et al., 2009). The GLDAS fields were aggregated from their native quarter-degree grid to the half-degree analysis grid by block averaging, and were converted to a common unit of centimeters of water equivalent.

$$\mathrm{GWSA}(t) = \mathrm{TWSA}'(t) - [\,\mathrm{SM}'(t) + \mathrm{SWE}'(t) + \mathrm{CI}'(t)\,] \tag{3}$$

where GWSA(t) is the groundwater storage anomaly (cm), TWSA′(t) is the total water storage anomaly, and SM′(t), SWE′(t) and CI′(t) are the anomalies of soil moisture (summed over the four

GLDAS layers), snow water equivalent and canopy interception (all cm), each defined as in Eq. 2, following the closed-column subtraction of  (Rodell et al., 2009).

A physical cap of plus or minus 100 centimeters was imposed on the isolated field to guard against isolated excursions produced by residual leakage or by short-lived model artifacts (Eq. 4). The cap was applied after differencing and clipped only a very small fraction of cell-months, which confirmed that it acted as a safeguard rather than as an active constraint on the recovered signal. Months without a valid GRACE solution were left as gaps and were not interpolated at this stage, so that the learning could later be restricted to genuine observations. Diagnostic checks over four regions with well-established behavior showed that the recovered signal declined over the intensively abstracted plains and remained near zero or positive over the humid tropics, which established that the isolation had produced a physically sensible field before any model was fitted.

$$\mathrm{GWSA(t)} \leftarrow \mathrm{clip(GWSA(t),\ -100,\ +100)\ cm} \tag{4}$$

where clip(·, −100, +100) limits the value to the interval [−100, +100] cm; the bounds are a physical safeguard adopted in this study and are not a fitted parameter.

To let the physics act on the correct budget, the removed stores were retained as a separate other-storage anomaly, and the identity linking the two components to the observed total was preserved exactly (Eq. 5). Because total water storage is the sum of groundwater and this other-storage term, the water balance governs the total rather than groundwater alone. The identity was verified to float precision on every uncapped cell-month before the model was allowed to proceed, which guaranteed that the atmospheric supply constraint introduced later acted on a quantity consistent with the GRACE observation itself.

$$\mathrm{TWSA(t)} = \mathrm{GWSA(t)} + \mathrm{OSA(t)}, \qquad \mathrm{OSA(t)} = \mathrm{SM}' + \mathrm{SWE}' + \mathrm{CI}' \tag{5}$$

where OSA(t) is the other-storage anomaly, the sum of the non-groundwater stores, so the identity holds by construction and the water balance governs TWSA rather than groundwater alone.

## 2.5 Harmonization and assembly of the model grid

All inputs were brought onto a common half-degree grid spanning April 2002 to December 2022, a total of 249 monthly steps. Fields of finer native resolution were aggregated to the target grid by averaging the source values falling within each destination cell, and the ERA5-Land accumulations, stored as mean daily rates, were converted to monthly totals using the number of days in each month. Evaporation was constrained to be non-negative, the groundwater-irrigated fraction was set to zero wherever the equipped-irrigation area was itself zero, and the vegetation index was filled in the small number of screened or missing months using a monthly climatology with an annual-mean fallback, with the observed status carried as a separate channel. Where isolated gaps in the groundwater field were filled, the filled months were flagged so that they could be excluded from every learning and evaluation split. No monitoring-well information of any kind entered the assembled predictor set, which kept the validation strictly downstream of the learning. The assembled and screened domain comprised the 52,427 cells retained after the surface-water exclusion.

## 2.6 Predictor screening

Six static fields (a seventh, slope, was derived from elevation) and six dynamic fields were assembled onto the model grid, and two forms of redundancy were then removed before learning. The static descriptors were screened for spatial multicollinearity using the variance inflation factor computed on standardized fields (O'brien, 2007). Nine candidate descriptors were tested, including the two geographic coordinates, and all returned low values well below the conventional cautionary threshold, with the highest inflation carried by the coordinates. The coordinates were then removed deliberately, so that the network could not learn position as a proxy for physical process and was instead required to rely on the measurable properties of each cell. Seven static descriptors were carried forward, six assembled directly and slope derived from the elevation field: permeability, porosity, elevation, slope, equipped-irrigation area, the groundwater-irrigated fraction, and urban fraction.

The dynamic fields required a targeted intervention. Precipitation and runoff were very strongly collinear, each returning a variance inflation factor more than twenty times the threshold of ten, because runoff is largely a lagged and routed expression of precipitation. Retaining both as separate inputs would have made the fitted relationships unstable, so the two were combined into a single net-moisture input defined as precipitation minus runoff (Eq. 6). After this substitution every dynamic predictor fell below the threshold, and the substituted field retained a physically interpretable spatial pattern, exceeding a hundred millimeters per month over the equatorial rainforest belts and approaching zero over the major deserts. Five dynamic predictors remained: the net-moisture input, evaporation, air temperature, the vegetation index, and shallow soil moisture. Shallow soil moisture was taken from ERA5-Land rather than from the GLDAS fields used in Eq. 3, so that the soil-moisture store removed from total water storage was not also supplied to the network as an input. The same water balance also defines the net atmospheric supply that later bounds the trend, namely the net-moisture input minus evaporation.

$$\frac{\mathrm{d\,TWSA}}{\mathrm{dt}} \approx \mathrm{P} - \mathrm{E} - \mathrm{R}, \qquad \text{with} \quad \mathrm{F} = \mathrm{P} - \mathrm{R} \tag{6}$$

where P, E and R are the precipitation, actual evaporation and total runoff of the cell (mm $\mathrm{month}^{-1}$), $\frac{\mathrm{d\,TWSA}}{\mathrm{dt}}$ is the rate of change of total storage, and F is the net-moisture input used as a predictor; the terrestrial water balance follows (Rodell et al., 2004).

The twelve retained predictors, seven static and five dynamic, formed the complete input to the learning model. Direct examination of each predictor against groundwater storage showed weak and strongly non-linear relationships, with the linear and information-based association measures diverging for several fields, which was the empirical justification for a non-linear sequence model in place of multiple linear regression.

## 2.7 The physics-constrained decomposition network

The reconstruction model was a recurrent decomposition network that predicted the groundwater storage anomaly for every cell from predictors alone, with no observation of the target ever presented as an input. Its structure is shown in Figure 2 and had four parts: an input representation, a shared sequence encoder, three additive output pathways, and a set of physical constraints imposed during training.

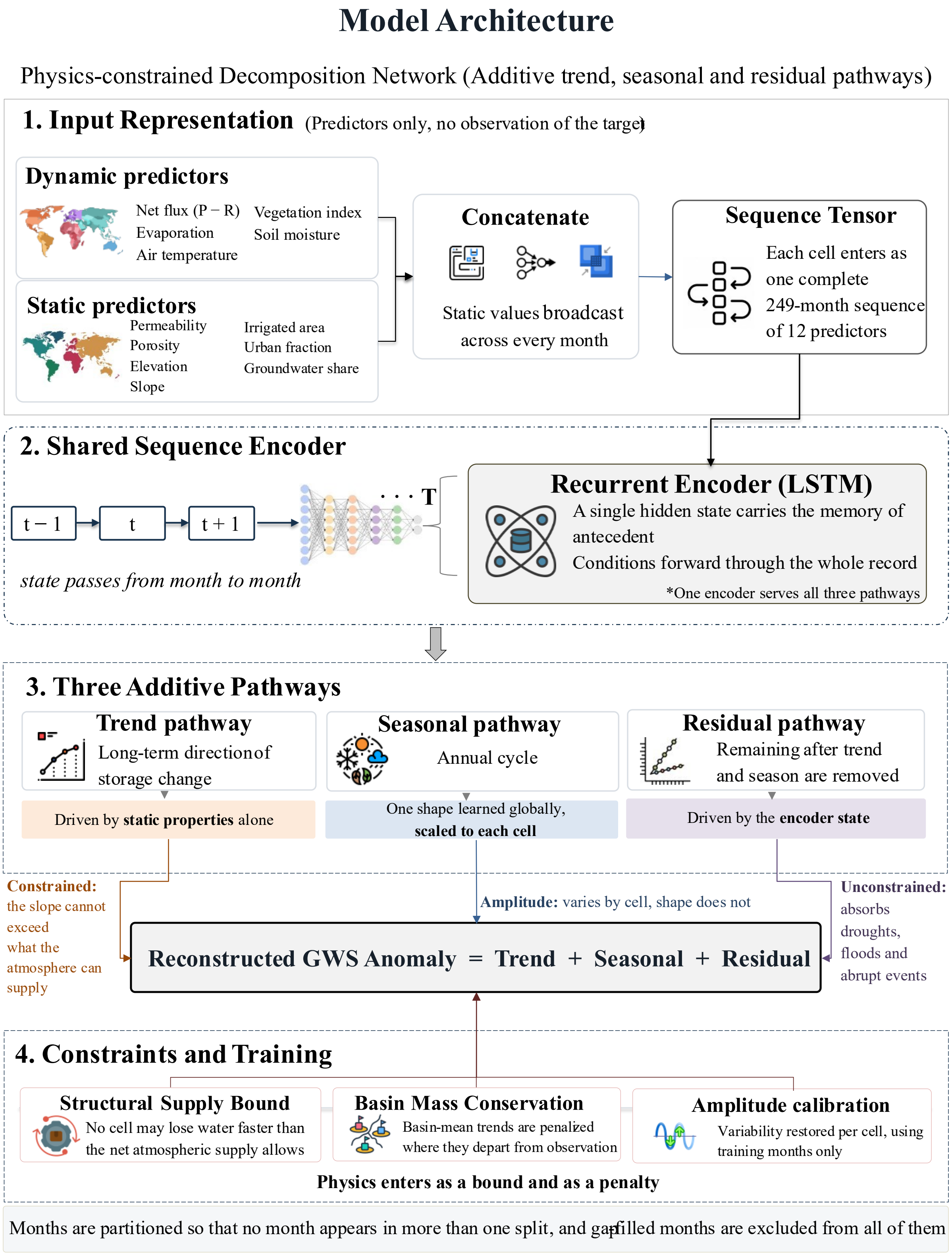

**Figure 2: Architecture of the physics-constrained decomposition network.** Each cell enters as one complete 249-month sequence of twelve predictors, with no observation of the target. A single

recurrent encoder feeds three additive pathways: a trend driven by static properties alone and bounded by the net atmospheric supply, a seasonal cycle of fixed global shape and per-cell amplitude, and an unconstrained residual that absorbs droughts, floods, and other transient events. Physics enters both as the structural bound on the trend and as a soft basin mass-conservation penalty applied during training.

### *2.7.1 Input representation and shared encoder*

Every predictor was standardized to zero mean and unit variance before entering the model (Eq. 7), so that fields of very different physical units contributed on a comparable scale.

$$\tilde{x} = \frac{x - \mu}{\sigma} \tag{7}$$

where x is a predictor value, μ and σ are its mean and standard deviation across the domain, and $\tilde{x}$ is the standardized predictor.

Each cell then entered the model as one complete sequence of 249 monthly steps carrying its twelve predictors, with the seven static descriptors broadcast unchanged across every month and concatenated to the five dynamic fields. The sequence was passed through a single recurrent encoder, a long short-term memory network with one hidden layer of sixty-four units (Eq. 8), whose hidden state carried the memory of antecedent conditions forward through the whole record and so represented the delayed response of storage to earlier moisture supply (Hochreiter & Schmidhuber, 1997). One encoder served all three output pathways, so the same learned representation of the cell history informed each component of the reconstruction.

$$h_i(t) = \mathrm{LSTM}\left(\left[\, \tilde{x}_i^{\mathrm{dyn}}(t)\,;\ \tilde{x}_i^{\mathrm{stat}} \,\right]\right) \tag{8}$$

where i indexes the cell, $h_i(t)$ is the hidden state of the recurrent encoder at month t, $\tilde{x}_i^{dyn}(t)$ is the vector of standardized dynamic predictors, $\tilde{x}_i^{stat}$ is the vector of standardized static predictors, [· ; ·] denotes concatenation, and LSTM(·) is the long short-term memory network of (Hochreiter & Schmidhuber, 1997) with sixty-four hidden units.

### *2.7.2 Three additive pathways*

The reconstructed storage anomaly was formed as the sum of three interpretable components (Eq. 9), so that each could be constrained and read on its own.

$$\hat{y}_i(t) = \mathrm{Tr}_i(t) + \mathrm{Se}_i(t) + \mathrm{Re}_i(t) \tag{9}$$

where ŷ_i(t) is the reconstructed groundwater storage anomaly of cell i in month t (cm), and $Tr_i(t)$, $Se_i(t)$ and $Re_i(t)$ are its trend, seasonal and residual components respectively; this additive decomposition is original to the present study.

The trend pathway represented the long-term direction of storage change and was driven by the static descriptors alone, which tied secular behavior to the physical setting of each cell rather than to any incidental feature of the record. The slope of each cell was produced by a small feed-forward network of the static predictors, passed through a bounded hyperbolic tangent, and multiplied by the centered time axis in years (Eq. 10), where the per-cell bound b is the structural supply term defined in Eq. 13. The seasonal pathway represented the annual cycle as a single twelve-value shape learned globally and then scaled by a per-cell amplitude, so that the phase of the cycle was shared across the domain while its magnitude varied by cell (Eq. 11). The residual pathway was a linear read-out of the encoder state (Eq. 12), left unconstrained so that it could absorb droughts, floods, and abrupt events without contaminating the secular term.

$$\mathrm{Tr_i}(t) = u_i \cdot \tau(t), \qquad u_i = b_i \cdot \tanh\left(f_\theta\left(\tilde{x}_i^{stat}\right)\right), \qquad \tau(t) = \frac{t - \bar{t}}{12} \tag{10}$$

where $u_i$ is the per-cell trend slope (cm $yr^{-1}$), $\tau(t)$ is the time axis centered on its mean $\bar{t}$ and expressed in years, $f_\theta(\cdot)$ is a feed-forward network of the static predictors with parameters θ, tanh(·) is the hyperbolic tangent that bounds the slope before scaling, and $b_i$ is the structural supply bound of Eq. 13; the formulation is original to this study.

$$\mathrm{Se_i}(t) = A_i \cdot g[m(t)], \qquad A_i = h_\omega\left(\tilde{x}_i^{stat}\right), \qquad g \in \mathbb{R}^{12} \tag{11}$$

where $A_i$ is the per-cell seasonal amplitude (cm), $h_\omega(\cdot)$ is a feed-forward network of the static predictors with parameters ω, $g \in \mathbb{R}^{12}$ is a single global twelve-value monthly shape shared by all cells, and g[m(t)] selects its entry for the calendar month m(t); the formulation is original to this study.

$$\mathrm{Re_i}(t) = \mathbf{w}_r^{\top} h_i(t) + c_r \tag{12}$$

where $Re_i(t)$ is the residual component (cm), $h_i(t)$ is the encoder hidden state of Eq. 8, $w_r$ is a learned weight vector with transpose $w_r^{\top}$, and $c_r$ is a learned bias; the residual is a linear read-out of the encoder state, introduced in this study.

### *2.7.3 The structural supply bound*

The trend pathway was bounded by physics rather than left free. The long-term rate of storage change in any cell cannot exceed the net atmospheric moisture that the cell receives, because water that is not supplied cannot be lost indefinitely. A per-cell supply term was formed from the time-mean net atmospheric supply and converted to centimeters per year, and the bound was set to a safety multiple of its magnitude plus a small additive floor (Eq. 13), with the safety multiplier equal to 1.5 and the floor equal to 1 centimeter per year. The bounded tangent in Eq. 10 then held the slope inside this envelope by construction, so that no cell could drift faster than its atmospheric supply allowed. The bound had a median across the domain of about eight centimeters per year, and because it was structural it held for every configuration of the model and was never traded against fit.

$$S_i = \frac{12}{10} \cdot \langle F_i - E_i \rangle_t, \qquad b_i = \nu |S_i| + b_0 \tag{13}$$

where $S_i$ is the mean net atmospheric supply of cell i (cm yr$^{-1}$), ⟨·⟩_t denotes the average over all months, $F_i$ and $E_i$ are the net-moisture input and evaporation (mm month$^{-1}$, the factor 12/10 converting mm month$^{-1}$ to cm yr$^{-1}$), $b_i$ is the supply bound (cm yr$^{-1}$), $\nu = 1.5$ is a dimensionless safety multiplier, and $b_0 = 1$ cm yr$^{-1}$ is an additive floor; the bound is original to this study.

### *2.7.4 Basin mass conservation and weight selection*

A second, softer constraint enforced agreement with observed storage change at the basin scale. Once per training epoch, the mean predicted trend was computed over the whole domain for each basin carrying a conservation target, and its departure from the observed basin-mean trend was penalized in the mean square (Eq. 14), with 233 basins carrying such a target. Because a basin's cells are scattered across the domain, this term was evaluated over all of a basin's cells at once rather than within random training batches, which was what allowed the constraint to bind rather than dissolve into noise.

$$L_{basin} = \frac{1}{B} \sum_b \left( \frac{1}{|C_b|} \sum_{i \in C_b} u_i - T_b^{obs} \right)^2 \tag{14}$$

where $L_{basin}$ is the basin mass-conservation penalty, $B = 233$ is the number of basins carrying a target, $C_b$ is the set of cells in basin b and $|C_b|$ its size, $u_i$ is the trend slope of Eq. 10, and $T_b^{obs}$ is the observed mean groundwater trend of basin b (cm yr$^{-1}$); the penalty is original to this study.

The weight on this penalty was not assumed but selected by a sensitivity sweep over five values, each trained for a short run, with held-out error, efficiency, the fraction of drifting cells, and the basin-closure error recorded for every value. The response was not monotonic, so the choice was made on the empirical evidence of the sweep. A weight of one half was adopted because it was the only value that improved all four diagnostics at once, and it was retained on the strength of its basin-closure improvement, which was the largest and most consistent of the four responses.

### *2.7.5 Training protocol*

Learning was performed only on genuine observations. The observed months were partitioned into training, validation, and test sets in a seventy, fifteen, fifteen split, with no month appearing in more than one partition, and the gap-filled months were excluded from all three so that skill was assessed against real data alone. The data-fidelity term was an observation-weighted mean squared error evaluated on the training months (Eq. 15), and it was minimized together with the basin-conservation penalty under a single objective (Eq. 16) using the Adam optimizer with gradient clipping and a small weight decay (Goodfellow et al., 2016). The weight w carries the observation mask, taking the value one on observed training months and zero elsewhere.

$$L_{data} = \frac{\sum_{i,t} w_{it} \left(\hat{y}_i(t) - y_i(t)\right)^2}{\sum_{i,t} w_{it}} \qquad (15)$$

where $L_{data}$ is the data-fidelity loss, $y_i(t)$ is the observed groundwater storage anomaly, $\hat{y}_i(t)$ is the reconstruction of Eq. 9, and w_it is the observation mask, equal to one on observed training months and zero otherwise.

$$L = L_{data} + \lambda L_{basin}, \qquad \lambda \in \{0, 0.1, 0.25, 0.5, 1.0\}, \qquad \lambda^{\star} = 0.5 \qquad (16)$$

where L is the total training objective, $\lambda$ is the weight on the basin penalty, and $\lambda\star = 0.5$ is the value adopted after the sensitivity sweep; the objective was minimized with the Adam optimizer under gradient clipping and a small weight decay (Goodfellow et al., 2016).

At the selected weight the model was trained to convergence over a longer run with early stopping, and the validation loss fell monotonically to its minimum with no upturn, which indicated that training was stopped before overfitting began. Predictor influence was then quantified by permutation importance, in which each field in turn was shuffled across cells and the resulting increase in held-out error was recorded, with the physically primary drivers emerging as the most important.

## 2.8 Amplitude calibration and the final trend field

Training under a squared-error loss is known to compress the amplitude of variability, and the reconstruction indeed underestimated the standard deviation of storage relative to the observations. A per-cell variance-matching calibration was therefore applied. For each cell a multiplicative factor was set to the ratio of the observed to the reconstructed standard deviation, with an additive offset chosen to preserve the mean, and both were fitted exclusively on the observed training months so that the withheld months remained independent of the correction (Eq. 17). A global fallback was used for the small number of cells with too few observations to fit a stable per-cell factor. The median expansion factor was close to one and a third, consistent with a modest and largely uniform amplitude deficiency.

$$a_i = \frac{\sigma_i^{obs}}{\sigma_i^{rec}}, \qquad c_i = \mu_i^{obs} - a_i\,\mu_i^{rec}, \qquad \tilde{y}_i(t) = a_i\,\hat{y}_i(t) + c_i \tag{17}$$

where $a_i$ and $c_i$ are the per-cell multiplicative and additive calibration coefficients, $\sigma_i^{obs}$ and $\sigma_i^{rec}$ are the standard deviations of the observed and reconstructed series over the training months, $\mu_i^{obs}$ and $\mu_i^{rec}$ their means, and $\tilde{y}_i(t)$ is the amplitude-calibrated reconstruction (cm); the variance-matching calibration is this study's own.

Because each multiplicative factor is positive, the calibration preserves the sign of every cell's trend by construction, and the spatial pattern of the trends remained strongly correlated across the correction. The calibration therefore rescaled amplitude without altering the direction or the spatial structure of long-term change. Trends were then estimated on the calibrated series using the Theil-Sen slope, which is robust to outliers and to short gaps (Eq. 18) (Sen, 1968; Theil, 1992); and confidence intervals were obtained from two hundred block-bootstrap replicates resampled over whole years, so that the temporal correlation within each year was respected (Efron & Tibshirani, 1994; Künsch, 1989). A cell was flagged as carrying a significant trend where its bootstrap interval excluded zero. This calibrated, trend-resolved field was frozen as the citable product, and no monitoring well was permitted to modify it.

$$T_i = \underset{j<k}{\text{median}}\left[\frac{\tilde{y}_i(t_k) - \tilde{y}_i(t_j)}{t_k - t_j}\right] \tag{18}$$

where $T_i$ is the Theil-Sen trend estimate of cell i (cm $yr^{-1}$), taken as the median slope over all time-step pairs ($t_j$, $t_k$) with $j < k$ (Sen, 1968; Theil, 1992).

## 2.9 Validation strategy

Validation was kept strictly downstream of the learning, and no well observation entered the predictor set at any point. Three complementary tests were applied, each drawing on a different well population defined by the statistic it supported. Trend validation required wells with a robust long-term record, temporal validation required only enough observations to estimate a correlation, and benchmarking compared the reconstruction against two independent gridded products. The monitoring source was the GROW compilation, adopted for its wider spatial coverage than the alternative candidate network (Bäthge et al., 2026; Jasechko et al., 2024).

For trend validation a storage trend was derived for each qualifying well, with negative values denoting storage loss, which was required to carry an annual record spanning at least eight years, and the corresponding product trend was extracted over exactly the interval spanned by that well's own record rather than over a common period, so that every comparison was window matched. Wells were retained where they achieved at least twenty-four months of overlap with the reconstruction, and aggregation polygons were scored only where at least three wells were present. Agreement was evaluated at four nested scales, from the individual cell through aquifer and basin polygons to the largest basins, using inverse-variance weighting. Significance was deliberately not used as a screen, because filtering on it would have biased the retained sample toward large magnitudes. Skill was summarized by the Pearson correlation and the root mean squared error, alongside the mean bias (product minus well), the percentage of directional sign agreement, the Nash-Sutcliffe efficiency and the Kling-Gupta efficiency defined below (Eqs. 19 and 20; Nash & Sutcliffe, 1970; Gupta et al., 2009).

$$\mathrm{NSE} = 1 - \frac{\sum_t (\hat{y}_t - y_t)^2}{\sum_t (y_t - \bar{y})^2} \tag{19}$$

where NSE is the Nash-Sutcliffe efficiency, $y_t$ and $\hat{y}_t$ are the observed and reconstructed values, and $\bar{y}$ is the mean of the observations (Nash & Sutcliffe, 1970).

$$\mathrm{KGE} = 1 - \sqrt{(r-1)^2 + (\alpha-1)^2 + (\beta-1)^2}, \qquad \alpha = \frac{\sigma_{sim}}{\sigma_{obs}}, \qquad \beta = \frac{\mu_{sim}}{\mu_{obs}} \tag{20}$$

where KGE is the Kling-Gupta efficiency, r is the Pearson correlation between the reconstructed (sim) and observed (obs) series, α is the ratio of their standard deviations, and β the ratio of their means (Gupta et al., 2009).

The same procedure was applied to the two benchmark products: a physically based global hydrological model that uses no GRACE information and is therefore fully independent (Müller Schmied et al., 2021), and a GRACE data-assimilation field that shares the observational source and was treated throughout as a processing-consistency check rather than as independent validation (Gou & Soja, 2024). For the direct comparison between the two benchmarks, both fields were re-windowed to their common overlap so that neither was penalized for a shorter record.

For temporal validation, well water-level records were converted to storage using an external global specific-yield field that resolved every well in the validation set (T. Sun et al., 2026), aggregated within each polygon, and correlated month by month against the area-averaged reconstruction. This pathway required only six observations per well and at least three wells and eight overlapping months per polygon, which admitted a far larger population than the trend test. The individual sites within the best-performing polygons were confirmed to be spatially distributed rather than clustered, which established that the reported agreement rested on a network rather than on a few dominant sites.

## 2.10 Uncertainty decomposition

A full uncertainty budget was constructed by propagating three independent components of the trend error separately: an observation term, a model term, and a storage-property term. Each was derived from data or from a stated and tested assumption rather than from an arbitrary inflation factor.

The observation term was taken from the GRACE mascon uncertainty field, with a floor set to the lower quartile of the positive values applied to any cell reporting zero, so that no cell was assigned an implausibly perfect measurement. The model term was represented as a monthly error with a standard deviation of 7.05 cm, marginally above the final held-out root mean squared error of the calibrated reconstruction (6.973 cm, Section 3.7), so that the model component of the budget is slightly conservative. The storage-property term expressed the uncertainty in converting a satellite-derived change in water-equivalent thickness into an aquifer storage change, and was applied as a relative error scaled by the magnitude of the local trend (Eq. 21), taking the smaller value in cells containing at least one high-quality well and the larger value elsewhere. Both values were adopted as assumptions, with the imputed value placed inside the published range for proxy-derived specific yield.

$$\sigma_i^{Sy} = |T_i| \cdot \varepsilon_i, \qquad \varepsilon_i = 0.15 \text{ (well-constrained) or } 0.40 \text{ (imputed)} \tag{21}$$

where $\sigma_i^{Sy}$ is the specific-yield contribution to the trend uncertainty (cm yr$^{-1}$), $T_i$ is the trend of Eq. 18, and $\varepsilon_i$ is the assumed relative error of the specific yield, taking 0.15 where at least one high-quality well constrains the cell and 0.40 otherwise; the parameterization is this study's own.

The observation and model terms were propagated jointly through a Monte Carlo procedure. In each of two hundred realizations, spatially correlated but temporally independent Gaussian noise was added to the calibrated series and the trend was refitted by ordinary least squares (Eq. 22). The spread of the refitted trends across realizations gave the combined observation-and-model trend uncertainty, which was then partitioned into its observation and model parts in proportion to their respective variances (Eq. 23).

$$y_i^{(k)}(t) = \tilde{y}_i(t) + \left(\sigma_i^{obs}(t) + \sigma^{mod}\right) \cdot n_i^{(k)}(t) \tag{22}$$

where $y_i^{(k)}(t)$ is the k-th Monte Carlo realization of the series, $\tilde{y}_i(t)$ is the calibrated reconstruction, $\sigma_i^{obs}(t)$ is the GRACE observational uncertainty of the cell-month (cm), $\sigma^{mod}$ is the monthly model-error standard deviation set slightly above the held-out RMSE, and $n_i^{(k)}(t)$ is a unit-variance noise field, spatially correlated at the range of Eq. 26 and independent in time.

$$\sigma_i^{obs} = \sigma_i^{(o+m)}\sqrt{\phi_i}, \qquad \sigma_i^{mod} = \sigma_i^{(o+m)}\sqrt{1-\phi_i}, \qquad \phi_i = \frac{\left\langle \left(\sigma_i^{obs}\right)^2 \right\rangle}{\left\langle \left(\sigma_i^{obs}\right)^2 \right\rangle + \left(\sigma^{mod}\right)^2} \tag{23}$$

where $\sigma_i^{(o+m)}$ is the combined observation-and-model trend uncertainty from the spread of the Monte Carlo trends, $\phi_i$ is the observational variance share, and $\sigma_i^{obs}$ and $\sigma_i^{mod}$ are the observation and model parts obtained by partitioning $\sigma_i^{(o+m)}$ in proportion to their variances; the partition is this study's own.

The three components were then combined in quadrature to give the total trend uncertainty (Eq. 24), from which a signal-to-noise ratio was formed for every cell, and a trend was declared significant where that ratio reached the conventional threshold (Eq. 25).

$$\sigma_i = \sqrt{\left(\sigma_i^{obs}\right)^2 + \left(\sigma_i^{mod}\right)^2 + \left(\sigma_i^{Sy}\right)^2} \tag{24}$$

where $\sigma_i$ is the total trend uncertainty of cell i (cm yr$^{-1}$), combining the observation, model and specific-yield terms in quadrature.

$$\mathrm{SNR}_i = \frac{|T_i|}{\sigma_i} \geq 1.96 \tag{25}$$

where $SNR_i$ is the signal-to-noise ratio, $T_i$ the trend, $\sigma_i$ the total uncertainty, and 1.96 the two-sided ninety-five per cent normal threshold above which a trend was declared significant.

The spatial correlation of the noise was not assumed but measured. An empirical variogram of the trend field (Eq. 26) returned a correlation range of about six degrees, equivalent to roughly six hundred and sixty kilometers and to an effective correlation area of about a hundred and forty cells (Eq. 27). This length scale governed both the Monte Carlo noise and every later aggregation to basin and global totals (Cressie, 1993).

$$\gamma(h) = \frac{1}{2N(h)} \sum_{d_{ij} \approx h} \left(V_i - V_j\right)^2 \tag{26}$$

where γ(h) is the empirical semivariance at lag distance h, N(h) is the number of cell pairs separated by approximately h, and V_i and V_j are the trend values of the paired cells; the range at which γ(h) reaches its sill defines the correlation length (Cressie, 1993).

$$N_{corr} = \left(\frac{L}{\Delta}\right)^2, \qquad \varepsilon^* = \frac{1}{1.96} \approx 0.51 \tag{27}$$

where $N_{corr}$ is the effective number of cells per correlation area, L is the variogram range (about 6°), $\Delta = 0.5°$ is the grid spacing, and $\varepsilon^*$ is the specific-yield relative error at which a storage-property-dominated cell crosses the significance threshold; both relations are derived in this study. Because the storage-property term dominated the budget wherever the trend was large, and because it was set by assumption, its adopted value was tested directly rather than accepted. The significance of the trends was recomputed across a dense sweep of the imputed relative error, which revealed a sharp transition near the value of about 0.51 that equals the reciprocal of the significance threshold (Eq. 27). For a storage-property-dominated cell the signal-to-noise ratio reduces to the reciprocal of the relative error and is therefore almost independent of the trend magnitude, so that such cells crossed the threshold together at that point. This behavior is a property of the significance test rather than of the data.

The assessment was then repeated at the correlation-area scale by aggregating cells into blocks matched to the six-degree range (Eq. 28) and re-evaluating significance for each block. Two explicit bounds were carried. The conservative bound treated the storage-property error as fully correlated within a block, so that it did not average down (Eq. 29). The independent bound treated it as uncorrelated within a block, so that it averaged with the square root of the number of cells (Eq. 30). The observation and model error were held fully correlated within a block in both cases, so that only the specific-yield assumption differed between the two curves. The true significant fraction was bracketed by these bounds and lay nearer the independent one, because a lithology-based proxy varies from cell to cell within any six-hundred-kilometer block. The significance of the depletion at the correlation-area scale was therefore stable across the whole plausible range of the storage-property assumption, even where it was sensitive at the individual-cell scale.

$$T_B = \frac{\sum_{i \in B} w_i \; T_i}{\sum_{i \in B} w_i}, \qquad w_i = sin\theta_i \tag{28}$$

where $T_B$ is the area-weighted mean trend of block B, $T_i$ is the cell trend, and $w_i$ is the area weight, equal to the sine of the colatitude $\theta_i$ of cell i (equivalently the cosine of its latitude).

$$\sigma_B^{Sy,cons} = |T_B| \cdot \bar{\varepsilon}_B, \qquad \bar{\varepsilon}_B = \frac{\sum w_i \, \varepsilon_i}{\sum w_i} \tag{29}$$

where $\sigma_B^{Sy,cons}$ is the specific-yield uncertainty of block B under the conservative assumption that the error is fully correlated within the block, and $\bar{\varepsilon}_B$ is the area-weighted mean relative error of its cells; this bound is defined in the present study.

$$\sigma_B^{Sy,ind} = \frac{\sqrt{\sum_{i \in B} (w_i \, \varepsilon_i |T_i|)^2}}{\sum_{i \in B} w_i} \tag{30}$$

where $\sigma_B^{Sy,ind}$ is the specific-yield uncertainty of block B under the independent assumption that the error is uncorrelated between cells, so that it averages down within the block; this bound is defined in the present study.

Every cell was finally assigned to one of two reliability tiers, a well-constrained tier where at least one high-quality well allowed the storage coefficient to be checked, and a physics-extrapolated tier elsewhere. The tier flag was carried in the released product, together with the per-cell significance flag, so that users could restrict analysis to the more strongly constrained subset if they chose.

## 2.11 Derived global diagnostics

The calibrated trend field was translated into three global consequences, each computed from the same per-cell mass change but answering a distinct question. The mass rate of each cell was obtained from its trend, its area, and the density of water (Eq. 31).

$$\dot{m}_i = -\,\rho_w \, A_i \, \dot{h}_i \tag{31}$$

where $\dot{m}_i$ is the mass rate of change of cell i (kg yr$^{-1}$), $\rho_w = 1000$ kg m$^{-3}$ is the density of water, $A_i$ is the cell area (m$^2$), and $\dot{h}_i$ is the storage trend expressed as a rate of water-height change (m yr$^{-1}$); the negative sign makes depletion a mass loss (Konikow, 2011).

Two volumetric quantities were then distinguished and never used interchangeably (Eq. 32). Gross depletion was defined as the sum of volume loss over cells carrying a significant declining trend and expressed the extent and severity of aquifer loss. Net storage change was defined as the sum over all valid cells, allowing recharging cells to offset depleting ones, and was the physically correct quantity for the sea-level budget. The net exchange was converted to a sea-level equivalent by distributing it over the ocean area (Eq. 33). The uncertainty on these totals was aggregated under the six-degree correlation length rather than under an assumption of independent cells, which was untenable given the variogram result, and the larger, correlated figure was reported.

$$D_{gross} = \sum_{i:sig,\,T_i<0} A_i \left|\dot{h}_i\right|, \qquad D_{net} = \sum_{i:valid} A_i\ \dot{h}_i \tag{32}$$

where $D_{gross}$ is the gross depletion, summed over cells with a significant declining trend, here "significant" denotes cells whose Theil–Sen trend has a block-bootstrap confidence interval excluding zero. The stricter full-budget significance of section 3.11 is carried separately as the per-cell reliability flag and $D_{net}$ is the net storage change, summed over all valid cells (both km$^3$ yr$^{-1}$); the two are kept distinct throughout.

$$SLE = -\frac{D_{net}}{A_{ocean}} \tag{33}$$

where SLE is the sea-level equivalent (mm yr$^{-1}$), $D_{net}$ is the net storage change, and $A_{ocean}$ is the global ocean area, so that a net loss of land water produces a positive sea-level contribution (Konikow, 2011).

The same mass field was converted into a perturbation of the Earth rotational pole. The per-cell mass rate was accumulated into the two off-diagonal products of inertia (Eq. 34), where a is the Earth radius and the colatitude and longitude of each cell are written as theta and phi. These were converted to a polar-motion excitation using the standard rotational-deformation constants (Eq. 35), and the excitation was then reduced to a secular drift rate and azimuth (Eq. 36). The adopted constants were the Earth radius of 6.371 by ten to the sixth meters, the term C minus A of 2.6398 by ten to the thirty-fifth kilogram meters squared, and the load Love number of minus 0.301 (Adhikari & Ivins, 2016; J. M. Wahr, 1985).

$$\dot{I}_{13} = -a^2 \sum_i \dot{m}_i \, sin\theta_i cos\theta_i cos\varphi_i, \qquad \dot{I}_{23} = -a^2 \sum_i \dot{m}_i \, sin\theta_i cos\theta_i sin\varphi_i \tag{34}$$

where $\dot{I}_{13}$ and $\dot{I}_{23}$ are the rates of change of the two off-diagonal products of inertia (kg $m^2$ $yr^{-1}$), a is the Earth's radius, $\dot{m}_i$ is the cell mass rate of Eq. 31, $\theta_i$ is the colatitude and $\varphi_i$ the longitude of cell i (J. M. Wahr, 1985).

$$\frac{d\chi_{1,2}}{dt} = \left[\frac{(1 + k_2{'})\kappa}{C - A}\right] \dot{I}_{13,23}, \qquad \kappa = \left(1 - \frac{k_2}{k_s}\right)^{-1} \approx 1.43 \tag{35}$$

where $d\chi_1/dt$ and $d\chi_2/dt$ are the polar-motion excitation rates, $k_2' = -0.301$ is the load Love number of the induced deformation, $k_2$ and k_s are the fluid and secular Love numbers whose ratio defines $\kappa \approx 1.43$, and C and A are the Earth's polar and equatorial moments of inertia with $C - A = 2.6398 \times 10^{35}$ kg $m^2$ (Adhikari & Ivins, 2016; J. M. Wahr, 1985).

$$\dot{p} = a\sqrt{\left(\frac{d\chi_1}{dt}\right)^2 + \left(\frac{d\chi_2}{dt}\right)^2}, \qquad \psi = arctan\left(\frac{d\chi_2/dt}{d\chi_1/dt}\right) \tag{36}$$

where $\dot{p}$ is the secular drift speed of the rotational pole, $a = 6.371 \times 10^6$ m is the Earth's radius, and $\psi$ is the drift azimuth measured east of the Greenwich meridian (Adhikari & Ivins, 2016).

Three masks were evaluated to test robustness, spanning depletion-only, significant-net, and all-valid definitions, and the inferred direction was insensitive to the choice, so the significant-net mask was adopted as the physically appropriate one for a mass-redistribution calculation. The secular estimate was verified against a month-by-month simulation of the pole path, computed from the same field, which agreed closely in both rate and azimuth, and was compared against an earlier pole-drift study of the same phenomenon in both direction and magnitude (Seo et al., 2023).

## 2.12 Reconstruction workflow

The reconstruction was built in a fixed order that follows the flow of information. Time-varying and static inputs were first acquired and brought onto a common grid. A reliability mask was applied to remove cells where the gravimetric signal does not represent liquid groundwater. The groundwater component was then separated from total water storage, the predictors were screened for redundancy, and a physics-constrained learning model was trained on the observed months alone before its amplitude was calibrated. The product was compared against an independent hydrological model, a data-assimilation field, and a global network of monitoring wells that entered the analysis only at the validation stage. A full uncertainty budget was then propagated, and the recovered depletion was finally translated into its consequences for the sea-level budget and for the orientation of the Earth rotational pole. The complete workflow is shown in Fig. 3, and the internal structure of the learning model is shown previously in Fig. 2.

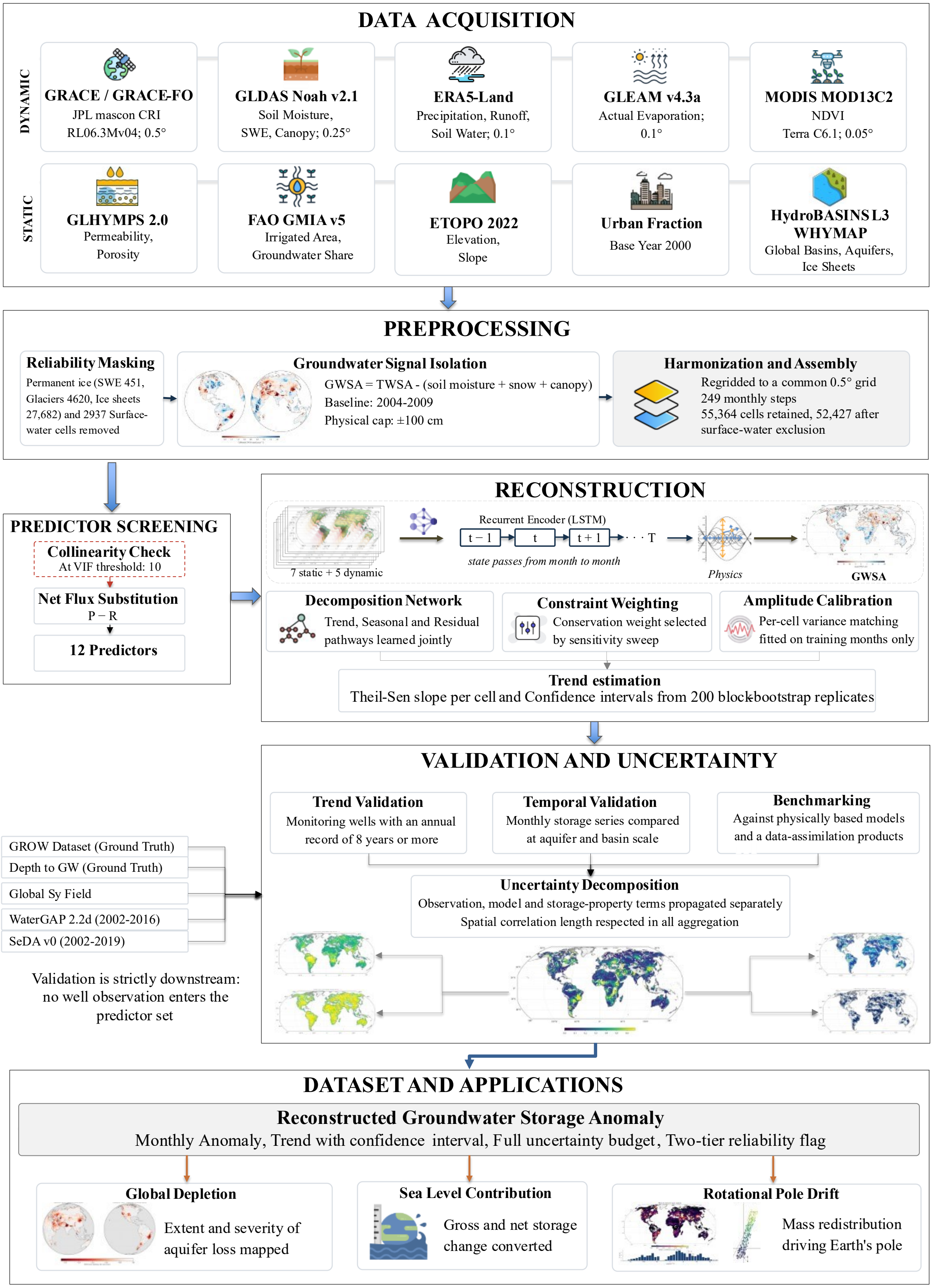

**Figure 3: Overview of the reconstruction workflow.** The pipeline proceeds from acquisition of the dynamic and static inputs, through reliability masking, groundwater-signal isolation, and harmonization onto a common half-degree grid, to predictor screening, the physics-constrained learning model, and the validation and uncertainty stages. Well observations enter only downstream of the learning, so no ground measurement can inform the reconstruction it is later used to test. The released product carries a monthly anomaly, a trend with confidence intervals, a full uncertainty budget, and a two-tier reliability flag.

# 3. Results

## 3.1 Reconstruction domain and observational basis

The reconstruction was performed on a global grid of 0.5 degree cells for April 2002 to December 2022, a span of 249 monthly steps. Of approximately 56,000 land cells, 55,364 held at least 60 months of usable GRACE observation and were retained. A reliability mask was then applied. Permanent ice was identified from three sources and produced 451 cells flagged by persistent snow water equivalent, 4,620 cells intersecting glacier outlines, and 27,682 cells lying within ice sheet polygons. The union of these covered 31.4 per cent of the land surface, but only 3,366 of the flagged cells fell inside the true GRACE analysis domain, so the effective cost of the ice mask was 4.85 per cent. A further 3,284 cells were flagged as permanently inundated surface water, of which 2,937 lay inside the modeling domain and were removed. A total of 52,427 cells remained for training and evaluation.

Spatial aggregation used two polygon layers. Basin boundaries were taken from the HydroBASINS level 3 map, of which 292 polygons exist globally and 233 intersected the domain and carried a mass conservation target. Aquifer outlines were taken from the WHYMAP Groundwater Resources of the World map.

Three distinct well populations were used and were not interchangeable. The catalogue population comprised every georeferenced site on file. The trend validation population was restricted to sites carrying an annual record of at least eight years, because a robust trend cannot be fitted to a shorter one. The temporal validation population required only six observations, because a correlation between two series can be estimated from far fewer points than a trend. The temporal population was therefore larger than the trend population despite being applied later in the analysis, and this ordering followed from the statistic being computed rather than from any inconsistency. Each population is stated explicitly in Table 2.

**Table 2: Well counts used at each validation stage.**

| Stage | Criterion applied | Wells retained |
|---|---|---|
| Catalogue, GROW timeseries | Georeferenced site on file | 204,292 |
| Catalogue, Jasechko et al. | Georeferenced site on file | 284,736 |
| Trend validation, entering | Finite coordinates and storage trend | 170,445 |
| Trend validation, eligible | Annual interval and span of 8 years or more | 113,591 |
| Trend validation, matched to this study (249 months) | 24 months or more of overlap | 94,686 |
| Trend validation, matched to WaterGAP (177 months) | As above, wider spatial coverage | 112,442 |
| Trend validation, matched to SeDA (178 months) | As above | 94,217 |
| Temporal validation | 6 or more observations | 149,637 |

## 3.2 Predictor fields

Six static (a seventh, slope, was derived from elevation) and six dynamic fields were assembled onto the model grid, and their spatial structure is shown in Figure 4 and Figure 5. The static fields located the physical setting of each cell. Permeability and porosity resolved the unconsolidated sedimentary basins, with the highest permeability coinciding with the Indo-Gangetic plain, the North China plain, and the North American mid-continent. The irrigation fields separated the regions in which groundwater was the dominant water source, with the groundwater irrigated fraction approaching 100 per cent across the Indo-Gangetic plain, the Arabian Peninsula, the North American High Plains, and northern Mexico.

The dynamic fields carried the long-term climatic signals over the analysis period. Precipitation declined over the eastern Amazon, the Congo basin, and southern Africa, and increased over the Sahel, northern Australia, and parts of central Asia. Air temperature increased almost everywhere and most strongly over the Arctic margins and the continental interiors. The vegetation index increased strongly over India, eastern China, and the Sahel, a pattern consistent with agricultural intensification, and shallow soil moisture declined over much of the mid-latitude interior. These fields therefore carried independent long-term signals and were not merely seasonal.

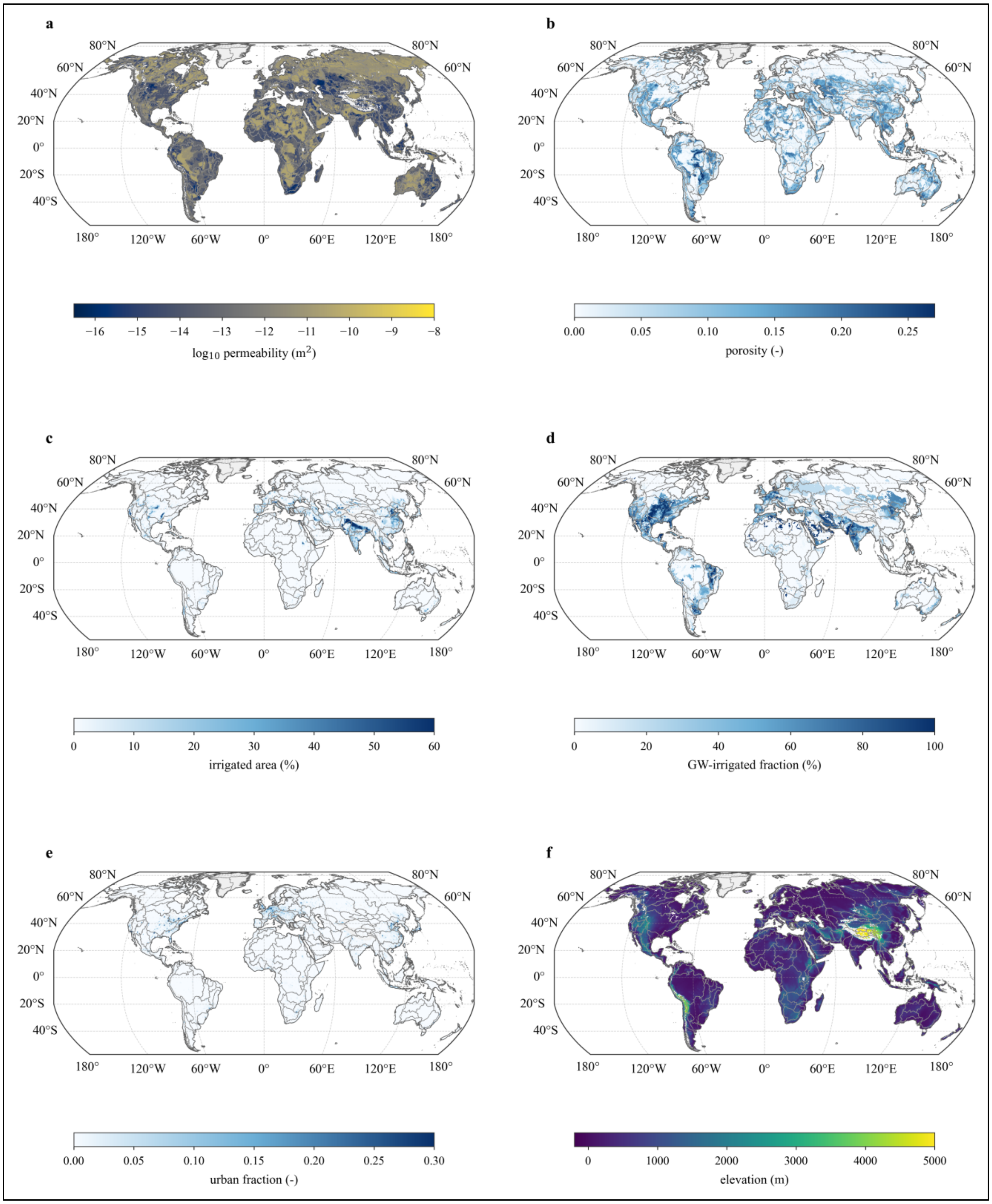

**Figure 4: Static predictor fields on the 0.5 degree model grid.** (a) Base ten logarithm of permeability in square meters, from GLHYMPS 2.0, with the highest values over unconsolidated sedimentary basins including the Indo-Gangetic plain, the North China plain, and the North American mid-continent. (b) Total porosity, showing the expected inverse relationship with permeability. (c) Area equipped for irrigation as a percentage of cell area, exceeding 50 per cent only in northern India, eastern China, and localized valleys of western North America. (d) Groundwater irrigated area as a percentage of the area equipped for irrigation, approaching 100 per cent across the Indo-Gangetic plain, the Arabian peninsula, the North American High Plains, and northern Mexico. (e) Urban fraction, rarely exceeding 0.30. (f) Surface elevation in meters from ETOPO 2022.

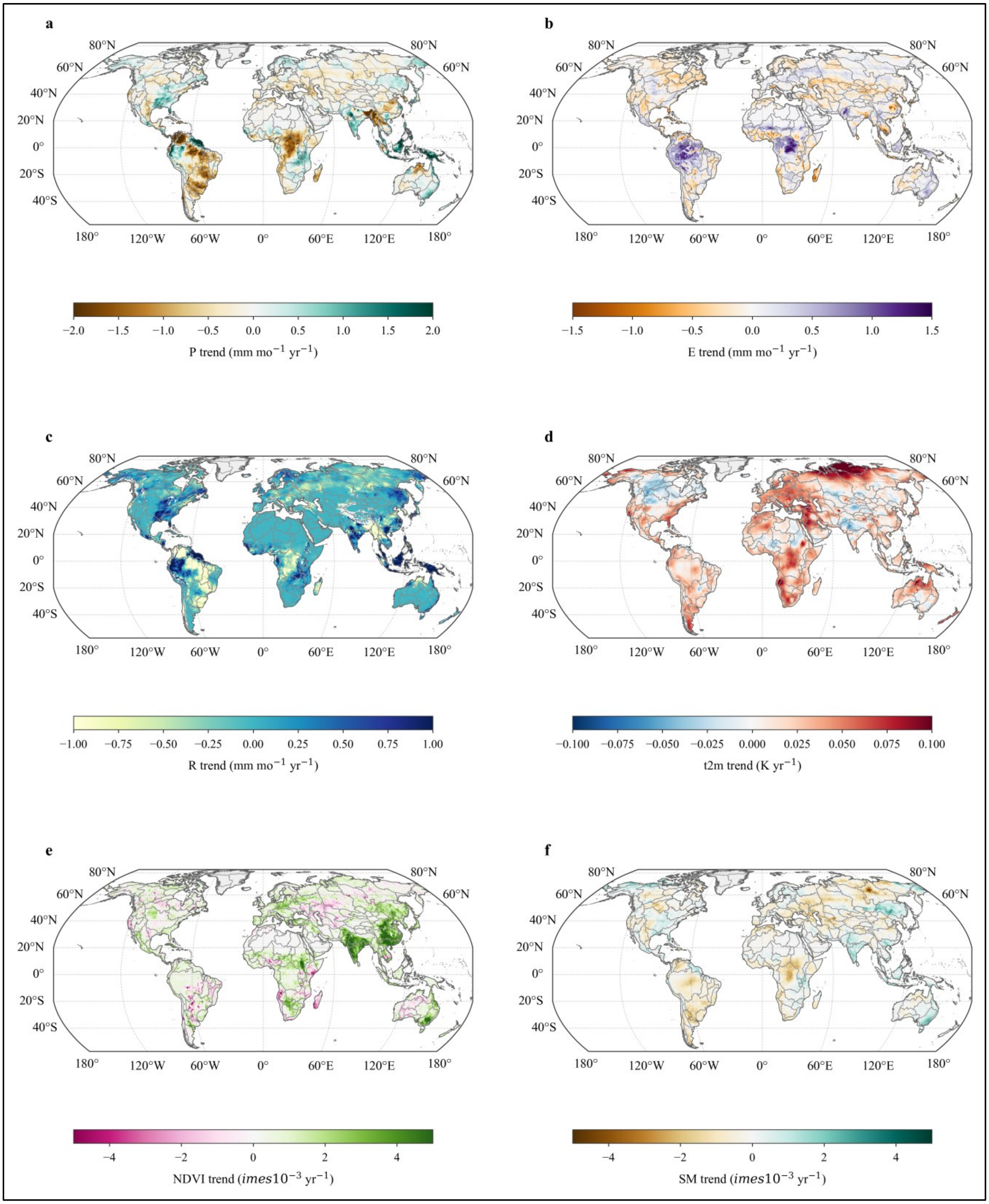

**Figure 5: Linear trends in the six dynamic predictors, 2002 to 2022.** (a) Precipitation in millimeters per month per year, drying over the eastern Amazon, the Congo basin, and southern

Africa and wetting over the Sahel, northern Australia, and parts of central Asia. (b) Actual evaporation, with a coherent increase over the humid tropics. (c) Total runoff, positive over most of the northern high latitudes. (d) Two-meter air temperature in kelvin per year, increasing almost everywhere and most strongly over the Arctic margins and the continental interiors. (e) Vegetation index, increasing over India, eastern China, and the Sahel. (f) Shallow soil moisture, declining over much of the mid-latitude interior. Color scales are saturated at the stated limits in every panel.

### 3.3 The GRACE-derived groundwater signal

The groundwater storage anomaly was isolated by subtracting the anomalies of four soil moisture layers, snow water equivalent, and canopy interception from total water storage anomaly, using a common 2004 to 2009 baseline and a physical cap of plus or minus 100 centimeters. Only 1,722 individual cell months required clipping, which confirmed that the cap acted as a safeguard against isolated excursions rather than as an active constraint.

Across the 55,364 cells the raw Theil-Sen trend ranged from minus 8.901 to plus 8.155 centimeters per year, with a first percentile of minus 3.774, a median of plus 0.021, and a ninety-ninth percentile of plus 2.091 centimeters per year. Almost exactly half of all cells, 48.3 per cent, declined. Four regions were examined as diagnostic checks and all four behaved as expected. North-west India returned minus 5.662 centimeters per year, the North China plain minus 1.451, the Arabian peninsula minus 0.893, and the Amazon plus 1.844. The trend field, the surface water balance residual, and the regional structure of the depletion centers are shown in Figure 6.

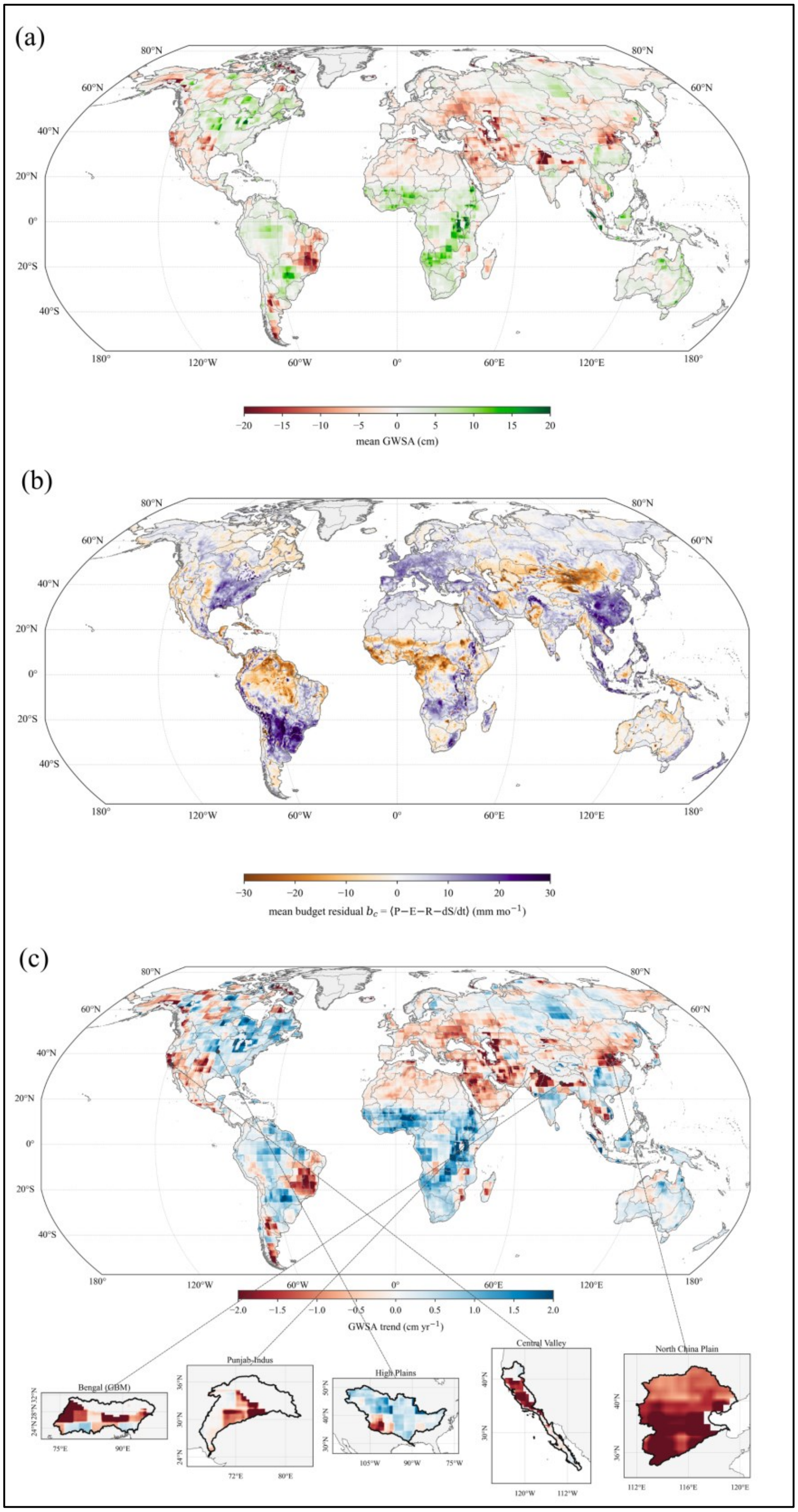

**Figure 6: The GRACE-derived groundwater signal before reconstruction.** (a) Mean groundwater storage anomaly in centimeters over the analysis period. Means below minus 20

centimeters were recovered over north-west India, the North China plain, the Middle East, the Caspian margin, and parts of Patagonia, and positive means occurred over the central Amazon, the Congo, and the Zambezi. (b) Mean residual of the surface water balance, defined as precipitation minus evaporation minus runoff minus the observed rate of storage change, in millimeters per month. A persistently negative residual indicates storage being drawn down faster than the atmosphere can replace it, which is the signature of abstraction rather than of climate, and is the quantity that the physics constraint bounds. (c) Raw Theil-Sen trend in centimeters per year with the five focus regions shown as insets. The color scale is saturated at plus or minus 2.0 while the field reaches plus or minus 8.9, so the insets rather than the color bar convey the magnitude within the depletion centers, and they confirm that the depletion signal was spatially coherent within each region rather than driven by isolated cells.

### 3.4 Predictor behavior within the depletion hotspots

The behavior of the predictors within the five focus depletion regions is shown in Figure 7, and it established that the inputs carried usable and physically interpretable information. A doubly peaked vegetation cycle was found in Punjab-Indus and the Central Valley, which is itself evidence of multiple cropping under irrigation and represents information the model could exploit. The static predictors separated the regions by aquifer setting, with systematically higher permeability in the alluvial Bengal, Punjab-Indus, and North China Plain systems than in the High Plains. The lagged cross correlation of each dynamic predictor against storage was strong and clearly periodic in Bengal, the High Plains, and the Central Valley, weaker but comparable in amplitude in Punjab-Indus, and almost absent at every lag in the North China Plain. The last result indicated storage there that was governed principally by abstraction rather than by climate, and it anticipated the weaker climatic predictability of the model in that region.

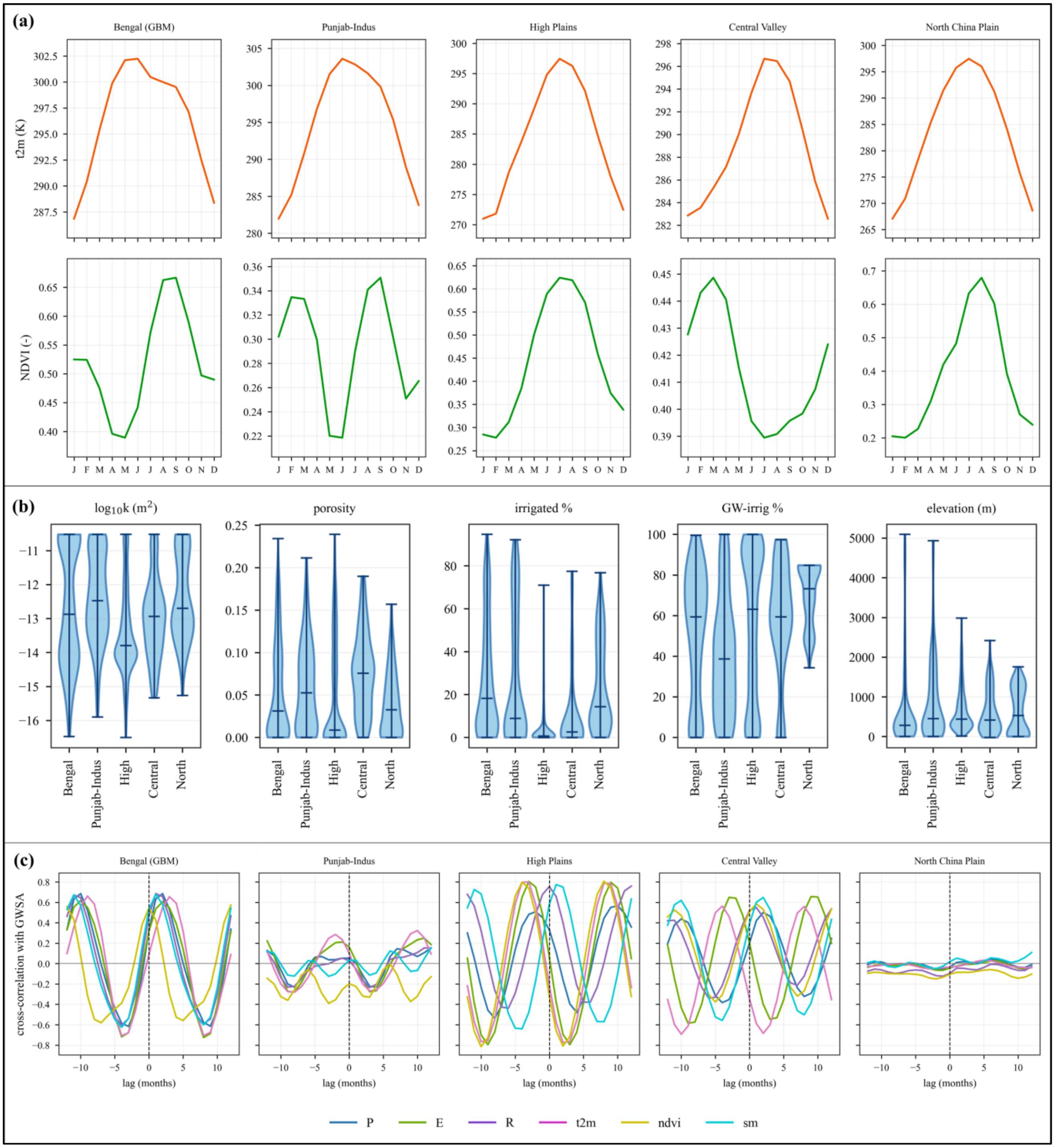

**Figure 7: Predictor behavior within the five focus depletion regions.** (a) Mean annual cycle of two-meter air temperature and vegetation index. Temperature followed a single summer maximum in all five regions, whereas the vegetation index showed a single growing season in Bengal, the North China Plain, and the High Plains and a double peak in Punjab-Indus and the Central Valley. (b) Distributions of the static predictors within each region, shown as violin plots with the median

marked. The groundwater irrigated fraction was strongly bimodal in the first four regions with a large mass near 100 per cent, whereas the North China Plain showed a narrower distribution centered near 70 per cent. (c) Lagged cross correlation of each dynamic predictor against groundwater storage anomaly for lags of minus 12 to plus 12 months. Bengal, the High Plains, and the Central Valley showed strong periodic structure with peak correlations above 0.6, and the North China Plain showed almost no correlation at any lag.

### 3.5 Predictor independence and the net flux substitution

Nine candidate static fields were screened for multicollinearity, and all returned variance inflation factors below 2.24, the highest being longitude and latitude at 2.23, followed by irrigated area at 1.85 and permeability at 1.64. None approached the conventional threshold of five. Seven fields were carried forward, namely permeability, porosity, elevation, slope, area equipped for irrigation, groundwater irrigated fraction, and urban fraction. The two geographic coordinates were excluded deliberately, so that the network could not learn position as a substitute for physical process and was required instead to rely on the measurable properties of each cell.

The dynamic predictors required intervention. Precipitation and runoff returned variance inflation factors of approximately 218 and 215, while all other dynamic predictors remained low. Retaining both as separate inputs would have made the fitted relationships unstable, so the two were combined into a single net moisture input defined as precipitation minus runoff. After the substitution every dynamic predictor fell below a threshold of ten, with air temperature at 6.8, evaporation at 5.1, soil moisture at 4.7, vegetation index at 4.3, and the net flux term at 3.2. The diagnosis and the spatial structure of the substituted field are shown in Figure 8, and the substituted field retained a physically interpretable pattern rather than being a purely statistical construct.

The relationship between each predictor and groundwater storage anomaly was then examined directly. Most relationships were weak and strongly non-linear when examined cell by cell, with the conditional mean varying only modestly across the range of each predictor and with the linear and information-based measures diverging for several predictors (Figure A.1, Figure A.2). This divergence was the central empirical justification for a non-linear sequence model in place of multiple linear regression.

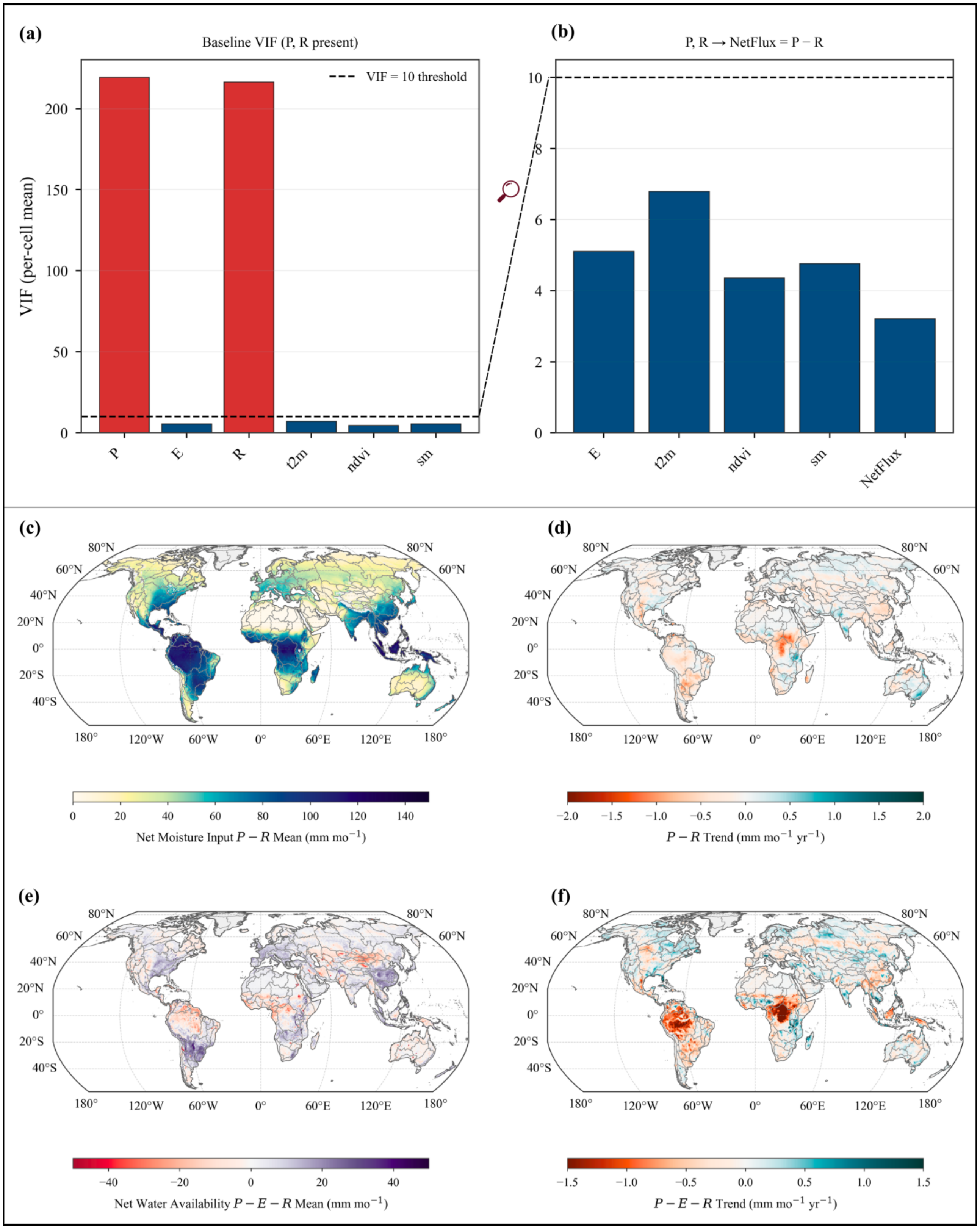

**Figure 8: Multicollinearity diagnosis and the net flux substitution.** (a) Variance inflation factors for the six dynamic predictors as originally specified, averaged per cell. Precipitation and runoff exceeded 200, more than twenty times the threshold of ten, while the remaining predictors sat safely below it. (b) Variance inflation factors after precipitation and runoff were replaced by their difference, the net moisture input. Every predictor then fell below the threshold, the highest being air temperature at 6.8. (c) Mean net moisture input in millimeters per month, exceeding 100 over the equatorial rainforest belts and approaching zero over the major deserts. (d) Trend in net moisture input, dominated by drying over the eastern Amazon and the Congo. (e) Mean net water availability, precipitation minus evaporation minus runoff. (f) Trend in net water availability. Panels c to f establish that the substituted variable retained physically interpretable spatial structure.

### 3.6 Model selection and reconstruction skill

The model received twelve predictors, five dynamic and seven static. It was trained on 150 months, validated on 32, and tested on 32, with no month appearing in more than one partition. The remaining 35 months were gap-filled values, which were excluded from all splits so that skill was assessed only against genuine observations. A structural supply bound was imposed with a median of 8.33 centimeters per year, and a soft basin mass conservation term was applied across the 233 basins.

The weight on the conservation term was selected by a sensitivity sweep across five values, each trained for 45 epochs (Table 3). A weight of 0.5 was adopted, being the only value to improve all four diagnostics at once, producing the lowest held-out error, the highest efficiency, the fewest drifting cells, and the best basin closure of any value tested. The response was not monotonic, since the intermediate weight of 0.25 performed worse than either neighbor, so the selection was made on the empirical evidence of the sweep rather than on the assumption that a stronger constraint is necessarily better. Each arm was trained from a single initialization, and the spread of held-out error across the sweep was 2.4 per cent, so a part of the difference between adjacent arms may reflect initialization variance. The selection was therefore justified primarily by the basin closure improvement, which was the largest and most consistent of the four responses.

Table 3: Sensitivity of the reconstruction to the basin conservation weight.

| Basin weight | Held-out RMSE (cm) | NSE | Cells drifting (%) | Basin closure RMSE (cm/yr) |
|---|---|---|---|---|
| 0.00 | 9.006 | 0.490 | 28.8 | 0.5285 |
| 0.10 | 8.999 | 0.491 | 28.6 | 0.5098 |
| 0.25 | 9.085 | 0.481 | 30.7 | 0.5046 |
| 0.50 | 8.789 | 0.515 | 26.7 | 0.4172 |
| 1.00 | 8.993 | 0.492 | 29.0 | 0.4793 |

Retraining at the selected weight for 120 epochs improved performance substantially. Held-out root mean squared error fell to 7.886 centimeters, Nash-Sutcliffe efficiency rose to plus 0.609, drifting cells fell to 22.5 per cent, and basin closure error fell to 0.2796 centimeters per year, an improvement of 47 per cent over the unconstrained arm after extended training. At matched training length (45 epochs, Table 3), the conservation term reduced basin-closure error by 21 per cent (0.5285 to 0.4172 centimeters per year) while held-out error fell by 2.4 per cent, a margin comparable to initialization variance. The constraint therefore improved physical consistency without degrading fit. The convergence, the held-out agreement, the trend bias, the basin closure, and the feature importance are shown in Figure 9.

Kling-Gupta efficiency at this stage was minus 0.763, and it was decomposed into a correlation of 0.78, a bias ratio of minus 0.74, and a variance ratio of 0.78. The bias component of this metric is the ratio of simulated to observed mean, and since groundwater storage anomaly has a median close to zero by construction, that ratio was poorly conditioned for this application. The negative efficiency was driven by that term rather than by the correlation, and the metric was therefore reported alongside rather than in place of the error, efficiency, correlation, and variance ratio.

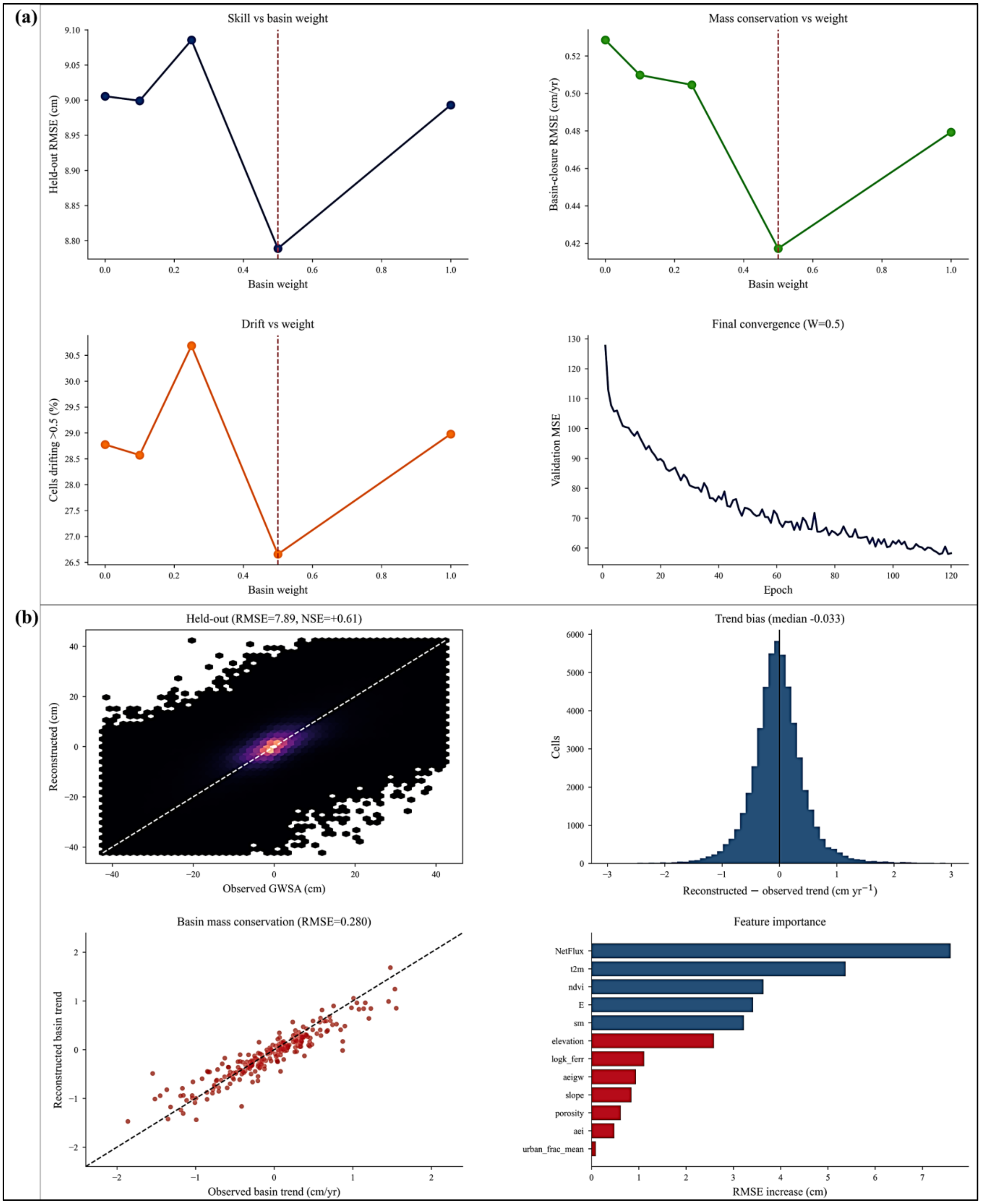

**Figure 9: Model selection and converged performance.** (a) Sensitivity of four diagnostics to the basin conservation weight, each arm trained for 45 epochs, with the adopted value of 0.5 marked,

together with the validation loss curve for the final 120 epoch run. The convergence curve fell from 128 at the first epoch to 58 at the last with no upturn, which indicated that training was stopped before overfitting began. (b) Performance of the converged model on withheld months. Upper left, reconstructed against observed values, density shaded, with a root mean squared error of 7.89 centimeters and a Nash-Sutcliffe efficiency of plus 0.61. Upper right, the difference between reconstructed and observed trend across all cells, sharply peaked with a median of minus 0.033 centimeters per year. Lower left, reconstructed against observed basin mean trend, with a root mean squared error of 0.280 centimeters per year. Lower right, permutation importance expressed as the increase in error when each predictor was shuffled, with net flux dominant at 7.59 centimeters and air temperature next at 5.37. The dominance of the physically primary driver confirmed that the model did not rely on incidental spatial covariates.

### 3.7 Amplitude correction

The variance ratio of 0.78 indicated that the reconstruction systematically underestimated the amplitude of storage variability, a known consequence of training under a squared error loss. Across the 52,427 cells the ratio of reconstructed to observed standard deviation had a lower quartile of 0.59, a median of 0.77, and an upper quartile of 0.98. Because the deficiency was reasonably uniform in space, a per-cell variance matching correction was applied with a median expansion factor of 1.30. The correction was fitted exclusively on training months, so that the withheld months remained independent of it.

Every measure of skill improved. Root mean squared error fell from 7.886 to 6.973 centimeters, Nash-Sutcliffe efficiency rose from plus 0.609 to plus 0.694, correlation rose from 0.78 to 0.85, and the variance ratio rose from 0.78 to 0.99. Kling-Gupta efficiency rose from minus 0.763 to plus 0.604. A cell-by-cell comparison of the trend field before and after correction showed the direction of the trend preserved in 100.00 per cent of the 52,427 cells, including every cell whose uncorrected trend exceeded 0.5 centimeters per year, with a pattern correlation of 0.9184. The correction therefore rescaled amplitude without altering the sign or the spatial structure of long-term change. Theil-Sen trends were then recomputed on the corrected series with confidence intervals from 200 block bootstrap replicates, and 39,686 of the 52,427 cells, or 76 per cent, returned intervals that excluded zero.

## 3.8 Comparison against independent products

Two comparison products were used. WaterGAP 2.2d is a physically based global hydrological model that uses no GRACE information and is fully independent. The GRACE-SeDA implied groundwater field is not independent, because it shares the GRACE source and was obtained by subtracting this study's own other storage components, and it was therefore treated throughout as a processing consistency check rather than as validation. A circularity diagnostic quantified the risk. The trend in the shared subtracted term correlated with this study at minus 0.210 and with the SeDA field at minus 0.201, while the two products correlated with each other at plus 0.410 across all cells over their common window. Since the shared term correlated only weakly with either product and with the opposite sign, the agreement between them was unlikely to be an artefact of the shared subtraction, although it remained no proof of accuracy.

The most informative feature of the comparison was that WaterGAP produced a near-zero trend across most of the land surface, with substantial signals confined to a small number of intensively irrigated regions. This followed directly from its construction, since it represents only the abstraction and recharge processes contained in its formulation and is quiescent wherever storage change is driven by climate variability rather than by abstraction. The three trend fields and their pairwise agreement are shown in Figure 10.

Temporal agreement was assessed separately using basin-averaged monthly series. This study correlated with WaterGAP at a median of plus 0.479 across 243 basins and with SeDA at a median of plus 0.841 across 231 basins. The median difference in the month of peak storage was one month against WaterGAP and zero months against SeDA, with 77 per cent of basins agreeing to within one month against SeDA. The reconstruction therefore reproduced the timing of the annual storage cycle as well as its long-term change.

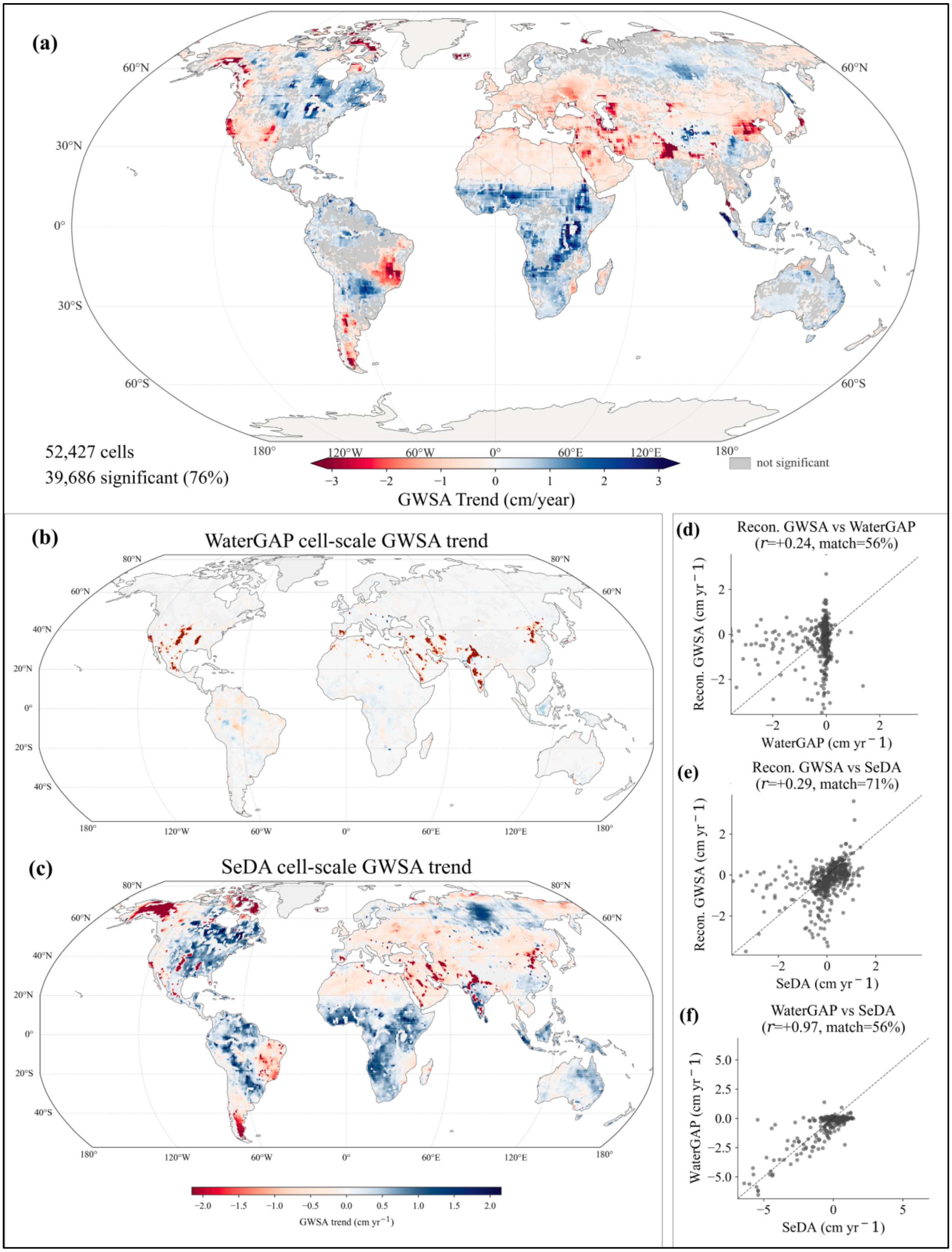

**Figure 10: Trend fields and inter-product agreement.** (a) Reconstructed groundwater storage anomaly trend in centimeters per year over 52,427 cells, of which 39,686, or 76 per cent, were significant by bootstrap confidence interval, with the remainder shown in grey. (b) WaterGAP 2.2d groundwater storage trend on a common scale, near zero across most of the land surface as a property of the model. (c) SeDA-implied groundwater storage trend on the same scale, with broad structure closer to that of the reconstruction, as expected from the shared observational origin. (d to f) Pairwise agreement between the three trend fields, each computed over the window the two products share and aggregated to aquifer polygons. The reconstruction against WaterGAP gave a correlation of plus 0.24 with 56 per cent directional agreement, against SeDA plus 0.29 with 71 per cent, and WaterGAP against SeDA plus 0.97 with only 56 per cent. The combination of high correlation and low sign agreement in panel f indicated that the two comparison products share a small number of very large signals in the intensively abstracted regions while disagreeing about the direction of change almost everywhere else.

## 3.9 Validation against monitoring wells

For each well the product trend was extracted over exactly the interval spanned by that well's own record rather than over a common period, so that comparisons were window matched. Of the 113,591 eligible wells, 94,686 achieved at least 24 months of overlap with the reconstruction and were retained, and polygons required at least three wells to be scored. The results across four aggregation scales are summarized in Table 4 and shown in Figure 11 and Figure 12.

**Table 4: Trend agreement against monitoring-well storage trends at four aggregation scales.** n is the number of cells or polygons compared; r is the Pearson correlation; RMSE is the root mean squared error; NSE is the Nash-Sutcliffe efficiency; bias is the mean of product minus well trend, so negative values indicate that a product overstates decline; sign agreement is the percentage of units in which product and wells agree in the direction of change. All trends are window-matched to each well's own record. Bold marks the best value at each scale.

| Scale | Product | n | r | RMSE (cm $yr^{-1}$) | NSE | Bias (cm $yr^{-1}$) | Sign agreement (%) |
|---|---|---|---|---|---|---|---|
| Cell | This study | 2,983 | +0.31 | **2.03** | **0.00** | **+0.02** | **60.5** |
| | WaterGAP | 3,379 | +0.31 | 3.41 | −2.17 | −0.51 | 54.7 |
| | SeDA | 2,964 | **+0.38** | 2.97 | −1.13 | −0.28 | 57.2 |
| Aquifer | This study | 503 | **+0.31** | **1.94** | **+0.06** | **+0.02** | **66.0** |
| | WaterGAP | 504 | +0.10 | 3.38 | −1.86 | −0.50 | 55.2 |
| | SeDA | 503 | +0.13 | 3.43 | −1.94 | −0.29 | 58.4 |
| Basin | This study | 72 | **+0.34** | **1.50** | **0.00** | **+0.04** | **63.9** |

| Scale | Product | n | r | RMSE (cm yr⁻¹) | NSE | Bias (cm yr⁻¹) | Sign agreement (%) |
|---|---|---|---|---|---|---|---|
| | WaterGAP | 72 | +0.01 | 2.19 | −1.15 | −0.12 | 61.1 |
| | SeDA | 70 | +0.29 | 1.62 | −0.14 | −0.05 | 62.9 |
| Basin ≥ 200,000 km² | This study | 59 | **+0.37** | **1.56** | **+0.05** | **+0.02** | 62.7 |
| | WaterGAP | 59 | +0.01 | 2.39 | −1.22 | −0.20 | 61.0 |
| | SeDA | 58 | +0.31 | 1.71 | −0.11 | −0.16 | **65.5** |

The pattern across scales was the substantive result. This study retained its skill as the aggregation scale increased and improved slightly at the largest scale, whereas WaterGAP lost essentially all skill above cell scale. This study also achieved the lowest error against wells at every scale, by a factor of between 1.1 and 1.8. It was also the only product with a near-zero bias (+0.02 to +0.04 centimeters per year) and a Nash-Sutcliffe efficiency close to or above zero at every scale, whereas WaterGAP (efficiency −1.15 to −2.17) and SeDA (−0.11 to −1.94) both overstated decline at well locations (bias −0.05 to −0.51 centimeters per year) and performed worse than the mean of the well trends. The confidence interval on the aquifer-scale correlation, from plus 0.21 to plus 0.40, excluded zero, whereas that of WaterGAP, from minus 0.02 to plus 0.24, did not. The spatial distribution of the residuals is shown in Figure A.3, and this study produced visibly smaller and more spatially confined residuals across the Indian subcontinent than either comparison product.
Agreement varied with well density but not monotonically, rising from plus 0.19 in cells holding three to five wells to plus 0.57 in cells holding 25 to 50, with a dip at five to ten wells and a fall to plus 0.25 in the highest density tier. The decline at high density was attributable to the concentration of such cells in a small number of heavily abstracted regions where within-cell heterogeneity is greatest. This density dependence had a quantitative explanation established independently. Within cells containing three or more wells, the spread of well trends had a median robust standard deviation of 0.90 centimeters per year, and the ratio of that spread to the magnitude of the cell trend had a median of 1.88. In 69.7 per cent of such cells the internal spread equalled or exceeded the signal being estimated. The wells therefore did not agree with one another within a single 0.5 degree cell any better than they agreed with the products, and the correlation attainable by any gridded product against this network was bounded by the representativeness of the network rather than by the skill of the product. A moderate correlation was the expected outcome.

A related diagnostic bore on the interpretation of the well trends themselves. Wells whose trend was statistically significant had a median of minus 1.088 centimeters per year with 71.5 per cent declining, whereas wells whose trend was not significant had a median of minus 0.024 with 53.6 per cent declining. Filtering on significance would have shifted the median by minus 1.064 centimeters per year and produced a sample biased toward large magnitudes. Significance was therefore not used as a screen, and inverse variance weighting was applied in its place. One asymmetry should be noted. Because product trends were extracted over each well's own record clipped to the availability of each product, and because WaterGAP terminates in December 2016, WaterGAP trends were in general fitted over shorter and therefore noisier intervals, so a part of the difference in agreement may be attributable to record length rather than to model skill. The inter-product comparison of the previous section does not carry this asymmetry, since both fields were re-windowed to their common overlap before comparison.

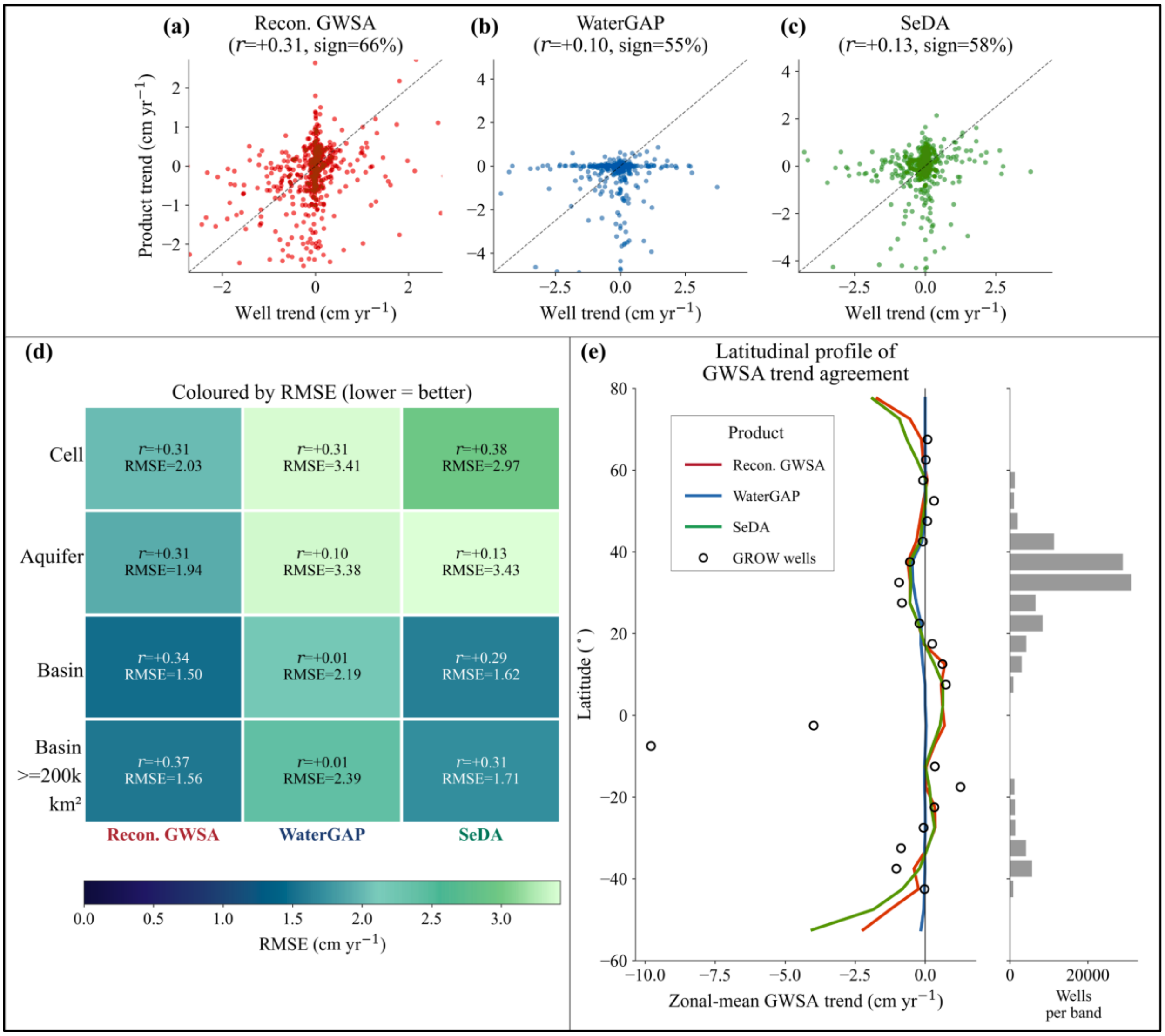

**Figure 11: Multi-scale validation against monitoring well storage trends.** (a to c) Aquifer-scale scatter of product trend against well-derived storage trend for this study, WaterGAP, and SeDA, with the one-to-one line dashed and all trends window matched to each well's own record. (d) Correlation and root mean squared error for all three products at four aggregation scales, with cells shaded by error so that darker indicates better. This study achieved the lowest error at every scale and was the only product whose skill was retained under aggregation. (e) Latitudinal profile of zonal mean trend for the three products against the well observations, with the number of wells per latitude band on the right. All three tracked the wells closely between 20 and 60 degrees north, where most wells are located, and diverged in the tropics and the southern hemisphere where the network is sparse.

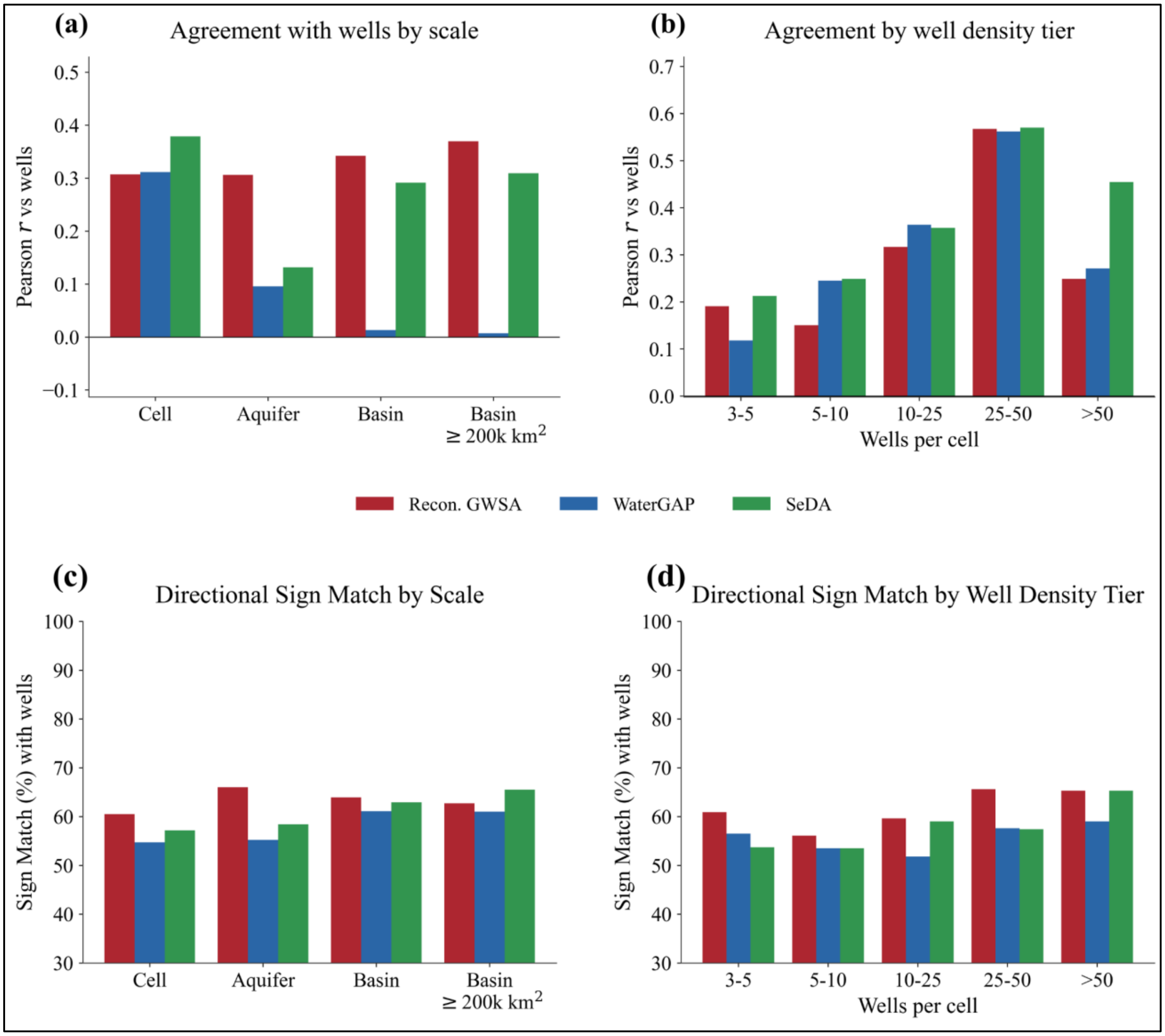

**Figure 12: Agreement with wells by aggregation scale and by well density.** (a) Pearson correlation with well trends at four aggregation scales. The divergence between this study and WaterGAP grew with scale, from parity at cell scale to plus 0.37 against plus 0.01 for the largest basins. (b) Correlation by the number of wells contributing to a cell, rising from plus 0.19 in the sparsest tier to plus 0.57 at 25 to 50 wells per cell, with a dip at five to ten wells and a fall in the highest tier. (c, d) Directional sign agreement by scale and by density. This study led at cell, aquifer, and basin scale, and was marginally exceeded by SeDA for basins above 200,000 square kilometers.

### 3.10 Temporal agreement at aquifer and basin scale

Temporal agreement was assessed by converting well level records to storage using an external global specific yield field, which resolved every well in the validation set, aggregating within each polygon, and correlating against the area-averaged reconstruction. This pathway required only six observations per well and retained 149,637 sites, and each polygon required at least three wells and eight overlapping months. Agreement was high in the best performing polygons. The ten best basins returned correlations between plus 0.73 and plus 0.91 and the ten best aquifers between plus 0.91 and plus 0.97 (Figure A.4, Figure A.5). Positive agreement was obtained across the great majority of polygons for which sufficient well data existed, with negative agreement confined to a small number of polygons in eastern Asia (Figure A.10). The individual sites within the highest performing polygons were spatially distributed rather than clustered, which confirmed that the correlations rested on a network rather than on a few sites (Figure A.7, Figure A.8, Figure A.9). Performance within the five focus depletion regions was more variable, and the variability was itself informative (Figure A.6). The North China Plain returned plus 0.91, Bengal plus 0.76, the High Plains plus 0.71, the Central Valley plus 0.56, and Punjab-Indus plus 0.43. These values must be read together with the number of overlapping months available in each, which ranged from 249 in the High Plains and the Central Valley to 19 in Bengal and Punjab-Indus and only nine in the North China Plain. The highest correlation of the five was therefore obtained from the shortest record and the smallest well count and is indicative rather than robust, whereas the two regions with full-length records returned plus 0.71 and plus 0.56, and these are the values that carry evidential weight.

### 3.11 Uncertainty decomposition

A full uncertainty budget was constructed by propagating three components separately. Observation uncertainty was taken from the GRACE mascon error field, with a median of 2.737 centimeters and a floor of 1.779 centimeters applied to cells reporting zero. Model uncertainty was propagated through a 200 member Monte Carlo ensemble. Storage-property uncertainty, which expresses the uncertainty in relating a satellite-derived change in water-equivalent thickness to the behavior of the aquifer, was applied as a relative error scaled by the magnitude of the local trend, taking a value of 0.15 in the 1,459 cells containing at least one high-quality well, where the local storage coefficient could be confirmed against observation, and 0.40 elsewhere. Both values were adopted as assumptions rather than derived from measurement. This term does not enter the reconstruction, which is expressed throughout in centimeters of water equivalent and requires no storage coefficient, but it governs the confidence with which the product may be interpreted as aquifer storage. Because the term is multiplicative on the trend while the observation and model terms are additive and small, with a combined median of 0.096 centimeters per year, the storage-property term necessarily dominated wherever the trend itself was large.

Spatial correlation was treated explicitly. An empirical variogram of the trend field returned a correlation range of approximately 6 degrees, equivalent to about 666 kilometers and to an effective correlation area of roughly 144 cells, and this length scale was used when aggregating uncertainty to basin and global totals. Total trend uncertainty had a lower quartile of 0.118, a median of 0.173, and a ninety-fifth percentile of 0.675 centimeters per year. The decomposition of variance was the most consequential result of this section. Averaged across all cells, observation uncertainty contributed 3.0 per cent, model uncertainty 35.6 per cent, and storage-property uncertainty 61.3 per cent. The storage-property term dominated in 35,461 cells, or 67.6 per cent of the domain, model error in 16,966 cells, and observation error in no cell at all (Figure 13, Figure A.11). Because the reconstruction is expressed in water-equivalent thickness, its volumetric totals do not depend on specific yield; the storage-property term instead governs the confidence with which the trends can be interpreted as changes in aquifer water level and compared with well observations. Improvement in the global characterization of specific yield would therefore sharpen that interpretation more than further improvement of the satellite record.

Because the storage-property term dominated the budget and was set by assumption, its adopted value was tested directly, and the result determines how the significance of the trends should be read (Figure A.12). When the imputed relative error was swept, the fraction of cells retaining significance at the individual grid scale fell from 70.6 per cent at an imputed error of 0.25 to 2.1 per cent at 0.55, with a sharp transition near 0.51. This transition is a property of the significance test rather than of the data, because for a storage-property-dominated cell the signal-to-noise ratio reduces to the reciprocal of the relative error and is therefore almost independent of the trend magnitude, so that such cells crossed the threshold of 1.96 together at a relative error of about 1 divided by 1.96, which is 0.51. The measured-tier error had almost no effect on this behavior, because only 2.8 per cent of cells carry a measured value. The adopted imputed error of 0.40 lay inside the published range of approximately 0.30 to 0.45 for proxy-derived specific yield and near the upper edge of that range, so the value was not selected to maximize the significant fraction.

At the correlation-area scale the significance was stable rather than fragile. When cells were aggregated to blocks matched to the 6 degree correlation length, the significant fraction at the adopted error was 61.4 per cent of the land area under the conservative assumption that the storage-property error is fully correlated within a block, and 76.7 per cent under the assumption that it is independent within a block, and the independent estimate remained near 77 per cent across the entire range of relative error tested. The true value lies between these two bounds and nearer the independent one, because a lithology-based proxy varies from cell to cell within any 600-kilometer block. The depletion finding therefore did not depend on the storage-property assumption. The trend field itself was unaffected by it, the storage-property term dominated the budget across the whole plausible range, and the significant land area at the correlation-area scale was robust. At the individual grid scale, 31,227 cells, or 59.6 per cent of the domain, retained significance at the adopted budget, and this figure was carried as the per-cell reliability flag, while the interpretation of the depletion rested on the aggregated and bracketed result. The 59.6 per cent figure was substantially more conservative than the 76 per cent significant under the trend fit alone, and the difference was the direct cost of admitting storage-property and model error into the assessment.

Cells were finally divided into two tiers. The 1,459 cells containing at least one high-quality well were classified as well constrained and returned a median signal-to-noise ratio of 3.85. The remaining 50,968 cells, classified as physics extrapolated, returned 2.12. Both exceeded the significance threshold, but the margin in the extrapolated tier was narrow, and the tier flag was carried in the released product so that users may restrict analysis to the constrained subset.

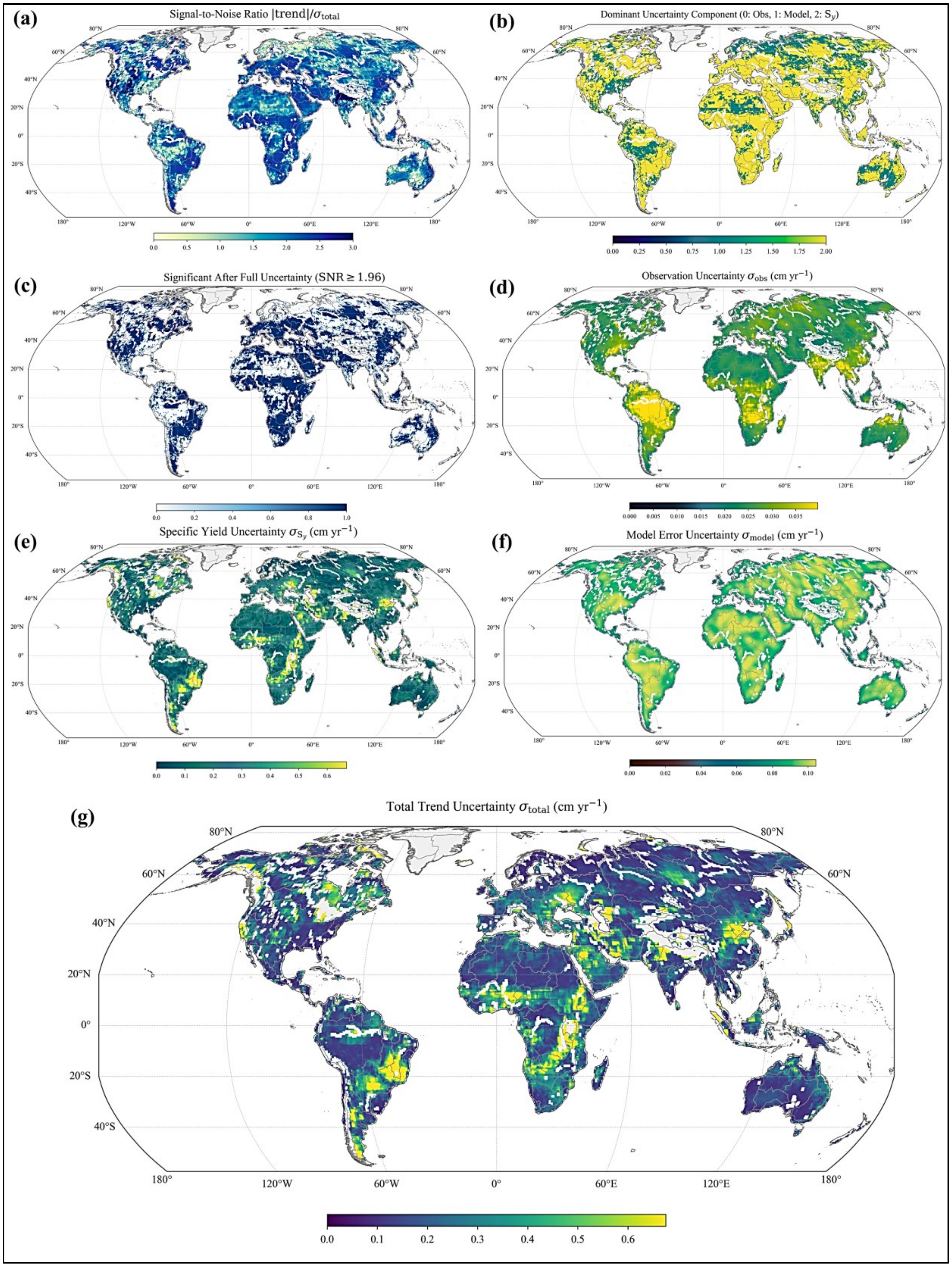

**Figure 13: Spatial structure of the uncertainty budget.** (a) Signal-to-noise ratio, the absolute trend divided by total uncertainty. (b) Dominant uncertainty component, coded as observation,

model, or storage property. (c) Cells retaining significance after the full budget was applied, comprising 59.6 per cent of the domain. (d to f) The three components separately, each in centimeters per year, with observation uncertainty spanning a range roughly twenty times smaller than the other two. (g) Total trend uncertainty, the quadrature sum of the three components.

### 3.12 Global depletion and sea level contribution

Two distinct volumetric quantities were computed, and the distinction between them was essential. Gross depletion was defined as the sum over cells carrying a significant negative trend and expresses the extent and severity of aquifer loss. Net storage change was defined as the sum over all valid cells, allowing recharging cells to offset depleting ones, and is the physically correct quantity for a sea level budget.

Gross depletion over 2002 to 2022 was estimated at 357.64 cubic kilometers per year across 21,632 cells. The associated uncertainty was 1.46 cubic kilometers per year under an assumption of independent cell errors and 11.6 cubic kilometers per year when the 6 degree correlation length was respected, and the larger figure was reported since independence between neighboring cells was untenable given the variogram result. This uncertainty conservatively includes the storage-property term of Section 3.11. Net storage change was minus 36.25 cubic kilometers per year across all valid cells and minus 46.64 across significant cells only, with correlated uncertainties of 17.82 and 16.79. Expressed as sea level equivalent, gross depletion corresponded to 0.9885 plus or minus 0.032 millimeters per year and net storage change to 0.1002 plus or minus 0.049 millimeters per year. The order-of-magnitude difference was the direct consequence of recharge elsewhere on the land surface offsetting losses in the depleting regions, and it is the reason the two figures must never be quoted interchangeably. The net figure of about 0.100 millimeters per year is the sea-level-relevant quantity, and the gross figure is a measure of extent and severity only. The extent of the depletion and its principal centers are shown in Figure 14.

The gross figure may be placed against two recent estimates. An earlier climate-model estimate, independently corroborated by observed pole drift, implied about 126.5 cubic kilometers per year, and a recent estimate corrected against monitoring wells reported 284 cubic kilometers per year before adjustment, rising to 550 afterward. The estimate of 357.64 cubic kilometers per year obtained here fell between these bounds. Restricting the sum to the 16,393 declining cells that also survived the full uncertainty budget of Section 3.11 (signal-to-noise ratio at least 1.96) lowered the total by only 6.6 percent, to 334.05 cubic kilometers per year, even though it removed almost a quarter of the cells; the removed cells carried marginal trends and contributed little volume, so the headline estimate does not depend on the choice of significance screen.

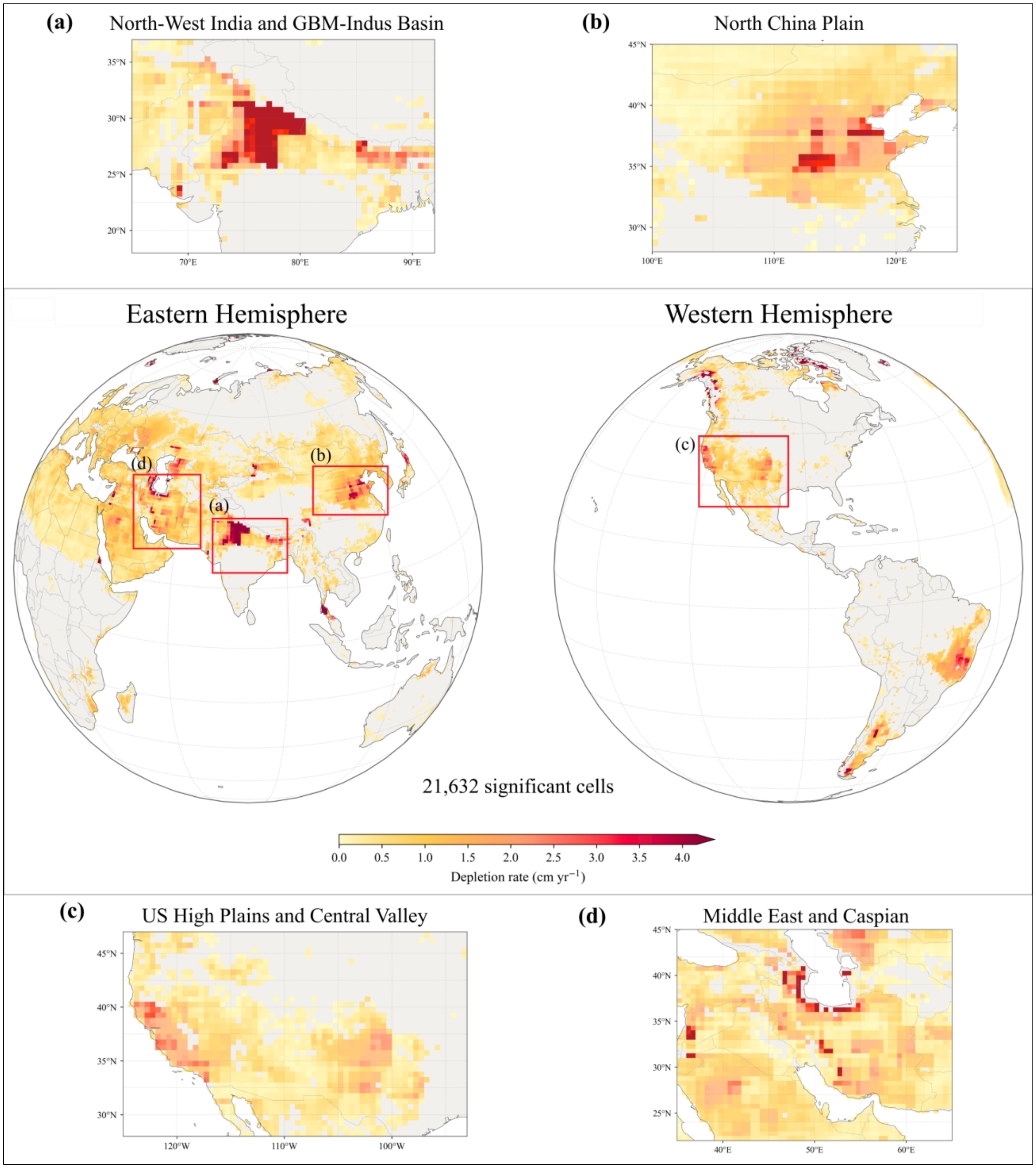

**Figure 14: Cells with a significant depleting trend.** The central globes show all 21,632 significantly depleting cells in the eastern and western hemispheres, colored by depletion rate in centimeters per year. Cells without a significant negative trend were not plotted, so the figure shows the extent of depletion rather than net storage change. (a) North-west India and the Ganges-Brahmaputra-Meghna and Indus basins, where rates exceeded 4 centimeters per year across a contiguous area spanning the Punjab and Haryana plains. (b) The North China Plain, with a coherent core above 3 centimeters per year across southern Hebei and northern Henan. (c) The western United States, with the California Central Valley and the southern High Plains as distinct centers. (d) The Middle East and Caspian margin, with the most intense signals along the western Caspian shore and across the Mesopotamian plain.

### 3.13 Contribution to Earth's rotational pole drift

The depletion field was converted into a perturbation of the Earth's rotational pole. The per-cell mass rate was obtained from the trend field and cell area, the off-diagonal inertia perturbations were accumulated over all cells, and these were converted to polar motion excitation using the standard rotational deformation constants (Figure 15).

Three masks were evaluated to test robustness. The depletion-only mask returned 6.216 centimeters per year toward 67.5 degrees east, the significant-net mask 6.916 toward 68.6 degrees east, and the all-valid-cell mask 6.971 toward 68.8 degrees east. The three azimuths spanned only 1.1 degrees, so the inferred direction was insensitive to the mask chosen, and the significant-net mask was adopted as the physically appropriate one for a mass redistribution calculation. The secular estimate was verified against a month-by-month simulation, computed from the same field, of the pole path, which returned 6.975 centimeters per year toward 69.1 degrees east and agreed to within 0.86 per cent in rate and 0.50 degrees in azimuth (Figure 16). Recomputing over the shorter 2002 to 2019 window returned 8.145 centimeters per year toward 67.97 degrees east, a shift of only 0.65 degrees, so the direction was not an artefact of the period selected although the rate was sensitive to it.

The comparison against the earlier pole-drift study required a clear statement of the discrepancy. That study reported 4.36 centimeters per year toward 64.16 degrees east, corroborating a climate-model depletion estimate of about 126.5 cubic kilometers per year (2,150 gigatonnes over 1993 to 2010). The azimuth obtained here differed by 4.5 degrees, which is close agreement in direction, while the rate was 1.59 times larger and the underlying depletion estimate approximately 2.8 times larger. A direct diagnostic identified the origin of the difference. Over the 178 month window in which this study and the SeDA product overlap, the annual total depletion rate averaged 337.0 cubic kilometers per year here against 214.6 in SeDA, and a regression over the storage curve gave 368.1 against 232.9. The ratio of 1.58 was almost exactly the ratio between the two pole drift rates. The two series were nevertheless very highly correlated, at plus 0.996 for annual storage and plus 0.441 for cell-wise trend over the 39,438 significant cells, with 80.8 per cent sign agreement. The two products therefore described the same spatial and temporal pattern of depletion and differed systematically in amplitude by a factor of approximately 1.6.

The most likely explanation was the amplitude correction applied in this study, which restored the storage variability that a squared error loss suppresses and whose median expansion factor of 1.30 was of the same order as the discrepancy. Independent support was available, since a recent well-corrected study found its estimate rose by a factor of approximately 1.94 after correction against monitoring wells. Two studies employing different gravimetric solutions, different land surface components, and different correction procedures therefore reached the same conclusion, that uncorrected satellite gravimetry understates the rate of groundwater depletion. The estimate presented here should accordingly be understood as an amplitude-restored one, and its difference from the earlier study should be read as a methodological consequence rather than as a disagreement about the underlying hydrology, given also that the two analyses span different periods. The agreement in azimuth to within 4.5 degrees indicated that the two studies agreed about where mass was being lost.

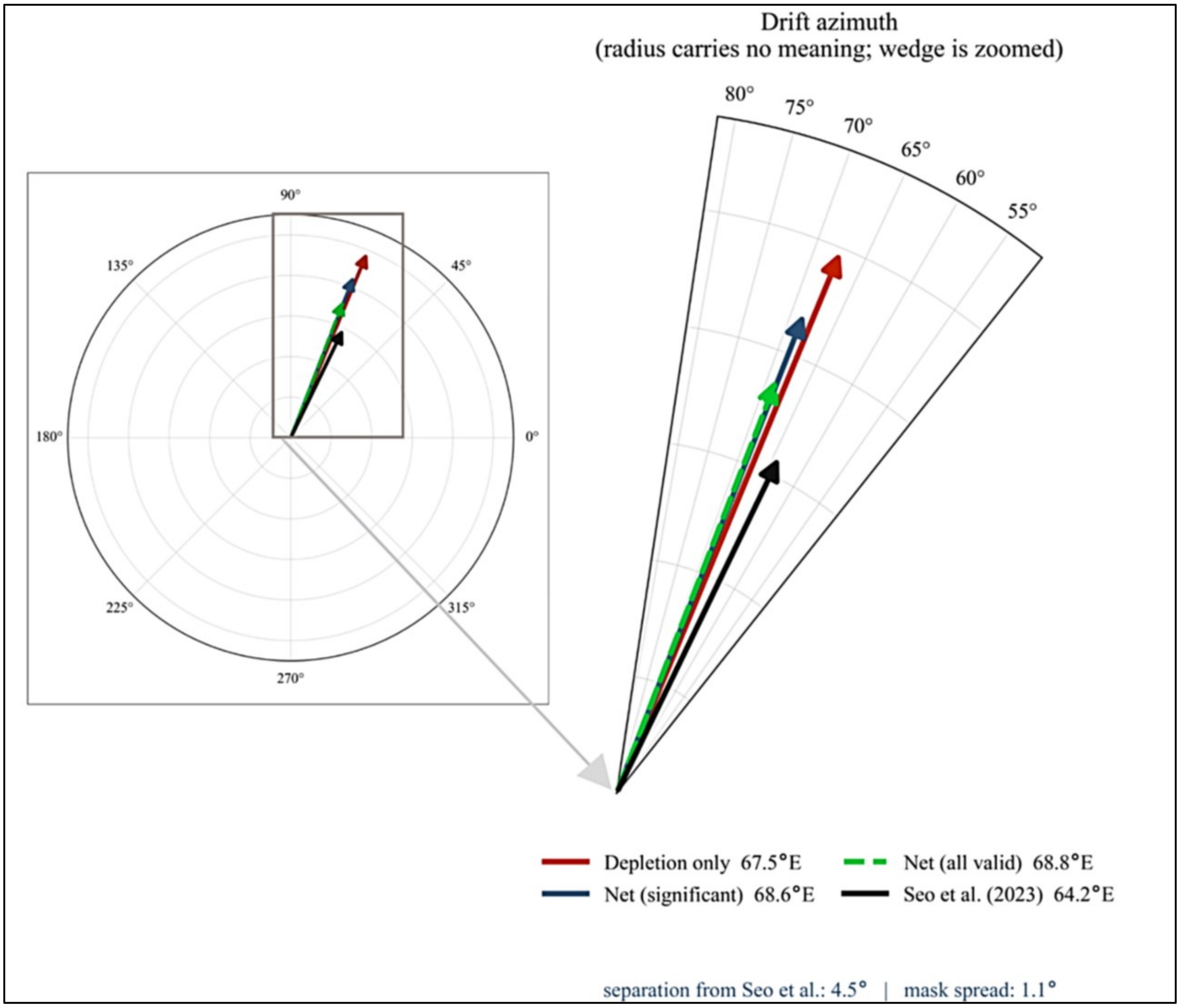

**Figure 15: Drift azimuth of the rotational pole.** Drift direction for the three masks of this study and for Seo et al. (2023), shown on a zoomed wedge; arrow length carries no meaning as titled above. The three masks span 1.1° and lie 4.5° from the published direction. The conversion from groundwater trend to polar-motion excitation was derived using Eqs 31–36.

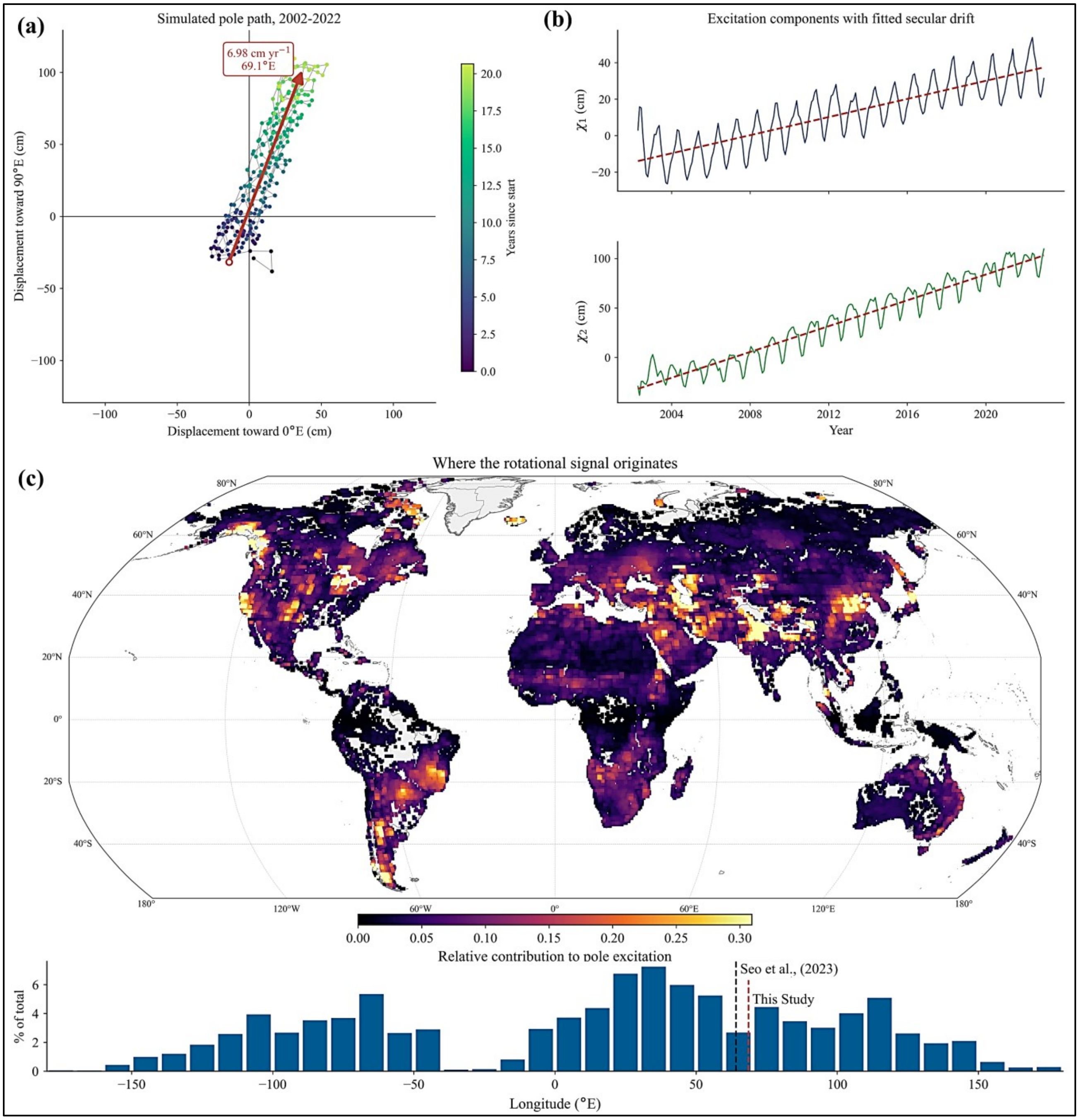

**Figure 16: Simulated pole path and its origin.** (a) Month-by-month simulated displacement of the rotational pole over 2002 to 2022, colored by time elapsed, with the fitted secular drift arrow labelled. The simulated drift of 6.98 centimeters per year toward 69.1 degrees east agreed with the secular estimate to within 0.86 per cent in rate and 0.50 degrees in azimuth, which confirmed the numerical consistency of the secular estimate. (b) The two excitation components with their fitted secular drift, showing a linear trend superimposed on a large annual oscillation, of which only the secular term was reported. (c) Relative contribution of each cell to the total pole excitation, with the longitudinal distribution below. Contributions concentrated between 20 and 60 degrees east, the interval containing the Indus, Ganges, and Middle Eastern systems, with secondary contributions from the North China Plain and western North America.

## 4. Discussion

Global groundwater storage is never measured directly, and the central difficulty of this work was to recover it as a continuous global field while keeping the learned reconstruction inside the limits that the physics of the water cycle imposes. Three questions had to be answered for that recovery to be credible. The groundwater component had to be separated from total water storage and reconstructed without gaps; the resulting trends had to remain trustworthy once their full uncertainty was exposed; and the recovered depletion had to be checked against evidence that owed nothing to the reconstruction itself. The findings answer all three, and their interest lies less in any single number than in the fact that one hidden quantity was constrained from several independent directions at once and still held together.

### 4.1 Synthesis of the principal findings

The first objective, the reconstruction itself, was met with a model that reproduced withheld GRACE observations with a root mean squared error of 6.973 centimeters and a Nash-Sutcliffe efficiency of plus 0.694, satisfied basin-scale mass conservation to within 0.280 centimeters per year, and its amplitude correction preserved the sign of every trend by construction. The physical constraint improved physical consistency without degrading fit: at matched training length, the basin-conservation term reduced basin-closure error by 21 per cent while held-out error changed by less than initialization variance, and extended training at the selected weight lowered closure error further, to 0.280 centimeters per year. Permutation importance placed the net atmospheric moisture input far ahead of every other predictor, at 7.59 centimeters against 5.37 for air temperature and almost nothing for urban fraction, confirming that the model had learned to rely on the physically primary driver rather than on incidental spatial covariates.

The second objective, validation and uncertainty, was met through an independent monitoring network and a decomposed error budget. Against the GROW well compilation the reconstruction achieved a lower trend error than both a physically based global hydrological model and a GRACE data-assimilation field at every spatial scale tested, and its correlation advantage over the hydrological model widened as aggregation increased. At aquifer scale the reconstruction returned a correlation of plus 0.31 whose confidence interval excluded zero, whereas the hydrological model returned plus 0.10 with an interval that did not, and the reconstruction carried the lowest error against wells at every scale by a factor of between 1.1 and 1.8, with near-zero bias where both comparison products overstated decline. The uncertainty budget separated the observational, model, and specific-yield contributions to the trend error and found specific yield, rather than the satellite observation or the learning model, to be the dominant term. Of the 52,427 cells, 39,686, or 76 per cent, carried trends whose bootstrap intervals excluded zero.

The third objective, the planetary consequences, was met by carrying the reconstruction outward into the mass budget and the Earth's rotation. Gross depletion over 2002 to 2022 was estimated at 357.64 plus or minus 11.6 cubic kilometers per year across 21,632 significantly declining cells, equivalent to 0.989 millimeters per year of sea-level rise, while net storage change, the physically correct quantity for the ocean budget, was minus 36.25 cubic kilometers per year, equivalent to 0.100 millimeters per year. The associated mass redistribution was found to drive a drift of the Earth's rotational pole of 6.92 centimeters per year toward 68.6 degrees east, a result that was verified against a month-by-month simulation of the pole path, computed from the same field, agreeing to within 0.86 per cent in rate and 0.50 degrees in azimuth.

## 4.2 The physical constraint as the central advance

The result that most repays interpretation is the behavior of the physical constraint, because it explains why a learned model was able to produce a field that can be defended as physically real rather than merely well fitted. Purely data-driven models can reproduce an observed record closely and yet extrapolate to states that no water balance would allow, a concern that has accompanied the rapid adoption of machine learning across the Earth sciences (Reichstein et al., 2019). The reconstruction guarded against this by holding the long-term trend of each cell within the net moisture that the atmosphere can supply and by holding the basin-mean trends to observation. The surface-water-balance residual, defined as precipitation minus evaporation minus runoff minus the observed rate of storage change, was found to be persistently negative over the irrigated belts of Asia and the arid subtropics, which is the signature of storage being drawn down faster than the atmosphere can replace it, and therefore of abstraction rather than of climate. This is precisely the quantity that the supply bound limits, so the constraint was not an abstract regularizer but a statement of the physics of the depletion itself.

That both fit and physical realism improved together places the work within the tradition of physics-informed and physics-guided learning, in which process knowledge is embedded so that predictions remain consistent with known physics (Jiang et al., 2020; Read et al., 2019). The choice of a recurrent encoder was equally deliberate, since the long short-term memory network carries the memory of antecedent conditions forward through time and is well suited to the delayed response of storage to earlier moisture supply that these networks capture in other hydrological settings (Kratzert et al., 2018). By separating the reconstruction into a trend, a seasonal cycle, and a residual, the decomposition made each component independently constrainable and interpretable, so that the physics could be applied to the secular term without distorting the seasonal and transient behavior that a single undifferentiated output would have entangled with it.

## 4.3 Reconciling the magnitude of global groundwater depletion

The estimate of gross depletion obtained here, 357.64 cubic kilometers per year, falls between two recent published values, and the way it does so is itself a finding. (Seo et al., 2023) tested a climate-model depletion estimate of about 126.5 cubic kilometers per year (1993–2010) against the observed drift of the pole, while (T. Sun et al., 2026) reported 284 cubic kilometers per year before adjustment against monitoring wells, rising to 550 cubic kilometers per year afterwards. The reconstruction presented here lies between these bounds. Because the estimate of (Seo et al., 2023) spans 1993 to 2010 while this reconstruction spans 2002 to 2022, the ratio between the two rates reflects both the amplitude restoration applied here and the different analysis windows, and is not a like-for-like comparison. A matched-window comparison was therefore made against the data-assimilation product: over their 178-month overlap, a regression over the storage curve gave 368.1 cubic kilometers per year for this study against 232.9 for that product, a ratio of 1.58. The two series were nonetheless very highly correlated, at plus 0.996 for annual storage and with 80.8 per cent sign agreement across significant cells, so they describe the same spatial and temporal pattern of depletion and differ mainly in amplitude.

The origin of that amplitude difference was identified as the correction applied in this study, which restored the storage variability that a squared-error training loss suppresses and whose median expansion factor of 1.30 is of the same order as the discrepancy. Independent support came from (T. Sun et al., 2026), who corrected a satellite-derived depletion field against wells by quantile mapping and found their estimate rise by a factor of roughly 1.94. Two studies, using different gravimetric solutions, different land-surface components, and different correction procedures, therefore converged on the same conclusion, that uncorrected satellite gravimetry understates the rate of groundwater depletion. The estimate presented here should accordingly be read as an amplitude-restored one, and its difference from (Seo et al., 2023) should be understood as a methodological consequence and a difference in analysis window rather than as a disagreement about the underlying hydrology, since the two studies agree on the direction of the mass loss to within 4.5 degrees of azimuth. This interpretation is consistent with the finding that global hydrological models underestimate the largest storage trends relative to GRACE (Scanlon et al., 2018), which was reproduced directly here in the near-zero trend field of the independent model (Müller Schmied et al., 2021) and in the shared observational structure of the data-assimilation product (Gou & Soja, 2024). Earlier statistical reconstructions of total water storage did not confront this amplitude question because they did not isolate groundwater or resolve its depletion centers (Humphrey & Gudmundsson, 2019).

## 4.4 The limits of validation and the role of specific yield

The validation against monitoring wells returned correlations that were modest in absolute terms, and their correct interpretation is important to the defensibility of the whole study. The substantive result was not the height of any single correlation but the pattern across scales: the reconstruction retained its skill as the aggregation scale increased, and improved slightly at the largest scale, whereas the independent hydrological model lost essentially all skill above the individual cell. The modest cell-scale values were shown to reflect a property of the reference data rather than a failure of the product. Within cells holding three or more wells, the robust spread of well trends had a median of 0.90 centimeters per year, and in 69.7 per cent of such cells that internal spread equaled or exceeded the magnitude of the cell trend being estimated. The wells therefore do not agree with one another inside a single half-degree cell any better than they agree with the products, which sets a ceiling on the correlation that any gridded field can attain and reframes the comparison as a test against an irreducibly noisy reference (Bäthge et al., 2026).

Specific yield sat at the center of both the validation and the uncertainty. It is the quantity that converts a measured water-level change into a stored volume, and it has long been recognized as one of the largest and least constrained sources of error in groundwater accounting (Long et al., 2016; Scanlon et al., 2012). Its emergence as the dominant term in the error budget was therefore expected rather than surprising, and the important step was not to remove it but to test it. When the significance of the depletion was re-evaluated at the correlation-area scale, it was found to be stable across the whole plausible range of the specific-yield assumption, even where it was sensitive at the scale of an individual cell. The depletion finding thus rests not on a particular value of a poorly known parameter but on a scale at which that parameter ceases to control the outcome, which is what allows it to be defended.

## 4.5 Planetary consequences of depletion

Two geophysical consequences of the reconstructed mass change were evaluated. The first was the sea-level budget. Net storage change corresponded to 0.100 millimeters per year of sea-level rise, within the span of published net land-water contributions to the ocean budget (Reager et al., 2016; Wada et al., 2017), whereas estimates based on gross depletion alone are larger (Konikow, 2011). Gross depletion measures the extent and severity of aquifer loss and is not a sea-level quantity; the order-of-magnitude difference between the two is the direct consequence of recharge elsewhere on the land surface offsetting losses in the depleting regions, so the two must never be quoted interchangeably or compared across studies.

The second consequence was the drift of the Earth's rotational pole. The inferred direction was robust, since three different mask definitions spanned only 1.1 degrees of azimuth, and a month-by-month simulation of the pole path computed from the same field reproduced the secular estimate to within 0.86 per cent in rate and 0.50 degrees in azimuth, confirming the numerical consistency of the calculation. The excitation was found to concentrate between 20 and 60 degrees east, the interval containing the Indus, Ganges, and Middle Eastern aquifer systems, with secondary contributions from the North China Plain and the western United States, consistent with the role of terrestrial water loss over Eurasia in recent polar motion (Adhikari & Ivins, 2016). The azimuth lay within 4.5 degrees of that inferred by (Seo et al., 2023) from an independent, model-based depletion estimate. That mass loss isolated at the scale of individual aquifers implies a pole drift in nearly the same direction as an independently derived estimate supports the interpretation of the reconstruction as a coherent, planetary-scale process, although the drift rate remains specific to the analysis window and to the amplitude restoration applied here.

## 4.6 Novelty and contribution

With the results in hand, the contribution of the work can be stated in terms of what was demonstrated rather than what the field had lacked. The first contribution was to show that hard physical constraints can be embedded within a learned reconstruction of isolated groundwater storage, and not merely of total water storage, in such a way that physical consistency improves without degrading fit. The constraint behaved as a source of physical consistency rather than as a regularizer that trades accuracy for plausibility, a property that recommends the approach for other underdetermined problems in the Earth sciences.

The second contribution was to resolve an apparent disagreement in the literature about the magnitude of global groundwater depletion. By tracing the difference between this study and a data-assimilation product to a single amplitude factor, and by finding that factor mirrored in an independent well-based correction, the work turned a spread of divergent published numbers into one coherent picture in which uncorrected gravimetry understates the true rate. A methodological insight of this kind, that the disagreement was about amplitude and not about pattern, is more durable than any single number.

The third contribution was a transparent and decomposed uncertainty budget that located the binding constraint on interpretation in the specific yield, tested that assumption directly, and aggregated errors under the measured spatial correlation length rather than under an untenable assumption of independence between neighboring cells. The fourth was to carry a single reconstruction from aquifer storage through to two planetary consequences and to validate the whole against an independent global monitoring network, uniting satellite gravimetry, land-surface physics, in-situ hydrology, and geodesy around one hidden quantity. Where the introduction framed these as gaps to be filled, the results establish them as a demonstrated and internally consistent contribution to how groundwater storage can be reconstructed and defended.

## 4.7 Limitations

Several limitations qualify these conclusions and should be stated plainly. The isolation of groundwater depends on subtracting modelled soil moisture, snow, and canopy storage, so errors in the land-surface model propagate into the recovered signal, a dependence that is shared by every GRACE-based groundwater estimate and is not peculiar to this study (Chen et al., 2016; Rodell et al., 2009). Specific yield is assumed rather than measured at the scale of a satellite footprint, and although its influence was tested and shown not to overturn the depletion finding at the correlation-area scale, it remains the dominant term in the budget and confines the most strongly constrained results to the cells that contain high-quality wells.

The amplitude correction, while fitted only on training months and shown to preserve the sign of the trend in every cell, is a per-cell rescaling calibrated to the observed variance, and the reconstruction inherits the coarse native resolution of the satellite measurement beneath the half-degree grid on which it is expressed. The selection of the basin-conservation weight rested on single-initialization training runs whose held-out error spread was 2.4 per cent, so part of the difference between adjacent weights could reflect initialization variance, and the choice was justified primarily by the basin-closure improvement. Monitoring coverage is uneven and concentrated in North America, India, Europe, and Australia, so validation is strongest in those regions and weakest across much of Africa, central Asia, and South America. Within the five principal depletion hotspots, the spatial correlation of trends against wells (pooled r = +0.37) was lower than that of WaterGAP (+0.58) and the data-assimilation product (+0.56). Cell-to-cell gradients inside a hotspot lie below the few-hundred-kilometer native resolution of GRACE, whereas WaterGAP prescribes abstraction cell by cell and the data-assimilation product is downscaled with high-resolution inputs. The strength of the reconstruction therefore lies in the magnitude and direction of change, where it carried the lowest error and bias at every aggregation scale, rather than in its sub-footprint spatial pattern. Finally, the inferred pole-drift rate was found to be sensitive to the analysis window, rising to 8.145 centimeters per year over the shorter 2002 to 2019 period, even though its direction was not, so the rate should be read as specific to the window from which it was derived.

### 4.8 Implications and recommendations

The practical implications follow from where the depletion is concentrated. The significantly declining signal was dominated by a small number of aquifer systems, the Indus and Ganges plains, the North China Plain, the Middle East and the Caspian margin, and the Central Valley and southern High Plains of the United States, where local rates exceeded several centimeters per year. Management effort directed at these systems would therefore address the large majority of the global signal. The finding that uncorrected satellite products understate the rate of loss has a direct bearing on water-security planning, because it implies that the pace of depletion in these regions has been more rapid than the uncorrected records suggest, which sharpens the case for demand management and for the protection of recharge in the most stressed basins (Bierkens & Wada, 2019; Famiglietti, 2014; Jasechko et al., 2024; Rodell et al., 2018; Taylor et al., 2013).

Three directions would most improve on the work. Measured or better-constrained specific-yield fields would reduce the dominant term in the uncertainty budget more than any other single change, and would tighten the conversion between the mass signal and aquifer storage. Denser and more even monitoring, particularly across Africa and the parts of Asia and South America where wells are sparse, would raise the ceiling that reference noise currently places on validation. Extending the record as the GRACE Follow-On mission continues, and bridging the gap between the two missions with the physically constrained framework developed here, would sharpen the long-term trends on which every downstream estimate depends. The framework itself, a decomposition network with structural physical bounds, is not specific to groundwater and could be transferred to other underdetermined components of the terrestrial water cycle.

## 5. Conclusion

This study reconstructed a continuous, physically bounded, and observationally consistent record of global groundwater storage change for April 2002 to December 2022, and in doing so answered the research question in the affirmative. A physics-constrained decomposition network isolated the groundwater signal from total water storage, held its trend within the limits of the atmospheric water supply, and reproduced withheld satellite observations while satisfying basin-scale mass conservation. The reconstruction outperformed an independent hydrological model and a data-assimilation product against an independent global well network, and its skill held as the aggregation scale increased. Its amplitude, once restored, reconciled a spread of divergent published estimates into a single account in which uncorrected gravimetry understates the true rate of loss. Quantitatively, the work placed global gross groundwater depletion at about 358 cubic kilometers per year, with a net storage change equivalent to 0.10 millimeters per year of sea-level rise and traced its fingerprint to a drift of the Earth's rotational pole of about 6.9 centimeters per year toward 68.6 degrees east, a value reproduced by a month-by-month simulation of the pole path. The uncertainty was decomposed rather than lumped, its dominant term in the specific yield was identified and tested, and the depletion finding was shown to survive across the plausible range of that assumption at the scale at which it is interpreted. Groundwater depletion is thus shown, from aquifer to atmosphere, to be a coherent, quantifiable, and planetary-scale process, and the framework developed here offers a defensible basis on which to continue monitoring it as the satellite record lengthens.

## Author contributions

A.S. conceived the study, developed the methodology, performed the formal analysis, curated the data, produced the visualizations and wrote the original draft. S.K.S. contributed to the conceptualization and methodology, supervised the work, guided the interpretation of the results and reviewed and edited the manuscript. M.H. contributed to the conceptualization, supervised the work, assisted with the evaluation of the manuscript. All authors read and approved the final manuscript.

## Acknowledgements

The authors thank NASA JPL PO.DAAC and GES DISC, ECMWF and the Copernicus Climate Change Service, the GLEAM team, the GROW consortium, the WaterGAP developers, and the developers of GRACE-SeDA for making their data openly available.

## Data availability

The groundwater storage reconstruction produced in this study, comprising the monthly groundwater storage anomaly field, the amplitude-calibrated trends with confidence intervals, the decomposed per-cell uncertainty budget, the two-tier reliability flag, and the derived depletion, sea-level-equivalent, and polar-motion products, on a 0.5° global grid for April 2002 to December 2022, has been deposited in Zenodo and is publicly available at https://doi.org/10.5281/zenodo.23909628 under a CC BY 4.0 license.

All input datasets are publicly available from their original providers, as listed in Table 1. The JPL GRACE and GRACE-FO CRI-filtered mascon solution (RL06.3 V4, short name TELLUS_GRAC-GRFO_MASCON_CRI_GRID_RL06.3_V4) is available from NASA JPL PO.DAAC. GLDAS Noah L4 monthly 0.25° v2.1 is available from NASA GES DISC at https://doi.org/10.5067/SXAVCZFAQLNO. ERA5-Land monthly averaged reanalysis (total precipitation, volumetric soil water layer 1, and total runoff) is available from the Copernicus Climate Data Store at https://doi.org/10.24381/cds.68d2bb30. ERA5 monthly averaged two-metre air temperature is available from the Copernicus Climate Data Store at https://doi.org/10.24381/cds.f17050d7. GLEAM v4.3a monthly actual evaporation is available from https://www.gleam.eu. MODIS MOD13C2 Collection 6.1 vegetation indices are available from NASA LP DAAC at https://doi.org/10.5067/MODIS/MOD13C2.061. GLHYMPS permeability and porosity are available from Borealis at https://doi.org/10.5683/SP2/TTJNIU. The FAO Global Map of Irrigation Areas v5 is available from the Food and Agriculture Organization at https://www.fao.org/aquastat. ETOPO 2022 (60 arc-second surface relief) is available from NOAA NCEI at https://doi.org/10.25921/fd45-gt74. The urban-fraction dataset is available at https://doi.org/10.1038/s41467-020-15788-7. HydroBASINS (level 3, v1c) is available from

https://www.hydrosheds.org. The WHYMAP Groundwater Resources of the World map (2008 edition, 1:25,000,000) is available from the Federal Institute for Geosciences and Natural Resources (BGR) at https://www.whymap.org (source attribution "Datenquelle: WHYMAP, © BGR Hannover & UNESCO Paris"). The Randolph Glacier Inventory v6.0 is available at https://doi.org/10.7265/N5-RGI-60. Aquifer boundaries and annual depth-to-groundwater records are available at https://doi.org/10.5281/zenodo.10003697, associated with https://doi.org/10.1038/s41586-023-06879-8. The GROW monitoring-well compilation is available at https://doi.org/10.5281/zenodo.15149480, associated with https://doi.org/10.1038/s41597-026-06966-1 (released under CC BY-NC-SA 4.0 and used under those terms). The global specific-yield field is available from figshare at https://doi.org/10.6084/m9.figshare.31669225, associated with https://doi.org/10.1038/s41467-026-73521-2. WaterGAP 2.2d (histsoc) output is available as described in https://doi.org/10.5194/gmd-14-1037-2021. The GRACE-SeDA total water storage product is available at https://doi.org/10.1038/s44221-024-00194-w (data: https://doi.org/10.3929/ethz-b-000648738). Coastlines and surface water are from Natural Earth (public domain; https://www.naturalearthdata.com), and country outlines are from the Esri World Countries (Generalized) layer, used under Esri terms. Full accession identifiers are given in Table 1.

## Code availability

The code supporting this study (version 1.0.0) is available at https://github.com/argho129/gwsa-pcdlstm and archived in Zenodo at https://doi.org/10.5281/zenodo.23964933 under the MIT license. The repository contains the complete pipeline: reliability masking, groundwater-signal isolation, predictor screening, the physics-constrained decomposition network and its training, amplitude calibration, the validation workflow, the uncertainty decomposition, and the derived depletion, sea-level, and polar-motion diagnostics, together with the scripts used to produce the figures and tables in this work. Analyses were performed in Python 3.11.1.

## Annex

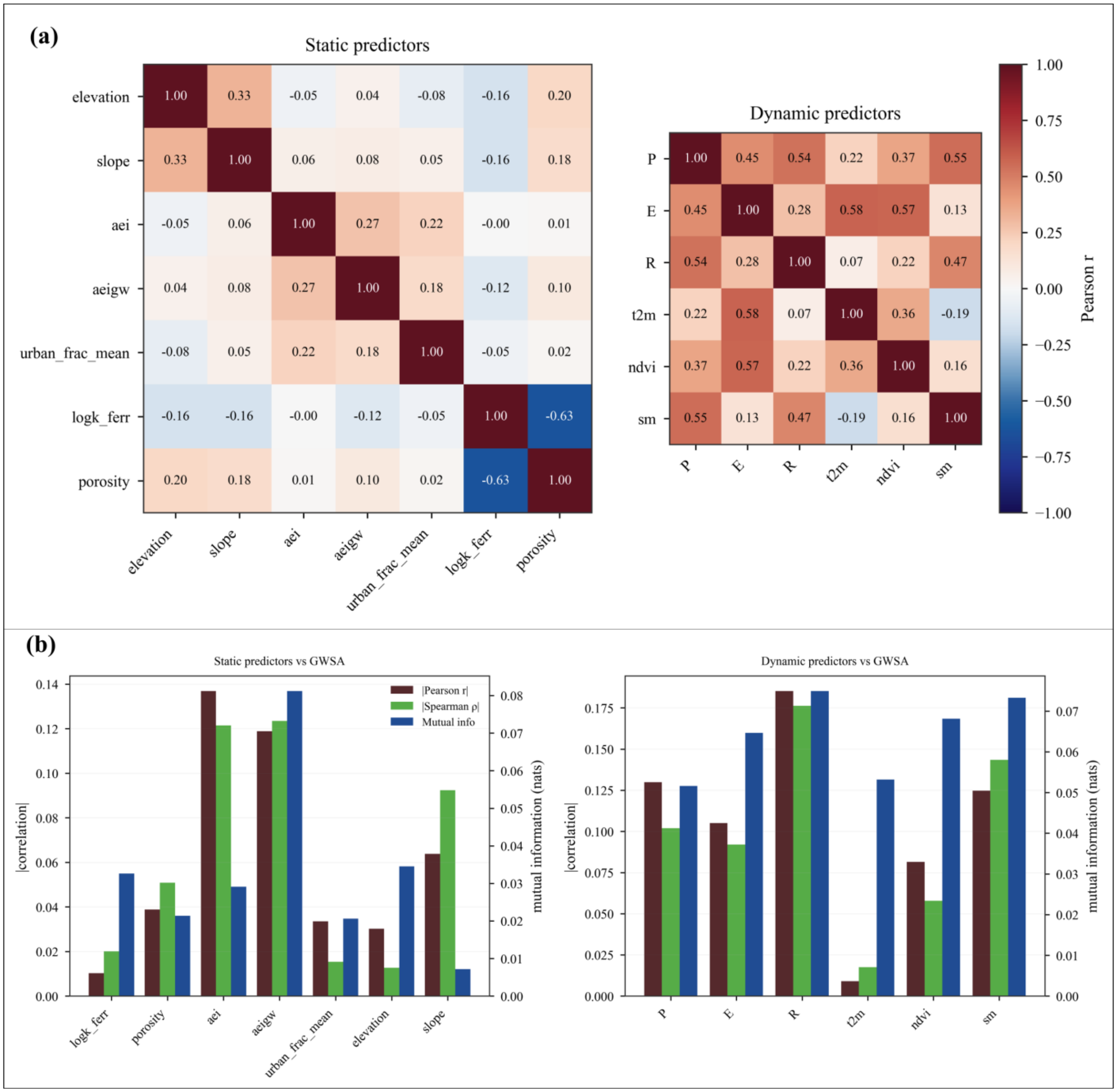

**Figure A. 1: Predictor correlation structure and association with the target.** (a) Pearson correlation matrices for the static predictors, left, and the dynamic predictors, right. Among the static fields the only substantial relationship was the expected inverse coupling of permeability and porosity at minus 0.63. Among the dynamic fields, evaporation correlated with air temperature at 0.58 and with the vegetation index at 0.57, and precipitation correlated with runoff at 0.54 and with soil moisture at 0.55. The static matrix shows the seven fields carried into the model. (b) Association of each predictor with groundwater storage anomaly measured as absolute Pearson

correlation, absolute Spearman correlation, and mutual information. Among the static predictors, irrigated area and groundwater irrigated fraction returned the strongest associations by all three measures. Among the dynamic predictors, runoff and soil moisture ranked highest, and air temperature returned the weakest linear correlation but a substantial mutual information value, which indicated a relationship that was real but not monotonic.

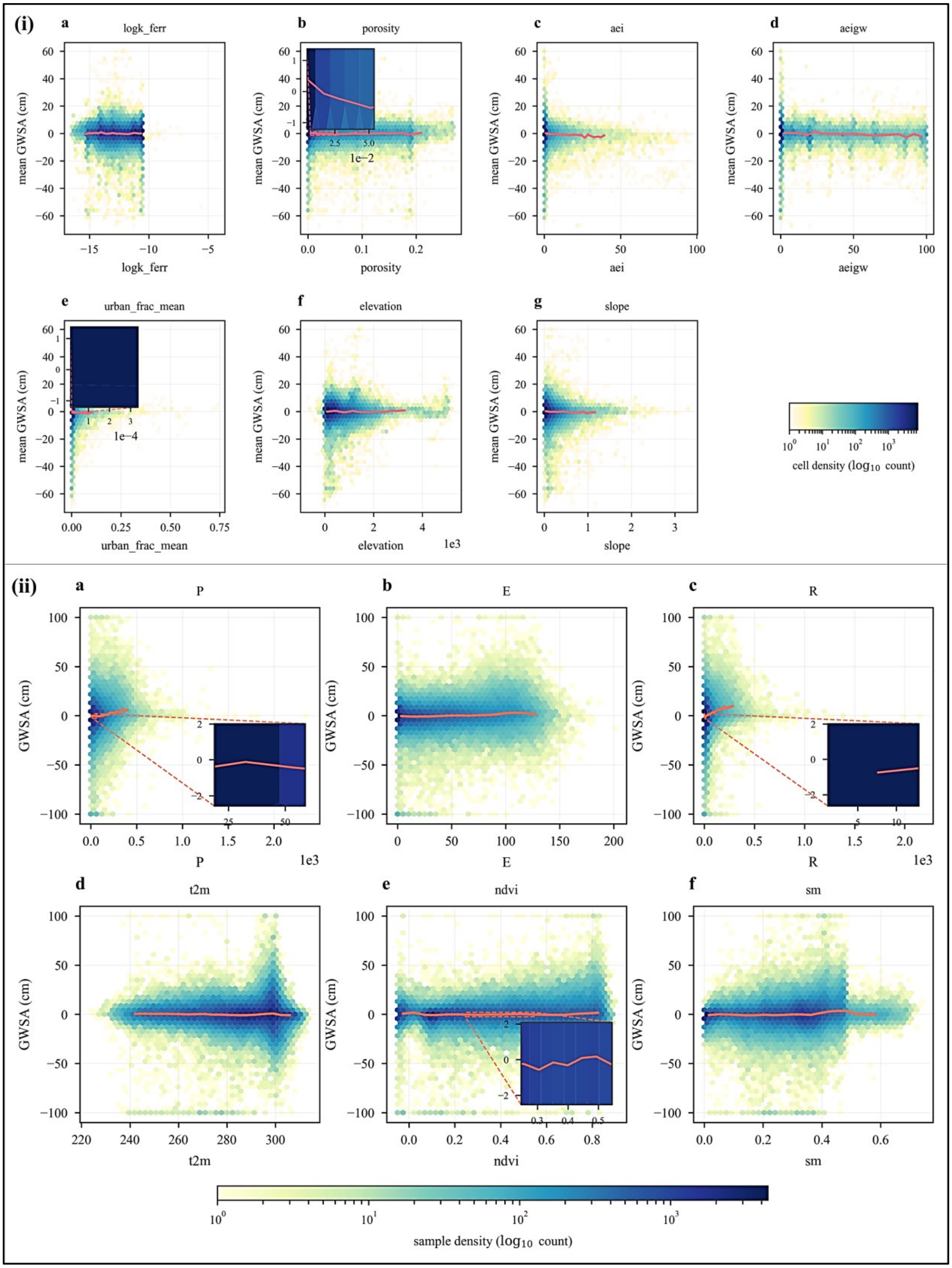

**Figure A. 2: Bivariate density of groundwater storage anomaly against each predictor.** (i) Static predictors, plotted as mean groundwater storage anomaly per cell against each static field. (ii) Dynamic

predictors, plotted for all cell months. Color gives sample density on a logarithmic scale and the red line traces the conditional mean, with insets magnifying the dense region near the origin. In every panel the conditional mean varied only modestly across the range of the predictor, and several relationships reversed direction within the plotted range. The absence of any strong monotonic relationship was the justification for the non-linear model.

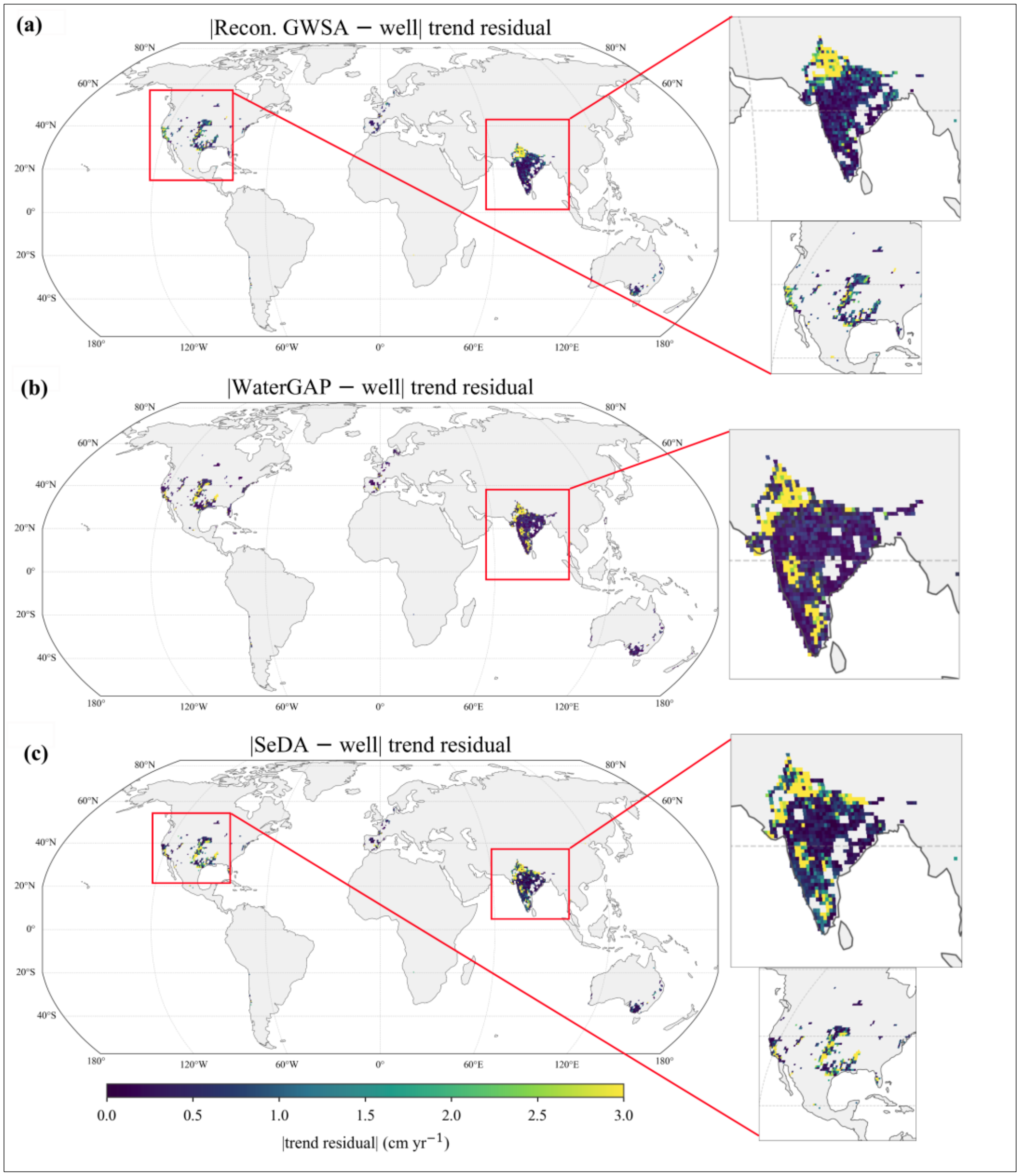

**Figure A. 3: Absolute trend residual against wells.** (a to c) Magnitude of the difference between product and well-derived trend for this study, WaterGAP, and SeDA, with the Indian subcontinent and North America magnified. Residuals concentrated in the two regions where both well density and trend magnitude were highest. This study produced visibly smaller and more spatially confined residuals across the Indian subcontinent than either comparison product, and WaterGAP in particular showed extensive residuals above 2.5 centimeters per year across the western Gangetic plain where this study did not.

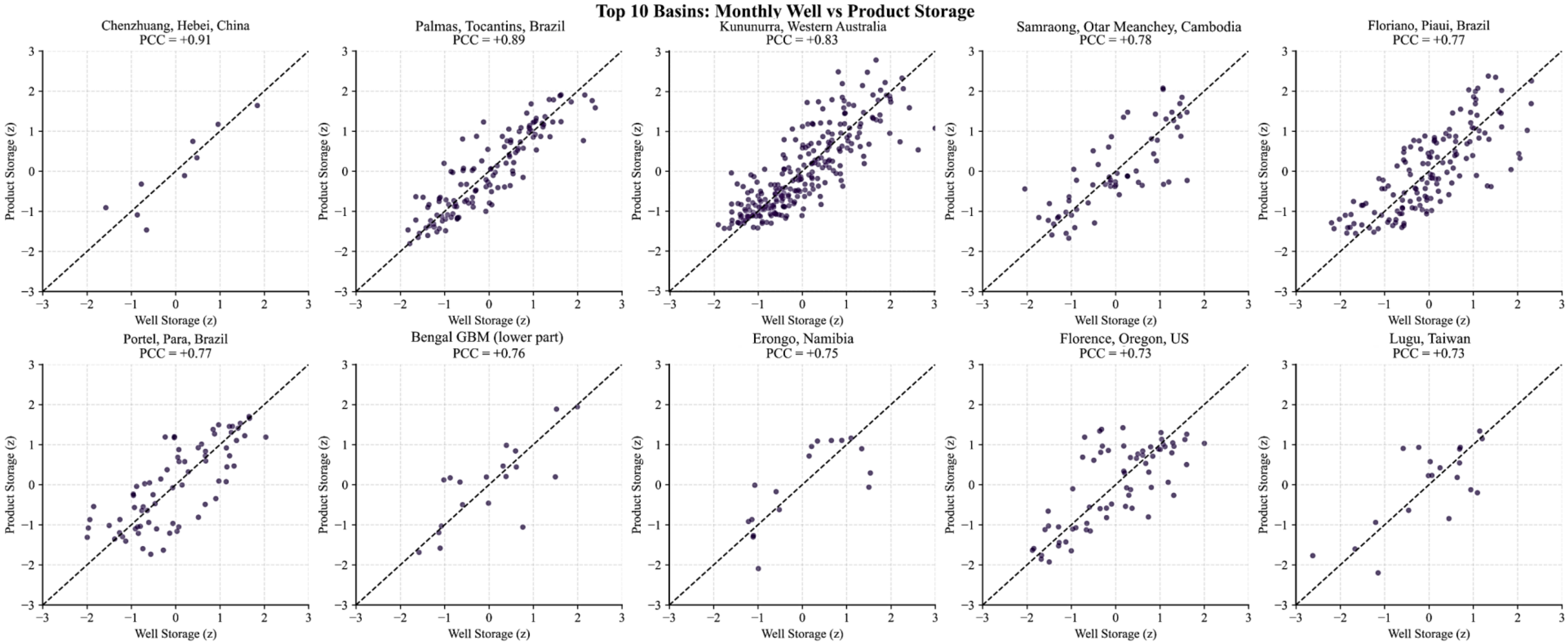

**Figure A. 4: Monthly well storage against reconstructed storage for the ten best performing basins.** Each panel plots standardized polygon-mean well storage against standardized reconstructed storage, with the one-to-one line dashed and the Pearson correlation stated. The ten best performing HydroBASINS level 3 basins returned correlations from plus 0.73 to plus 0.91. These correlations must be read together with the record lengths given in the text.

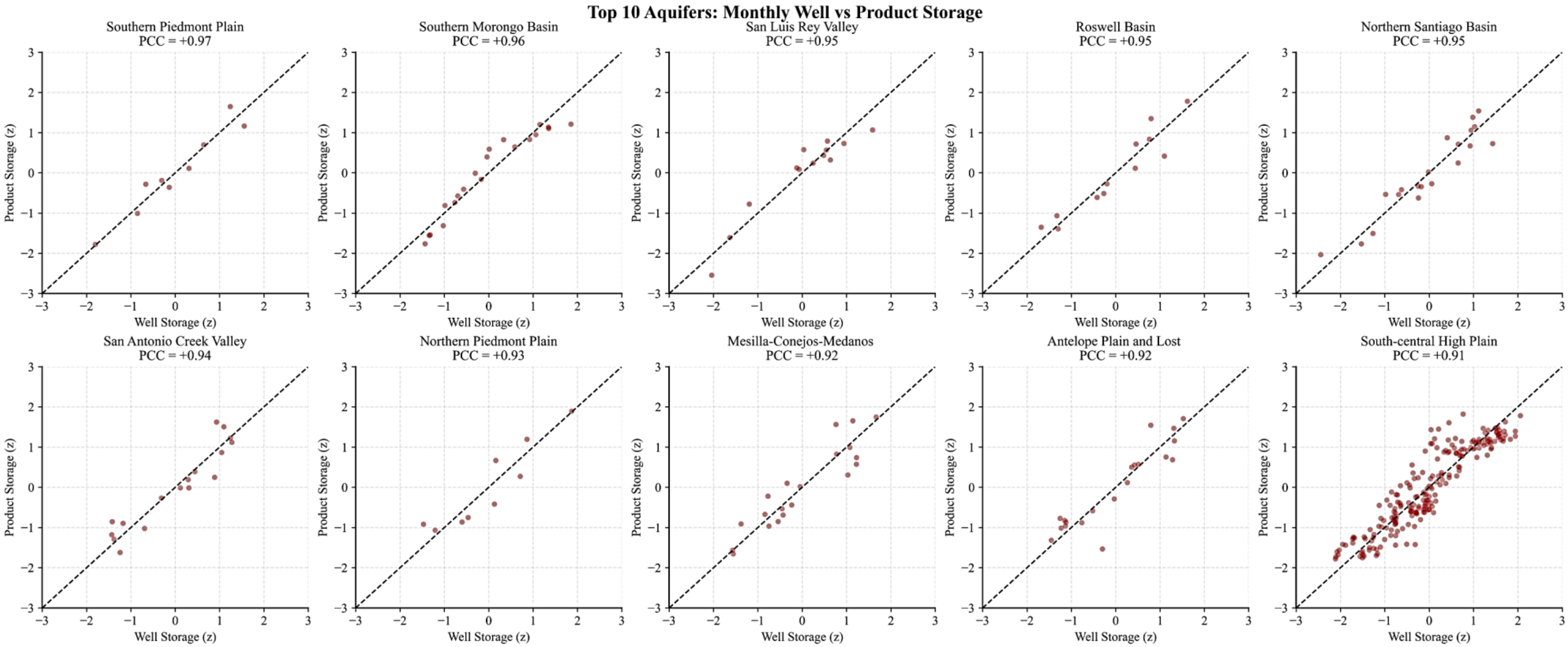

**Figure A. 5: Monthly well storage against reconstructed storage for the ten best performing aquifers.** Panels as in Figure A.4. The ten best performing aquifer polygons, irrespective of area, returned correlations from plus 0.91 to plus 0.97.

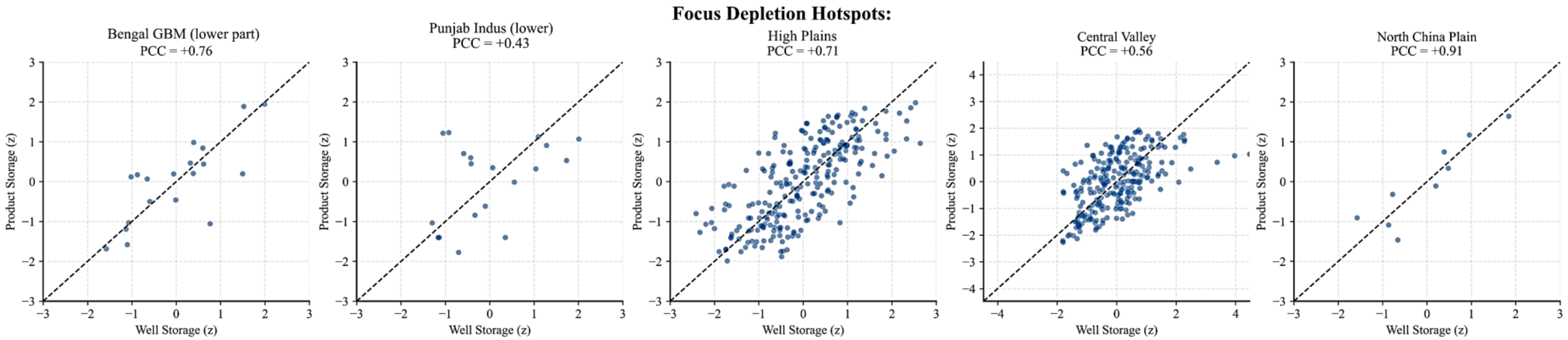

**Figure A. 6: Monthly well storage against reconstructed storage for the five focus depletion regions.** Panels as in Figure A.4. Correlations were plus 0.76 for Bengal, plus 0.43 for Punjab-Indus, plus 0.71 for the High Plains, plus 0.56 for the Central Valley, and plus 0.91 for the North China Plain. These values must be read together with the number of overlapping months, which ranged from 249 in the High Plains and the Central Valley to nine in the North China Plain.

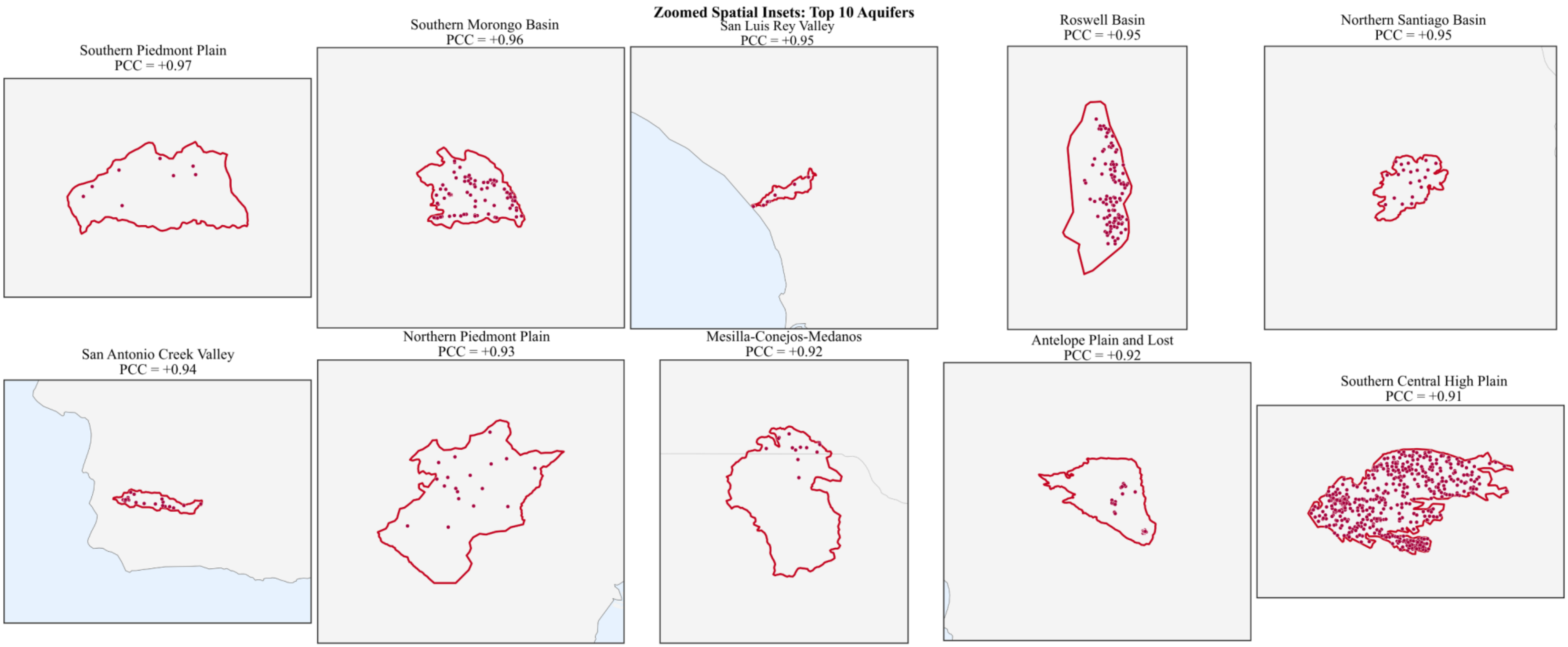

**Figure A. 7: Monitoring site distribution within the ten best performing aquifers.** Sites are plotted within each aquifer polygon of Figure A.5, in the same order. Distributions were spatially spread except where the text notes a sparse network, in which case the correlation should be treated accordingly.

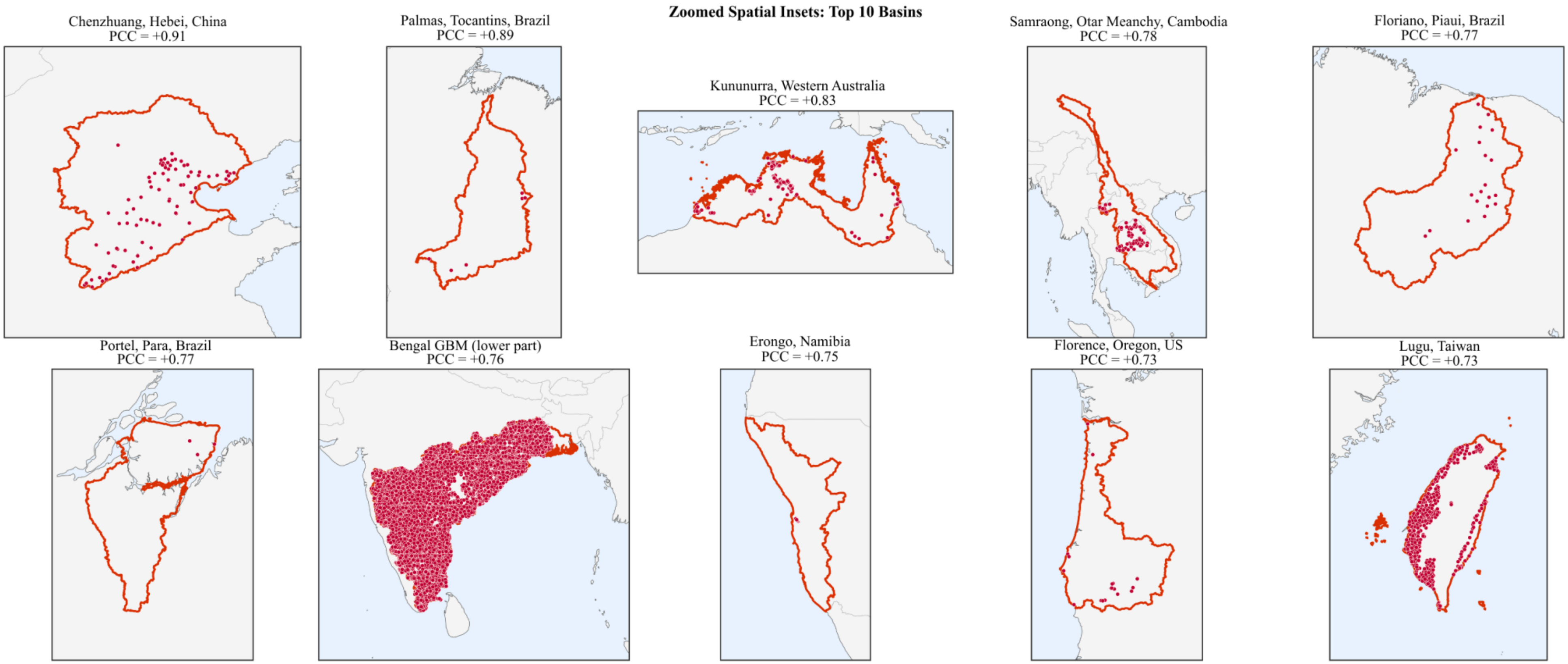

**Figure A. 8: Monitoring site distribution within the ten best performing basins.** Individual monitoring sites are plotted within each basin polygon of Figure A.4, in the same order. The sites were spatially distributed rather than clustered, which confirmed that the reported correlations were supported by a network rather than by a few sites.

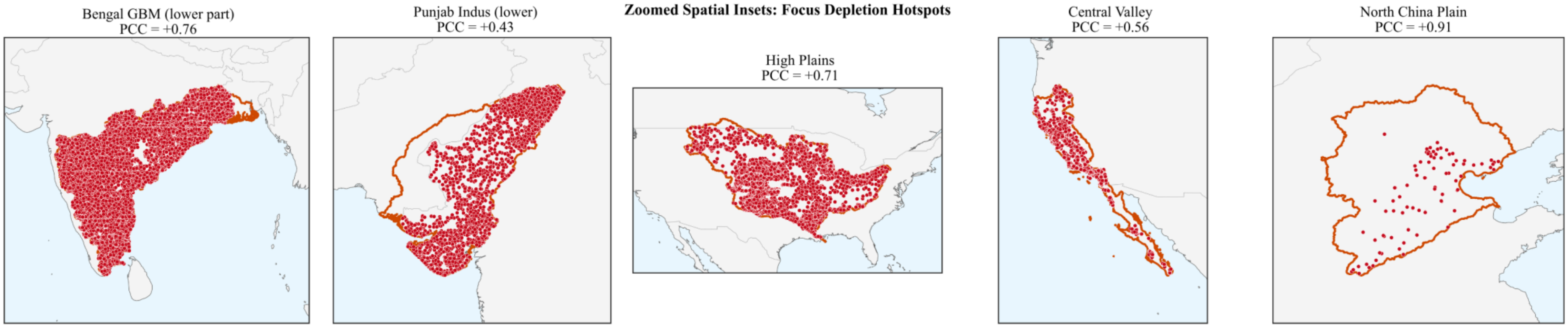

**Figure A. 9: Monitoring site distribution within the five focus depletion regions.** Sites are plotted within each focus polygon of Figure A.6, in the same order, and confirm that the correlations rested on distributed networks.

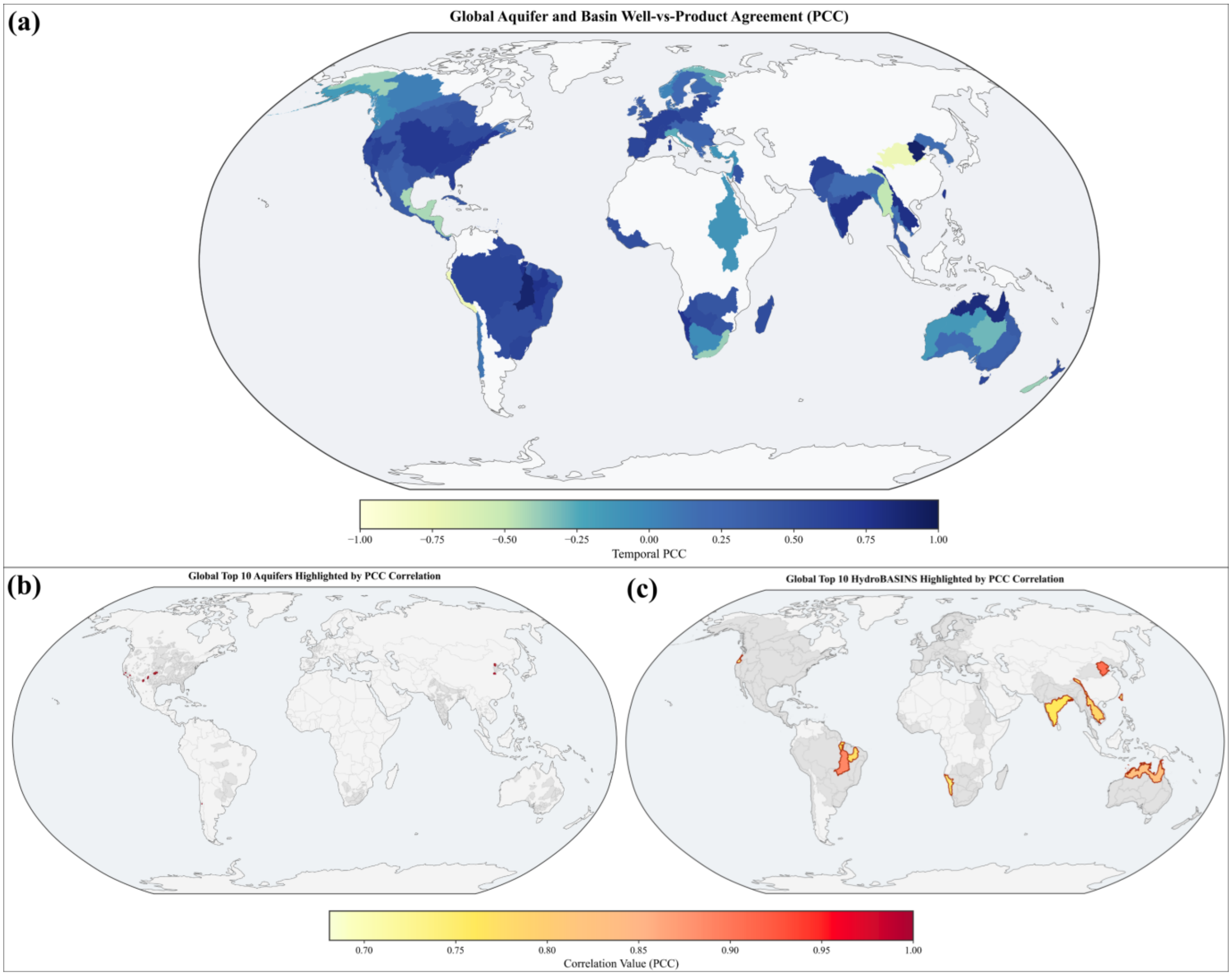

**Figure A. 10: Global distribution of temporal agreement.** (a) Temporal Pearson correlation between polygon-aggregated well storage and reconstructed storage, for every aquifer and basin polygon meeting the minimum well and month requirements. Positive agreement was obtained across the great majority of polygons, with negative values confined to a small number in eastern Asia. (b, c) Location of the ten best performing aquifers and basins, colored by correlation. Their spread across four continents indicated that high temporal agreement was not confined to any single hydrological or climatic setting.

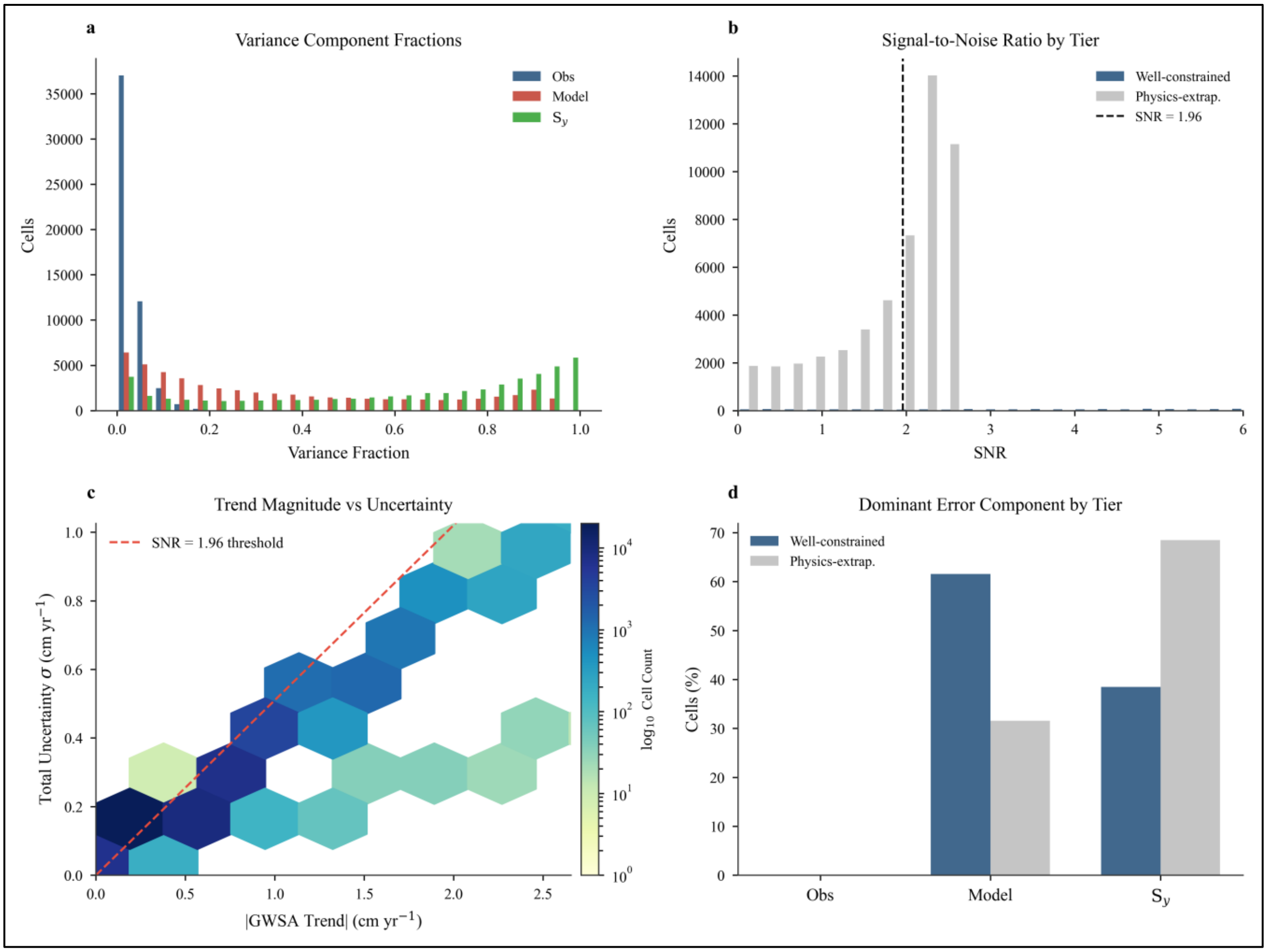

**Figure A. 11: Components and tier diagnostics of the uncertainty budget.** (a) Distribution across cells of the fraction of total trend variance contributed by each component, with the storage-property distribution concentrated at high fractions and the observation distribution near zero. (b) Distribution of the signal-to-noise ratio separated by tier, with the significance threshold of 1.96 marked. Both tiers had median values above the threshold, but the well-constrained tier at 3.85 sat well clear of it while the physics-extrapolated tier at 2.12 did not. (c) Total trend uncertainty against absolute trend magnitude, with the significance threshold shown as a diagonal. (d) Proportion of cells in which each component dominated, by tier. Model error dominated in the well-constrained tier where specific yield was measured, and the storage-property term dominated in the extrapolated tier where it was imputed, while observation error dominated nowhere.

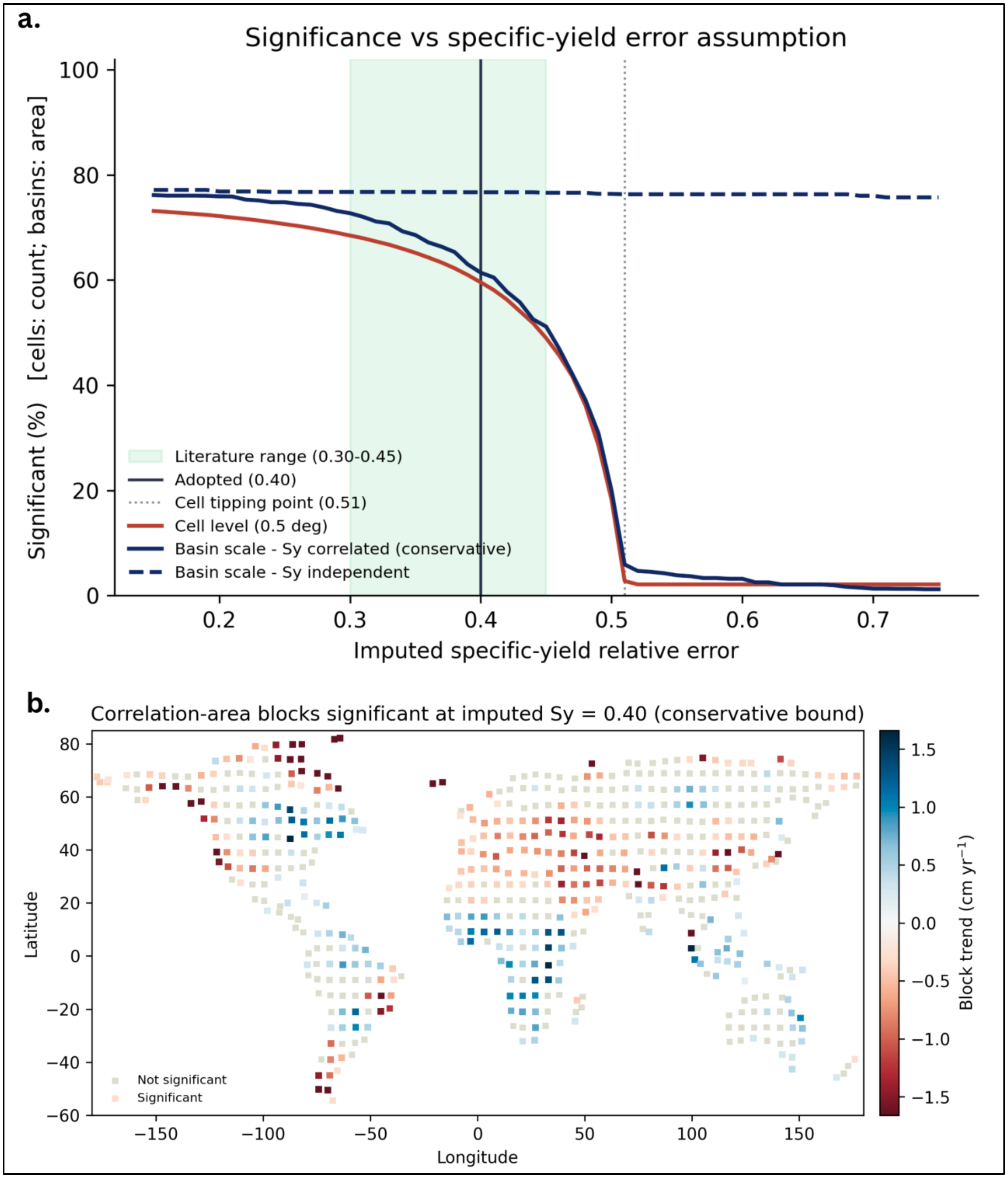

**Figure A. 12: Significance under varying specific-yield error assumption**. (a) Fraction of cells or land area retaining significance as a function of the imputed specific-yield relative error. The cell-level curve fell sharply near a relative error of 0.51, which equals the reciprocal of the significance threshold. The two basin-scale curves bracketed the correlation-area result, the conservative bound assuming the storage-property error to be fully correlated within a block and the independent bound assuming it to be uncorrelated. The adopted value of 0.40 lay inside the shaded literature range and near its upper edge. (b) Correlation-area blocks retaining significance at the adopted value under the conservative bound, colored by block trend.

**Table A. 1: Significant fraction under varying imputed specific-yield relative error.**

| Imputed relative error | Cell level (% of cells) | Basin, correlated (% of area) | Basin, independent (% of area) |
|---|---|---|---|
| 0.25 | 70.6 | 74.8 | 76.8 |
| 0.40 (adopted) | 59.6 | 61.4 | 76.7 |
| 0.55 | 2.1 | 3.8 | 76.3 |